\documentclass[a4paper,11pt]{article}
\pdfoutput=1
\usepackage{jheppub}
\usepackage{graphicx,color}
\usepackage{amsmath,autobreak}
\usepackage{array}
\usepackage{ulem}
\usepackage{float}

\newcolumntype{P}[1]{>{\centering\arraybackslash}p{#1}}
\allowdisplaybreaks
\RequirePackage{ifpdf}
\usepackage{amsmath}
\usepackage{mathtools}

\usepackage{jheppub}
\usepackage[final]{pdfpages}
\usepackage{ifpdf}
\usepackage{slashed}

\usepackage{hyperref}

\usepackage{color}
\usepackage{graphics}

\usepackage{etoolbox}
\usepackage{fixmath}

\usepackage{caption}
\usepackage{subcaption}
\usepackage{amsfonts}

\usepackage{multirow}
\usepackage{epstopdf}

\usepackage{relsize}
\usepackage{float}

\usepackage[table]{xcolor}
\usepackage{tabularx}

\newcommand{\xo}{\mbox{$x_1^0$}}
\newcommand{\xt}{\mbox{$x_2^0$}}

\newcommand{\gt}{\overline{\mathcal{G}}}

\usepackage{tikz}
\usetikzlibrary{positioning,arrows}
\usetikzlibrary{decorations.pathmorphing}
\usetikzlibrary{decorations.markings}
\usetikzlibrary{shapes.geometric}
\usepackage{endnotes}
\def\D{{\cal D}}

\def\ep{\epsilon}
\def\zo{\overline{z}_1}
\def\zt{\overline{z}_2}

\newcommand{\Pzz}{P_{gg}^{(0)}(z_l)}
\newcommand{\Poz}{\mbox{$P_{gg}^{(1)}(z_l)$}}
\newcommand{\Ptz}{\mbox{$P_{gg}^{(2)}(z_l)$}}

\tikzset{
    vector/.style={decorate, decoration={snake}, draw},
    provector/.style={decorate, decoration={snake,amplitude=2.5pt}, draw},
    antivector/.style={decorate, decoration={snake,amplitude=-2.5pt}, draw},
    fermion/.style={draw=black,
      postaction={decorate},decoration={markings,mark=at position .55
        with {\arrow[draw=black]{>}}}},
    fermionbar/.style={draw=black, postaction={decorate},
                       decoration={markings,mark=at position .55 with {\arrow[draw=black]{<}}}},
    fermionnoarrow/.style={draw=black},
    gluon/.style={decorate, draw=black,decoration={coil,amplitude=4pt, segment length=6pt}},
    scalar/.style={dashed,draw=black,
      postaction={decorate},decoration={markings,mark=at position .55
        with {\arrow[draw=black]{>}}}},
    scalarbar/.style={dashed,draw=black,
      postaction={decorate},decoration={markings,mark=at position .55
        with {\arrow[draw=black]{<}}}},
    scalarnoarrow/.style={dashed,draw=black},
    electron/.style={draw=black,
      postaction={decorate},decoration={markings,mark=at position .55
        with {\arrow[draw=black]{>}}}},
    bigvector/.style={decorate, decoration={snake,amplitude=4pt}, draw},
}
\title{Rapidity distribution of pseudo-scalar Higgs boson 
to $\rm{\textbf{NNLO}_A+\overline{\textbf {NNLL}}}$}

\author[a,b]{V. Ravindran,}
\author[a,b,c,d]{Aparna Sankar} 
\author[a,b,e]{and Surabhi Tiwari} 
\affiliation[a]{The Institute of Mathematical Sciences,\\
	IV Cross Road, Taramani, 
	Chennai 600 113, India
       
}

\affiliation[b]{
	Homi Bhabha National Institute,\\ 
	Training School Complex, Anushakti Nagar, 
	Mumbai 400085, India
}
\affiliation[c]{
Physik Department T31, James-Franck-Stra{\ss}e 1,\\ Technische Universit{\"a}t M{\"u}nchen,
D-85748 Garching, Germany
}
\affiliation[d]{
Max-Planck-Institut f{\"u}r Physik, F{\"o}hringer Ring 6,
80805 M{\"u}nchen, Germany
}
\affiliation[e]{

Institut f{\"u}r Theoretische Teilchenphysik, Karlsruhe Institute of Technology
(KIT), Wolfgang-Gaede Stra{\ss}e 1, 76128 Karlsruhe, Germany
}

\emailAdd{ravindra@imsc.res.in}
\emailAdd{aparna@mpp.mpg.de}
\emailAdd{surabhi.tiwari@kit.edu}

\preprint{IMSc-/2022/03, MPP-2023-189, TTP23-040, P3H-23-065}

\abstract{
We present the differential predictions for the rapidity distribution of pseudo-scalar Higgs boson through gluon fusion at the LHC. These results are obtained taking into account the soft-virtual (SV) as well as the next-to-soft virtual (NSV) resummation effects to next-to-next-to-leading- logarithmic ($\rm{\overline{NNLL}}$) accuracy and matching them to the approximate fixed order next-to-next-to-leading- order ($\rm{NNLO_A}$) computation. We perform the resummation in two dimensional Mellin space using our recent formalism \cite{Ajjath:2020lwb} by limiting ourselves to the contributions only from gluon- gluon ($gg$) initiated channels. The $\rm{NNLO_A}$ rapidity distribution of pseudo-scalar Higgs is obtained by applying a ratio method on the NNLO rapidity distribution of the scalar Higgs boson. We also present the first analytical results of $\rm{N^3LO}$  rapidity distribution of pseudo-scalar Higgs at SV+NSV accuracy. The phenomenological impacts of $\rm{{NNLO}_A+\overline{{NNLL}}}$ predictions for 13 TeV LHC are studied. We observe that, for $m_A$ =125(700) GeV, the SV+NSV
resummation at $\rm{ \overline{NNLL}}$ level brings about 14.76\% (11.48\%) corrections to the $\rm{NNLO}_A$ results at the central scale value of $\mu_R=\mu_F=m_A$.
Further, we find that the sensitivity to the renormalisation scale gets improved substantially by the inclusion of NSV resummed predictions at $\rm \overline{NNLL}$ accuracy.}
\begin{document}

\allowdisplaybreaks[4]
\unitlength1cm
\maketitle
\flushbottom
\let\footnote=\endnote
\renewcommand*{\thefootnote}{\fnsymbol{footnote}}

\section{Introduction}
Measurement of a variety of observables to very high precision is one of the thrust areas in the 
physics programme of the Large Hadron Collider (LHC). Precision studies based on these measurements provide 
crucial tests of the consistency of the Standard Model (SM) and any significant deviation can also hint towards new physics beyond SM. 
The discovery of the Higgs boson \cite{Aad:2012tfa,Chatrchyan:2012ufa}, one of the major milestones in particle physics,
led to a better understanding of the dynamics behind the electroweak symmetry breaking \cite{Higgs:1964ia,Higgs:1964pj,Higgs:1966ev,Englert:1964et,Guralnik:1964eu}, and in larger picture, opened up a plenty of opportunities to unravel hidden physics behind various phenomena.      
Despite this sucesses, the SM lacks in fronts in not providing satisfactory explanations for phenomena 
such as baryon asymmetry in the universe, the existence of the dark matter, the neutrino mass etc, and hence falls short of 
being a complete theory of fundamental interactions. Unravelling these phenomena demands one to go beyond the borderline of the SM. 
One of the possible extensions of the SM is the supersymmetric theories which provide an elegant solution to the 
above mentioned problems. Supersymmetric theories generally predict a richer Higgs sector than the Standard Model (SM). 
In the Minimal Supersymmetric Standard Model (MSSM) for instance one introduces two complex Higgs doublets, 
which originate five physical Higgs bosons: two CP-even Higgs bosons (h, H) two charged Higgs bosons (H$^{\pm}$) and, 
finally, a CP-odd (pseudo-scalar) Higgs boson (A)~\cite{Fayet:1974pd, Fayet:1976et, Fayet:1977yc,Dimopoulos:1981zb, Sakai:1981gr, Inoue:1982pi, Inoue:1983pp,Inoue:1982ej}. 

Ever since the Higgs boson  was discovered at the LHC ~\cite{Aad:2012tfa,
  Chatrchyan:2012xdj}, there exists curiosity among the high energy physics community
to understand whether it is the Higgs boson of the SM or not.  
This leads to a physics program aiming at probing
its interaction with other SM particles  with extreme precision that will determine its properties.
This can shed light on whether
the discovered Higgs boson is the scalar or pseudo-scalar Higgs bosons of extended models. 
Such a study requires precise predictions for their production cross sections and the decay rates. 
In particular, the production of CP-odd Higgs boson/pseudo-scalar at the LHC has been studied in detail, 
taking into account the higher order QCD radiative corrections, 
owing to similarities with its CP-even counter part. Among other channels, it is desirable to look for pseudo-scalar 
Higgs boson in the gluon fusion through heavy fermions due to its appreciable coupling in the small and moderate $\tan\beta$
in the minimal version of SUSY model, where  $\tan\beta$ is the ratio of vacuum expectation values $v_i, i=1,2$.  
Furthermore, the large gluon flux leads to an enhancement in the cross section.

Perturbative QCD (pQCD) provides the most successful framework to compute the observables that can be 
measured at the LHC. The production of a pseudo-scalar Higgs boson through gluon fusion at leading order suffers from 
large theoretical uncertainties, particularly due to the presence of renormalisation scale $\mu_R$ 
arising from the strong coupling constant. It also contains mild theoretical uncertainties which result 
from the factorisation scale $\mu_F$ in the parton distribution functions. In order to deal with these scale 
uncertainties as well as to uplift the accuracy of theoretical predictions, one has to go beyond the 
wall of leading-order (LO) computations. 

The QCD higher order corrections to the production of CP-even scalar as well as CP-odd pseudo-scalar bosons through gluon fusion are 
known for a long time in the literature. For the case of a scalar Higgs boson, results for the inclusive production cross section are 
available up to next-to-next-to-next-to-leading order(N$^{3}$LO) QCD~\cite{Anastasiou:2002yz,Harlander:2002wh,Ravindran:2003um,
 Anastasiou:2015ema}, 
within the framework of an effective theory that results from the integration of top quark degrees of freedom.  This leads to  
variety of new interactions of the Higgs boson directly with the gluons~\cite{Ellis:1975ap,Shifman:1979eb,Kniehl:1995tn}.  
On the other hand, for the pseudo-scalar case, only next-to-next-to-leading order(NNLO) QCD results~\cite{Kauffman:1993nv,Djouadi:1993ji,Harlander:2002vv,Anastasiou:2002wq,
  Ravindran:2003um} in the effective theory~\cite{Chetyrkin:1998mw} 
are known. The exact quark mass dependence for scalar and
pseudo-scalar production is known to next-to-leading order(NLO) QCD~\cite{Spira:1993bb,Spira:1995rr}. 
For N$^3$LO predictions of pseudo-scalar cross section, both three loop form factors and real emission contributions are 
required. The computation of form factors
is technically cumbersome\cite{Ahmed:2015qpa} as pseudo-scalar Higgs boson couples to SM fields 
through two composite operators that mix under renormalisation due to the axial anomaly and additionally, 
a finite renormalisation constant needs to be introduced in order to restore the chiral Ward identities. Moreover, these 
operators involve Levi-Civita tensor and $\gamma_5$ which are not very straightforward to define in dimensional 
regularisation. The three loop form factor obtained in \cite{Ahmed:2015qpa} was later combined with suitable soft distribution
function \cite{Ravindran:2005vv,Ravindran:2006cg,Ahmed:2014cla} and mass factorisation kernels for the computation 
of the soft plus virtual (SV) contribution at N$^3$LO in \cite{Ahmed:2015qda}. Later, in \cite{Ahmed:2016otz}, 
a new determination of approximate N$^3$LO pseudo-scalar boson cross section has been introduced, based 
on the N$^3$LO results of scalar Higgs boson \cite{Anastasiou:2016cez}.  

In addition to the the inclusive production cross section, 
the differential rapidity distribution is among the most important observables, which is expected to be 
measured in upcoming days. This demands for very precise theoretical predictions of this observable. 
The computation of the 
transverse momentum and rapidity distributions for the scalar Higgs boson up to NLO has been done 
in \cite{Ravindran:2004mb,deFlorian:1999zd} and for the pseudo-scalar Higgs boson in \cite{Field:2002pb}. 
Both inclusive cross section and differential rapidity distribution get large contributions from logarithms arising from 
certain kinematic regions, thus spoiling the reliability of the fixed order predictions. This usually occurs at 
the threshold region, namely when the mass of pseudo-scalar Higgs boson becomes equal to the partonic center of 
mass energy, due to the presence of soft gluons. Hence, the large logarithms resulting from soft gluons in the 
perturbative series need to be resummed to provide sensible predictions. In the pioneering works  
of Sterman \cite{Sterman:1986aj} and of Catani and Trentadue \cite{Catani:1989ne}, resummation of leading large logarithms 
for the inclusive rates in the Mellin space and also to  differential $x_F$ distribution \cite{Catani:1989ne} using 
double Mellin moments were achieved. Soft gluon resummation of the gluon fusion cross section has been performed 
to next-to-next-to-next-to-leading logarithmic(N$^3$LL) accuracy for the scalar Higgs case \cite{Catani:2003zt,Moch:2005ky,Ravindran:2005vv,Ravindran:2006cg,Idilbi:2005ni,Ahrens:2008nc,deFlorian:2009hc,Bonvini:2014joa,Catani:2014uta} and to next-to-next-to-leading logarithmic(NNLL) accuracy for 
the pseudo-scalar case~\cite{deFlorian:2007sr}. In \cite{Agarwal:2018vus}, the resummed transverse momentum distribution has been calculated up to $\rm {NNLO_A+ NNLL}$ for the pseudo-scalar Higgs boson.

A generic threshold resummation formula valid to N$^3$LL accuracy for 
colour-neutral final states was derived in~\cite{Catani:2014uta}, requiring only the virtual three-loop amplitudes as 
process-dependent input. Exploiting the factorization properties of differential cross sections as well as the renormalisation 
group (RG) invariance, an all order $z$-space formalism was also developed in \cite{Ravindran:2006bu}, to study  
the threshold-enhanced contribution to rapidity distribution of any colorless particle. In \cite{Banerjee:2017cfc}, 
the same formalism \cite{Ravindran:2006bu} was used to study the threshold resummation of rapidity distribution of Higgs boson 
and later  to the Drell-Yan (DY) process \cite{Banerjee:2018vvb}. For different approaches and their applications, 
see \cite{Laenen:1992ey,Sterman:2000pt,Mukherjee:2006uu,Bolzoni:2006ky,Becher:2006nr,Becher:2007ty,Bonvini:2014qga,
Ebert:2017uel,Cacciari:2001cw}.

The resummed predictions played a crucial role to understand the experimental data in the threshold regions.
Besides the threshold enhanced logarithms which are also called as the soft virtual (SV) logarithms, the subleading 
logarithms, called the next-to-soft virtual (NSV) logarithms, are also present in the partonic channels beyond leading order 
in perturbation theory. There have been a surge of interests in the community of theoretical physicists to understand 
the nature of these subleading logarithms by using various methods \cite{ Laenen:2008ux,Laenen:2010uz,Bonocore:2014wua,Bonocore:2015esa,Bonocore:2016awd,DelDuca:2017twk, Bahjat-Abbas:2019fqa,Soar:2009yh,Moch:2009hr,deFlorian:2014vta,Beneke:2018gvs,Beneke:2019mua,Beneke:2019oqx}. Recently, the well-established ideas of collinear factorisation and renormalisation group invariance have been implemented to understand the perturbative structure of NSV logarithms for inclusive processes in \cite{Ajjath:2022kyb,Ajjath:2020ulr}.
Following the same formalism of \cite{Ajjath:2022kyb,Ajjath:2020ulr}, in a series of articles \cite{Ajjath:2020sjk,Ajjath:2021lvg,Ajjath:2021bbm}, we studied variety of inclusive reactions to 
understand the impact of NSV logarithms and found a systematic way to sum them up to all orders in $z$ as well as in the Mellin $N$ spaces. 
We have also studied the perturbative structure of the NSV logarithms in the context of rapidity distributions of DY and Higgs productions 
in \cite{Ajjath:2020lwb} . In addition, for the first time, a procedure to resum them in a systematic manner in the 
double Mellin space beyond the SV accuracy has also been developed in \cite{Ajjath:2020lwb}. Further, we have studied the phenomenological relevance of the NSV resummation in the context of both DY and Higgs rapidity distributions in \cite{Ajjath:2021pre,Ravindran:2022aqr} respectively. For the pseudo-scalar Higgs, the resummed predictions including both SV and NSV are recently available to NNLO+$\rm{\overline{NNLL}}$ accuracy in \cite{Bhattacharya:2021hae} for the inclusive cross section case. However, similar predictions for the differential case is not available in the literature.    

In this article, we explore the role of SV and NSV resummed contributions for the differential rapidity distribution of pseudo-scalar Higgs boson in gluon fusion channel by employing the formalism developed in \cite{Ajjath:2020lwb}. In particular, we compute the SV and NSV resummed terms to $\rm{\overline{NNLL}}$ accuracy in the double Mellin $N$-space. Further, we study the phenomenological impact of adding these resummed predictions to the fixed order results through a matching procedure. 

The paper is structured as follows: The first section deals with the theoretical description for the interaction of a pseudo-scalar Higgs with the QCD particles. In the next section, we discuss the computation of the rapidity distribution at the fixed order level. We explicitly show the computational steps at NLO and discuss a procedure to obtain an approximate NNLO which is denoted as $\rm {NNLO_A}$.
In section \ref{sec:SV+NSV}, we review the framework given in \cite{Ajjath:2020lwb} for computing the SV+NSV contributions to the rapidity distribution of a pseudo-scalar Higgs in gluon fusion. The following subsections are devoted to the discussions on the ingredients that are required to compute the rapidity distribution at SV+NSV approximation to $\rm{{N^3LO}}$ accuracy. In section \ref{MI}, a relation between inclusive cross section and rapidity distribution has been exploited to determine the unknown coefficients of certain logarithms which contribute to the SV+NSV rapidity distribution. The results of SV+NSV rapidity distribution for pseudo-scalar Higgs in gluon fusion at the partonic level are presented up to $\rm{{N^3LO}}$ in section \ref{DNSV}. In section \ref{MorePhi}, we focus on the characteristic structure of the NSV soft-collinear function along with some of its peculiar features. In section \ref{RES}, we review the formalism to resum the NSV logarithms to $\rm{\overline{NNLL}}$ accuracy followed by the phenomenological studies of the resummed predictions in section \ref{NUM}. Finally, we conclude our findings in section \ref{CL}.

\section{Pseudo-scalar Higgs effective field theory}
\label{sec:effL}
We begin with setting up the theoretical framework for our analysis.  
The coupling of a pseudo-scalar Higgs boson with glouns occurs only indirectly through a virtual
heavy quark loop which can be integrated out in the infinite quark
mass limit. The interaction between pseudo-scalar Higgs boson $\chi^{A}$ and the
QCD particles in the infinitely large top quark mass limit can be described by an effective Lagrangian~\cite{Chetyrkin:1998mw} and it is given by
\begin{align}\label{eq:L} {\cal L}^{A}_{\rm eff} = \chi^{A}(x) \Big[ - \frac{1}{8}
  {C}_{G} O_{G}(x) - \frac{1}{2} {C}_{J} O_{J}(x)\Big]
\end{align}
where the two operators are defined as \\
\begin{equation}
  O_{G}(x) = G^{\mu\nu}_a \tilde{G}^{\rho
    \sigma}_a \equiv  \epsilon_{\mu \nu \rho \sigma} G^{\mu\nu}_a G^{\rho
    \sigma}_a\, ,\qquad
  O_{J}(x) = \partial_{\mu} \left( \bar{\psi}
    \gamma^{\mu}\gamma_5 \psi \right)  \,.
  \label{eq:operators}
\end{equation}
Here, $G_{\mu\nu}^a=\partial_\mu G_\nu^a-\partial_\nu G_\mu^a
+g_sf^{abc}G_\mu^bG_\nu^c$ is the colour-field-strength tensor and
$\tilde G^{a\mu\nu}=\epsilon^{\mu\nu\rho\sigma}G_{\rho\sigma}^a$ is its dual with $\epsilon_{\mu \nu \rho \sigma}$ being the Levi-Civita tensor;
$G_\mu^a$ are the gluon fields, $g_s=\sqrt{4\pi\alpha_s}$
is the QCD gauge coupling, and $f^{abc}$ are the structure constants of the
SU($N$) algebra. The symbols $\psi$ and  $\bar \psi$ represent fields related to quarks and anti-quarks, respectively.
The Wilson coefficients $C_{G}$ and $C_{J}$ of the two operators originate from 
integrating out the heavy quark loop in effective
theory. The coefficient $C_{G}$ does not receive any QCD corrections beyond one loop
due to Adler-Bardeen theorem~\cite{Adler:1969gk}, whereas $C_{J}$ starts only at
second order in the strong coupling constant.  These Wilson
coefficients are given by~\cite{Chetyrkin:1998mw}
\begin{align}
\label{eq:const}
  C_{G} &= -a_{s} 2^{\frac{5}{4}} G_{F}^{\frac{1}{2}}
          {\rm cot}\beta\,, 
          \nonumber\\
  C_{J} &= - \left[ a_{s} C_{F} \left( \frac{3}{2} - 3\ln
          \frac{\mu_{R}^{2}}{m_{t}^{2}} \right) + a_s^2 C_J^{(2)} + \cdots \right] C_{G}\, ,
\end{align}
where $a_s=g_s^2/16 \pi^2$.
Here, $G_{F}$ denotes the
Fermi constant, ${\rm cot}\beta$ is the mixing angle in the
Two-Higgs-Doublet model and $C_F$ is the quadratic Casimir in the fundamental representation of QCD. The symbols $m_{A}$ and $m_{t}$ stand for the masses of
the pseudo-scalar Higgs boson and top quark (heavy quark), respectively. 
The bare strong coupling constant in the regularized theory is denoted by $\hat {a_s}$ which is
related to its renormalised counter-part by
%
\begin{align}
  \label{eq:asAasc}
  {\hat a}_{s} S_{\epsilon} = \left( \frac{\mu^{2}}{\mu_{R}^{2}}  \right)^{\epsilon/2}
  Z_{a_{s}} a_{s}
\end{align}
where $S_\epsilon = \exp \big(\frac{\epsilon}{2}\big[\gamma_E -\ln(4 \pi) \big]\big)$ with $\gamma_E$ being the Euler-Mascheroni constant. In the above expression, $\mu_{R}$ is the mass scale at which the strong coupling constant is renormalised. The scale $\mu$ is introduced to keep the unrenormalised strong
coupling constant dimensionless in $d=4+\epsilon$ space-time
dimensions. The renormalisation constant $Z_{a_{s}}$ up to
${\cal O}(a_{s}^{3})$ is given by
\begin{align}
  \label{eq:Zas}
  Z_{a_{s}}&= 1+ a_s\left[\frac{2}{\epsilon} \beta_0\right]
             + a_s^2 \left[\frac{4}{\epsilon^2 } \beta_0^2
             + \frac{1}{\epsilon}  \beta_1 \right]
             + a_s^3 \left[\frac{8}{ \epsilon^3} \beta_0^3
             +\frac{14}{3 \epsilon^2}  \beta_0 \beta_1 +  \frac{2}{3
             \epsilon}   \beta_2 \right]\,.
\end{align}
The coefficient of the QCD $\beta$ function $\beta_{i}$ \cite{Tarasov:1980au} are given in Appendix \ref{App}.
%


\section{Fixed order formalism}
\label{sec:Fixed}
The rapidity distribution for the production of a pseudo-scalar Higgs boson at the hadron colliders can be computed using  
\begin{align*}

\frac{d}{dY}\sigma^A\left(\tau,m_A^2,Y\right) &= \sigma^{A,(0)}(\tau,m_A^2) W(\tau,m_A^2,Y) 
\end{align*}
with
\begin{align}
\label{eq:rapidity-dist}
W &= \sum_{a,b=q,{\bar q},g} \int_{0}^1 dx_1 \int_{0}^1 dx_2 \hat f_a\left(x_1 \right) \hat f_b\left(x_2 \right)   
    \int_0^1 dz\delta(\tau-z x_1 x_2) 
    \times \int  \left[ dPS_{1+m}\right] \left| \overline{\cal M}_{ab} \right|^2 \nonumber\\&
  \times   \delta\left( Y-\frac{1}{2} \ln \left( \frac{P_2 \cdot q}{P_1 \cdot q}\right)\right) \,.
\end{align}
Here, $\sigma^{A,(0)}(\tau,m_A^2)$ is the born cross section corresponding to the leading order(LO) process: $g(p1)+g(p2) \rightarrow A(q)$ at the parton level. The hadronic scaling varible $\tau$ is defined by $\tau \equiv m_A^2/S$ where $S$ is the square of hadronic centre of mass energy and the dimensionless variable $z$ is defined as $z \equiv m_A^2/\hat{s}$, where $\hat{s}$ is the square of partonic centre of mass energy. $\hat f_{a(b)}$ are the bare parton distribution functions (PDF) with $x_{1(2)}$ being the  fraction of the initial state hadronic momentum carried by the partons ($a,b$) that take part in the scattering at the partonic level. ${\overline{\cal M}}_{ab}$ denotes the scattering amplitude at the partonic level and the overline signifies the sum and average over all the quantum numbers for the final and initial state particles, respectively. $\left[ dPS_{1+m}\right]$ is the phase space element for the $A+m$-system where the integer $m$ depends on the number of radiated partons. The symbol $Y$ in Eq.(\ref{eq:rapidity-dist}) stands for the rapidity of the pseudo-scalar Higgs boson and it is defined as    
\begin{align}
\label{eq:rapidity}
    Y \equiv \frac{1}{2} \ln \left( \frac{P_2 \cdot q}{P_1 \cdot q}\right)\,,
\end{align}
where  $P_l~(l=1,2)$ is the momenta of incoming hadrons and $q$ denotes the momentum of pseudo-scalar Higgs boson. 
In order to define the threshold limit at the partonic level and to express the hadronic rapidity distribution in terms of the partonic one through convolution integrals, we choose to work
with the symmetric scaling variables $x_1^0$ and $x_2^0$ 
\begin{align}
\label{eq:tau-y-x0}
    Y \equiv \frac{1}{2} \ln\left( \frac{x_1^0}{x_2^0}\right) \quad \text{and} \quad \tau \equiv x_1^0 x_2^0\,.
\end{align}
In terms of these variables, the partonic contributions arising from the subprocesses are found to depend on the ratios 
\begin{align}
\label{eq:scaling-x}
    z_i \equiv \frac{x_i^0}{x_i},~~~ i=1,2 \,,
\end{align}
which play the role of scaling variables at the partonic level.  

The partonic rapidity distribution can be computed, within the framework of perturbative QCD, order by order in strong coupling constant. The contributions arising from beyond leading order contain the UV, soft and collinear divergences. Upon performing dimensional regularisation, the true nature of the UV divergences arises as poles in $\varepsilon$, and such divergences go away when the renormalisation of coupling, masses and fields are performed in modified minimal subtraction ($\overline {MS}$) scheme. 
The UV renormalised partonic rapidity distribution denoted by $\hat \Delta^A_{d,ab}$
is identified as,
\begin{align}
\label{eq:UVcrosssection}
    \frac{1}{x_1 x_2} \hat \Delta^A_{d,ab}\left(z_1,z_2,\hat{a_s},\mu^2,m_A^2,\mu^2_R \right) = \int  \left[ dPS_{1+m}\right] \int_0^1 dz \left| \overline{\cal M}_{ab} \right|^2 
    \times 
    \nonumber \\ \delta(\tau-z x_1 x_2)  
    \delta\left( Y-\frac{1}{2} \ln \left( \frac{P_2 \cdot q}{P_1 \cdot q}\right)\right) \,.
\end{align}
The soft and collinear divergences are collectively called as infrared (IR) divergences. The soft divergences come from zero momentum gluons in the loops of virtual contributions and real gluons in the gluon emission processes. The massless or light partons are responsible for collinear divergences. Due to the KLN theorem \cite{Kinoshita:1962ur,Lee:1964is}, soft and collinear divergences resulting from final state partons cancel independently after summing up contributions from all possible degenerate states. However, the collinear singularities arising from the collinear configurations involving initial state
particles remain. Those are removed at the hadronic level through the technique, known as mass factorisation. The infrared safe partonic rapidity distribution which is also termed as the partonic coefficient function (CF) can be obtained by using the factorisation formula given below 
\begin{align}
\label{eq:cross}
    \Delta^A_{d,ac}\left(z_1, z_2, {a_s(\mu_R^2)}, m_A^2, \mu_F^2, \mu^2_R \right) = \int_{z_1}^1 \frac{dy_1}{y_1} \int_{z_2}^1 \frac{dy_2}{y_2} \Gamma^{-1}_{a a^{'}} \left(\hat{a_s}, \mu^2, \mu^2_F, y_1,\epsilon \right)   \times 
     \nonumber \\  \hat \Delta^A_{d,a^{'}c^{'}} \left(\frac{z_1}{y_1}, \frac{z_2}{y_2}, \hat a_s, \mu^2, m_A^2, \mu^2_R, \epsilon \right)  
    \times \Gamma^{-1}_{c c^{'}} \left(\hat{a_s}, \mu^2, \mu^2_F, y_2,\epsilon \right) \,,
\end{align}
where $\Gamma_{a b}$ are the Altarelli-Parisi \textcolor{black}{(AP) \cite{Altarelli:1977zs}} kernels which essentially absorb the initial state collinear singularities.
The factored out initial state collinear divergences get absorbed into the bare PDFs to give a finite renormalised value at the scale $\mu_F$. Using Eq.(\ref{eq:cross}) and substituting it in Eq.(\ref{eq:rapidity-dist}), we obtain the expression for the rapidity distribution in terms of renormalised PDFs and finite CF as given below.
\begin{align}\label{sighad}
{d \sigma^A\over dY } &=
\sigma^{A,(0)}(\tau, m_A^2) 
\sum_{a,b=q,\overline q,g}
\int_{x_1^0}^1 {dz_1 \over z_1}\int_{x_2^0}^1 {dz_2 \over z_2}~ 
f_{a}\left({x_1^0 \over z_1},\mu_F^2\right)
f_{b}\left({x_2^0\over z_2}, \mu_F^2\right)
\nonumber \\&  \times \Delta^A_{d,ab} (z_1,z_2,m_A^2,\mu_F^2,\mu_R^2).
\end{align}
The perturbative expansion of the infrared safe CF in powers of strong coupling constant read as,
\begin{eqnarray}
\label{expD}
\Delta^A_{d,c}(z_1, z_2, m_A^2, \mu_F^2,\mu_R^2)  &=&
\sum_{i=0}^\infty a_s^i(\mu_R^2) \Delta^{A,(i)}_{d,c}(z_1, z_2, m_A^2, \mu_F^2,\mu_R^2) \,, 
~~ c=q,\bar q, g.
\end{eqnarray}
In this article, we have computed ${d \sigma^A\over dY }$ in Eq. \eqref{sighad} explicitly up to next-to leading order (NLO) for the production of pseudo-scalar Higgs boson in gluon fusion channel, which will be discussed in the following sub-section. We will also discuss how an approximate NNLO which is denoted as $\rm{NNLO_A}$ can be obtained by using the NNLO results of a scalar Higgs boson.
\subsection{Fixed order results at NLO and $\rm{\textbf{NNLO}_A}$} \label{sec:NLO}
Here, we begin with the computation of the rapidity distribution $d \sigma^A\over dY$ given in Eq. \eqref{sighad} at NLO accuracy. Note that the NLO contribution arises from the one loop virtual corrections to born process $g+g \rightarrow A$ and from the real emission processes, namely $g+g \rightarrow A+g$, $q+g \rightarrow A+q$ and $q+\bar{q} \rightarrow A+g$. In order to obtain $d \sigma^A\over dY$ at NLO, the first step is to calculate $\hat \Delta^{A,(i)}_{d,ab}$ given in Eq. \eqref{eq:UVcrosssection} for $i=1$ for the aforementioned processes. One of the ingredients to find  $\hat \Delta^{A,(1)}_{d,ab}$ is the phase space element $dPS_{1+m}$ which is relevant for the processes at NLO.  
For the virtual contributions to the born process, $dPS_{1+m}=dPS_1$ which is given as
\begin{align}
 dPS_1 = \frac{2 \pi}{m_A^2} \delta(1-z_1) \delta(1-z_2)   \,,
\end{align}
and for the real emission processes, we have two body phase space element $dPS_{1+m}=dPS_2$, which takes the following form:
\begin{align}
 dPS_2 = \frac{1}{8 \pi x_1 x_2} \frac{1}{\Gamma(1+{\epsilon \over 2})} \Bigg(\frac{m_A^2}{4 \pi} \Bigg)^{\epsilon/2} \frac{2 z_1 z_2 (1+ z_1 z_2)}{(z_1+z_2)^{2-\epsilon}} \Big( (1-z_1^2) (1-z_2^2) \Big)^{\epsilon/2}\,.
\end{align}
Next, we need to determine the square of the matrix element $|\overline {\cal M}_{ab}|^2$ which is averaged over spin, polarisation and color for the NLO processes mentioned above.  
For the one loop correction to the LO process, we obtain
\begin{align}
\begin{autobreak}

   | \overline{\cal M}|^2_{g+g \rightarrow A-\rm one~loop} =

      a_s \frac{C_A G_F^2 Q^4}{ (N^2-1)} \Bigg[ 
      
      \frac{1}{\epsilon^2}   \bigg\{
          - 2
          \bigg\}

      + \frac{1}{\epsilon}   \bigg\{
          - 1
          - L_{QF}
          \bigg\}

      +  \bigg\{
           \frac{3}{2}
          + \frac{7}{4} \zeta_2
          - \frac{1}{2} L_{QF}
          - \frac{1}{4} L_{QF}^2
          \bigg\}

       +  \epsilon   \bigg\{
          - \frac{5}{4}
          - \frac{7}{12} \zeta_3
          + \frac{7}{8} \zeta_2
          + \frac{3}{4} L_{QF}
          + \frac{7}{8} L_{QF} \zeta_2
          - \frac{1}{8} L_{QF}^2
          - \frac{1}{24} L_{QF}^3
          \bigg\}

       +  \epsilon^2   \bigg\{
           \frac{7}{8}
          - \frac{7}{24} \zeta_3
          - \frac{21}{16} \zeta_2
          - \frac{73}{320} \zeta_2^2
          - \frac{5}{8} L_{QF}
          - \frac{7}{24} L_{QF} \zeta_3
          + \frac{7}{16} L_{QF} \zeta_2
          + \frac{3}{16} L_{QF}^2
          + \frac{7}{32} L_{QF}^2 \zeta_2
          - \frac{1}{48} L_{QF}^3
          \bigg\}

       +  \epsilon^3   \bigg\{
          - \frac{9}{16}
          + \frac{7}{16} \zeta_3
          + \frac{35}{32} \zeta_2
          + \frac{1}{6} \zeta_2 \zeta_3
          - \frac{73}{640} \zeta_2^2
          + \frac{7}{16} L_{QF}
          - \frac{7}{48} L_{QF} \zeta_3
          - \frac{21}{32} L_{QF} \zeta_2
          - \frac{73}{640} L_{QF} \zeta_2^2
          - \frac{5}{32} L_{QF}^2
          - \frac{7}{96} L_{QF}^2 \zeta_3
          + \frac{7}{64} L_{QF}^2 \zeta_2
          + \frac{1}{32} L_{QF}^3
          + \frac{7}{192} L_{QF}^3 \zeta_2
          \bigg\}

       + \epsilon^4   \bigg\{
           \frac{5}{16}
          - \frac{35}{96} \zeta_3
          + \frac{37}{144} \zeta_3^2
          - \frac{49}{64} \zeta_2
          + \frac{1}{12} \zeta_2 \zeta_3
          + \frac{219}{1280} \zeta_2^2
          - \frac{39}{640} \zeta_2^3
          - \frac{9}{32} L_{QF}
          + \frac{7}{32} L_{QF} \zeta_3
          + \frac{35}{64} L_{QF} \zeta_2
          + \frac{1}{12} L_{QF} \zeta_2 \zeta_3
          - \frac{73}{1280} L_{QF} \zeta_2^2
          + \frac{7}{64} L_{QF}^2
          - \frac{7}{192} L_{QF}^2 \zeta_3
          - \frac{21}{128} L_{QF}^2 \zeta_2
          - \frac{73}{2560} L_{QF}^2 \zeta_2^2
          - \frac{5}{192} L_{QF}^3
          - \frac{7}{576} L_{QF}^3 \zeta_3
          + \frac{7}{384} L_{QF}^3 \zeta_2
          \bigg\} \Bigg]\,,
\end{autobreak}
\end{align}
where $L_{QF} = \log(\frac{Q^2}{\mu_F^2})$ and $\zeta_i$ are the Riemann zeta functions.
Here, $Q^2=m_A^2$.

For the real emission process $g+g \rightarrow A+g$, we find
\begin{align}
\begin{autobreak}

   | \overline{\cal M}|^2_{g+g \rightarrow A+g} =

        {a_s \over (\mu_R)^{\epsilon/2}} \frac{ \pi^2 C_A G_F^2}{(N^2-1)} \Bigg[ \frac{1}{D_1 D_2}  \Big(
           8  s^3
         \Big)

       + Q^2  \Big(
           32 
         \Big)

       + \frac{Q^2 }{ D_2}  \Big(
           16  s
         \Big)

       +  \frac{Q^2}{ D_1}  \Big(
           16  s
         \Big)

       + Q^4  \Big(
          -   \frac{16 }{s}
         \Big)

       +  \frac{Q^4}{ D_2}  \Big(
          - 8 
         \Big)

       +  \frac{Q^4}{ D_1}  \Big(
          - 8 
         \Big)

       +  \frac{Q^6}{ D_2}  \Big(
             \frac{8 }{s}
         \Big)

       +  \frac{Q^6}{ D_1}  \Big(
             \frac{8 }{s}
         \Big)

       + D_1  \Big(
          - 8 
         \Big)

       + D_1 Q^2  \Big(
             \frac{8 }{s}
         \Big)

       + D_1^2  \Big(
          -   \frac{8 }{s}
         \Big)

       - 16  s

    +   \epsilon \Bigg\{ \frac{1}{D_1 D_2}  \Big(
           4  s^3
         \Big)

       + Q^2  \Big(
           8 
         \Big)

       +  \frac{Q^2}{ D_2}  \Big(
           8  s
         \Big)

       +  \frac{Q^2}{ D_1}  \Big(
           8  s
         \Big)

       + Q^4  \Big(
          -   \frac{8 }{s}
         \Big)

       +  \frac{Q^4}{ D_2}  \Big(
          - 4 
         \Big)

       +  \frac{Q^4}{ D_1}  \Big(
          - 4 
         \Big)

       +  \frac{Q^6}{ D_2}  \Big(
             \frac{4 }{s}
         \Big)

       +  \frac{Q^6}{ D_1}  \Big(
             \frac{4 }{s}
         \Big)

       + D_1  \Big(
          - 4 
         \Big)

       + D_1 Q^2  \Big(
             \frac{4 }{s}
         \Big)

       + D_1^2  \Big(
          -   \frac{4 }{s}
         \Big)

       - 8  s
         \Bigg\}

       +   \epsilon^2 \Bigg\{ \frac{1}{D_1 D_2}  \Big(
          - 2  s^3
         \Big)

       + Q^2  \Big(
          - 4 
         \Big)

       +  \frac{Q^2}{ D_2}  \Big(
          - 4  s
         \Big)

       +  \frac{Q^2}{ D_1}  \Big(
          - 4  s
         \Big)

       + Q^4  \Big(
             \frac{4 }{s}
         \Big)

       +  \frac{Q^4}{ D_2}  \Big(
           2 
         \Big)

       +  \frac{Q^4}{ D_1}  \Big(
           2 
         \Big)

       +  \frac{Q^6}{ D_2}  \Big(
          -   \frac{2 }{s}
         \Big)

       +  \frac{Q^6}{ D_1}  \Big(
          -   \frac{2 }{s}
         \Big)

       + D_1  \Big(
           2 
         \Big)

       + D_1 Q^2  \Big(
          -   \frac{2 }{s}
         \Big)

       + D_1^2  \Big(
             \frac{2 }{s}
         \Big)

       + 4  s
         \Bigg\}

         +  \epsilon^3 \Bigg\{  \frac{1}{D_1 D_2}  \Big(
            s^3
         \Big)

       + Q^2  \Big(
           6 
         \Big)

       +  \frac{Q^2}{ D_2}  \Big(
           2  s
         \Big)

       +  \frac{Q^2}{ D_1}  \Big(
           2  s
         \Big)

       + Q^4  \Big(
          -   \frac{2 }{s}
         \Big)

       - \frac{Q^4}{ D_2} 

       - \frac{Q^4}{ D_1}  
       
       +  \frac{Q^6}{ D_2}  \Big(
           \frac{1}{s}
         \Big)

       +  \frac{Q^6}{ D_1}  \Big(
           \frac{1}{s}
         \Big)

       - D_1  

       +  D_1 Q^2 \Big(
           \frac{1 }{s}
         \Big)

       + D_1^2 \Big(
          - \frac{1 }{s}
         \Big)

       - 2  s
         \Bigg\}

        +  \epsilon^4 \Bigg\{ \frac{1}{D_1 D_2}  \Big(
          - \frac{1}{2}  s^3
         \Big)

       + Q^2  \Big(
          - 5 
         \Big)

       +  \frac{Q^2}{ D_2}  \Big(
          -  s
         \Big)

       +  \frac{Q^2}{ D_1}  \Big(
          -  s
         \Big)

       + Q^4  \Big(
           \frac{1}{s}
         \Big)

       +  \frac{Q^4}{ D_2}  \Big(
          \frac{1}{2} 
         \Big)

       +  \frac{Q^4}{ D_1}  \Big(
           \frac{1}{2} 
         \Big)

       +  \frac{Q^6}{ D_2}  \Big(
          - \frac{1}{2 s} 
         \Big)

       +  \frac{Q^6}{ D_1}  \Big(
          - \frac{1}{2 s} 
         \Big)

       + D_1  \Big(
           \frac{1}{2} 
         \Big)

       + D_1 Q^2  \Big(
          - \frac{1}{2 s} 
         \Big)

       + D_1^2  \Big(
           \frac{1}{2 s} 
         \Big)

       +  s \Bigg\}   \Bigg]
         \,,

\end{autobreak}
\end{align}

with $D_1 = \frac{-Q^2}{z} (1-z) ~(1-y)$ and $D_2=\frac{-Q^2}{z} (1-z)~ y$ , where
\begin{eqnarray}
z={\xo\xt \over x_1 x_2}\,, \quad \quad
y={x_2 x_2^0 (x_1+x_1^0)(x_1-x_1^0)
 \over (x_1 x_2^0+x_2 x_1^0) (x_1 x_2-x_1^0 x_2^0)}\,.
\end{eqnarray}
For the real emission process $q+g \rightarrow A+q$, we obtain
\begin{align}
\begin{autobreak}

 | \overline{\cal M}|^2_{q+g \rightarrow A+q} =

       {a_s \over (\mu_R)^{\epsilon/2}} \frac{\pi^2 C_F G_F^2}{(N^2-1)} \Bigg[ 

       \big(
          - 4  s
        \big)

      +  \frac{1}{D_1}\big(
          - 8  s^2
        \big)

      +  Q^2 \big(
          4 
        \big)

      +  Q^2 \frac{1}{D_1}\big(
          8  s
        \big)

      +  Q^4 \frac{1}{D_1}\big(
          - 4 
        \big)

      +  D_2 \big(
          4 
        \big)

      + \epsilon \bigg\{ \frac{1}{D_1}\big(
          - 4  s^2
        \big)

      +  Q^2 \big(
          4 
        \big)

      +  Q^2 \frac{1}{D_1}\big(
          4  s
        \big)

      +  Q^4 \frac{1}{D_1}\big(
          - 4 
        \big)

      +  D_2 \big(
           4 
        \big) \bigg\}

      + \epsilon^2 \bigg\{ \big(
           2  s
        \big)

      +  \frac{1}{D_1}\big(
           2  s^2
        \big)

      +  Q^2 \frac{1}{D_1}\big(
          - 2  s
        \big) \bigg\}

      +   \epsilon^3 \bigg\{\Big(
          -  s
        \big)

      +  \frac{1}{D_1}\big(
          -  s^2
        \big)

      +  Q^2 \frac{1}{D_1}\big(
            s
        \big) \bigg\}

      +  \epsilon^4 \bigg\{\bigg(
           \frac{1}{2}  s
        \bigg)

      +  \frac{1}{D_1}\bigg(
           \frac{1}{2}  s^2
        \bigg)

      +  Q^2 \frac{1}{D_1}\bigg(
          - \frac{1}{2}  s
        \bigg) \bigg\} \Bigg]\,,
         
\end{autobreak}
\end{align}
and for $g+q \rightarrow A+q$, we have
\begin{align}

    | \overline{\cal M}|^2_{g+q \rightarrow A+q}  =
         | \overline{\cal M}|^2_{q+g \rightarrow A+q}|_{D_1 \leftrightarrow D_2}\,.
  
\end{align}

Finally, for the real emission process $q+\bar q \rightarrow A+g$, we find 
\begin{align}
\begin{autobreak}

 | \overline{\cal M}|^2_{q+\bar q \rightarrow A+g} =

          {a_s \over (\mu_R)^{\epsilon/2}} \frac{ \pi^2 C_F G_F^2}{N} \Bigg[ 

\big(
           4  s
         \big)

       + Q^2  \big(
          - 8 
         \big)

       + Q^4  \bigg(
           \frac{4 }{s}
         \bigg)

       + D_1  \big(
           8 
         \big)

       + D_1 Q^2  \bigg(
          -   \frac{8 }{s}
         \bigg)

       + D_1^2  \bigg(
             \frac{8 }{s}
         \bigg)

       + \epsilon \bigg\{ \big(
          6  s
         \big)

       + Q^2  \big(
          - 12 
         \big)

       + Q^4  \bigg(
            \frac{6 }{s}
         \bigg)

       + D_1  \big(
           8 
         \big)

       + D_1 Q^2  \bigg(
          -   \frac{8 }{s}
         \bigg)

       + D_1^2  \bigg(
           \frac{8 }{s}
         \bigg) \bigg\}

       +  \epsilon^2 \bigg\{\big(
           2  s
         \big)

       + Q^2  \big(
          - 4 
         \big)

       + Q^4  \bigg(
           \frac{2 }{s}
         \bigg) \bigg\} \bigg] \,.
\end{autobreak}
\end{align}

Now, we can compute the NLO hadronic rapidity distribution ${d\sigma^A \over dY}$ in Eq. \eqref{sighad} by performing the convolution of $\Delta_{d,ab}^A$ for all the contributing processes at NLO discussed above with the corresponding PDFs. Note that $\Delta_{d,ab}^A$ can be obtained by substituting for  $\hat \Delta_{d, a' b'}^A$ in the factorisation formula \eqref{eq:cross} which requires the phase space element as well as the matrix element square computed above according to Eq. \eqref{eq:UVcrosssection}. We express the NLO rapidity distribution at the hadronic level as


\begin{align}
\frac{d\sigma^{A,\rm NLO}}{dY} = \sigma^{A,(0)} (\tau, m_A^2) \bigg\{\frac{d\sigma_{gg}^{A,(0)}}{dY} +a_s \frac{d\sigma_{gg}^{A,(1)}}{dY} +a_s \frac{d\sigma_{qg}^{A,(1)}}{dY} + a_s \frac{d\sigma_{gq}^{A,(1)}}{dY} + a_s \frac{d\sigma_{q \bar q}^{A,(1)}}{dY}   \bigg\} \,,
\end{align}
where
\begin{align}
 \frac{d\sigma_{gg}^{A,(0)}}{dY} = H_{gg}(x_1^0, x_2^0 , \mu_F^2) = f_g(x_1^0,\mu_F^2)~ f_g(x_2^0,\mu_F^2)\,.
\end{align}
The full result of ${d\sigma^{A,\rm NLO} \over dY}$ is provided in Appendix \ref{App:NLORes}. Here, we would like to mention one interesting observation of the NLO result of pseudo-scalar Higgs rapidity distribution. As already noted in \cite{Anastasiou:2002qz}, the NLO result of rapidity distribution of pseudo-scalar is related to that of scalar Higgs at partonic level as follows: 
\begin{eqnarray}
\Delta_{d,gg}^{A,(1)} =& \Delta_{d,gg}^{H,(1)}  + a_s~ 8 C_A~ \delta(1-z_1)\delta(1-z_2) \,, \nonumber \\ &
\Delta_{d,qg}^{A,(1)} = \Delta_{d,qg}^{H,(1)}  \,, \quad 
\Delta_{d,q \bar q}^{A,(1)} =\Delta_{d,q \bar q}^{H,(1)} \,,
\end{eqnarray}
where $H$ stands for the scalar Higgs boson. From a detailed analysis of the above results, it has been found that the difference in the rapidity distribution of scalar and pseudo-scalar Higgs at NLO for $gg$-channel arises only from the one loop virtual contribution to the respective born processes which is also termed as one loop form factor(FF). Note that the FFs of both scalar and pseudo-scalar Higgs bosons are different due to the presence of different Wilson coefficients corresponding to $ggA$(pseudo-scalar) and $ggH$(scalar) vertices in the Higgs effective field theory\cite{Chetyrkin:1998mw}. However, we note that the full hadronic NLO rapidity distribution of pseudo-scalar Higgs can be correctly obtained from that of scalar Higgs by using the formula given below
\begin{align}\label{eq:AH}
 \frac{d\sigma^{A,\rm NLO}}{dY} = {\cal R_{AH}} \times  \bigg( \frac{\sigma^{A,(0)}}{\sigma^{H,(0)}} \bigg) \times \frac{1}{C_H^2} \frac{d\sigma^{H,\rm NLO}}{dY} \,,
\end{align}
with
$\sigma^{H,(0)}$ being the born cross section for scalar Higgs and $C_H$ is the Wilson coefficient for $ggH$ effective vertex \cite{Chetyrkin:1997iv}. In the above formula, ${\cal R_{AH}}$ is the ratio of the modulus square of the finite form factors ($FF$) corresponding to pseudo-scalar and scalar Higgs \cite{Anastasiou:2015vya, Ahmed:2015qpa}, that means ${\cal R_{AH}}=\frac{|FF_A|^2}{|FF_H|^2}$. We provide the expression of the ratio factor ${\cal R_{AH}}$ up to $a_s^3$ below
\begin{align}\label{eq:Ratio}
 {\cal R_{AH}} &=
     \Bigg[ 1 + a_s \Big\{8 C_A \Big\} 

       + a_s^2  \bigg\{ n_f C_F   \bigg(  - 31 + 12 L_{rmt}- 4 ~ L_{qr} \bigg)

       +  C_A n_f   \bigg(  - \frac{2}{3}
      - \frac{4}{3} L_{qr} \bigg)
 \nonumber\\& 
       +  C_A^2   \bigg( \frac{215}{3} - \frac{20}{3} L_{qr} \bigg) \bigg\}

       + a_s^3  \bigg\{ n_f   \bigg(  - 4 C_J^{(2)} \bigg)

       +n_f C_F^2   \bigg( \frac{763}{9}   
       + 4 L_{qr} + \frac{32}{3} \zeta_3 \bigg)
\nonumber\\& 
       +  n_f^2 C_F   \bigg( \frac{4520}{81} - \frac{328}{9} L_{qr} + 16~ L_{qr} ~ L_{rmt}  
       - \frac{8}{3} L_{qr}^2 + 16
         \zeta_2 \bigg)

       +  C_A n_f C_F   \bigg(  - \frac{67094}{81} 
       \nonumber\\&  + 96 L_{rmt}+ \frac{1492}{9} L_{qr}  
       - 88 L_{qr} L_{rmt}
          + \frac{44}{3} L_{qr}^2 + \frac{224}{3} \zeta_3 - 88 \zeta_2 \bigg)

       +  C_A n_f^2   \bigg(  - \frac{631}{81}   \nonumber\\& 
       + \frac{134}{27} L_{qr} - \frac{8}{9} L_{qr}^2 + \frac{16}{3} \zeta_2 \bigg)

       +  C_A^2 n_f   \bigg( \frac{1973}{81} - \frac{838}{27} L_{qr} + \frac{4}{9} L_{qr}^2    
       - 16 \zeta_3 - \frac{8}{3}
         \zeta_2 \bigg)
\nonumber\\& 
       +  C_A^3   \bigg( \frac{68309}{81} - \frac{6028}{27} L_{qr} + \frac{220}{9} L_{qr}^2 - \frac{208}{3} \zeta_3 -
         \frac{440}{3} \zeta_2 \bigg) \bigg\} \Bigg] \,,

\end{align}
where $L_{qr} = \ln \Big(  \frac{q^2}{\mu_R^2}\Big)$ and $L_{rmt} = \ln \Big(  \frac{\mu_R^2}{m_t^2}\Big)$. 

Now we ask the following question: can this ratio factor ${\cal R_{AH}}$ be used for computing the rapidity distribution of pseudo-scalar from that of scalar Higgs beyond NLO accuracy? In \cite{Ahmed:2016otz}, one of the authors of this article had studied the applicability of this ratio method in obtaining the inclusive cross section of pseudo-scalar from that of scalar Higgs beyond NLO. In \cite{Ahmed:2016otz}, it has been established that an approximate NNLO result can be obtained for the inclusive cross section of pseudo-scalar from that of scalar Higgs by using this ratio method. In that case, the difference between the exact and approximate results are found to be in terms of next-to-next-to-soft contributions which are suppressed by $(1-z)^2$ with respect to the leading soft terms that means they vanish in the threshold limit $z \rightarrow 1$. In addition, there are no $\log(1-z)$ terms present in the difference of exact and approximate NNLO results. This suggests that one can compute the inclusive cross section of pseudo-scalar Higgs from that of scalar Higgs by employing the ratio method up to next-to soft terms or NSV terms correctly. In addition, in \cite{Ahmed:2016otz}, it has been shown that the approximate NNLO results provide an excellent approximation to the exact one where the discrepancy is at most $~ 2\%$ for high mass region whereas it is around $1\%$ for the low mass case.             

Drawing inspiration from the above observation for the inclusive case at NNLO, in this article, we attempt to go beyond NLO for uplifting the theoretical accuracy of the predictions for pseudo-scalar Higgs rapidity distribution. We begin with computing the approximate NNLO rapidity distribution of the pseudo-scalar from the exact NNLO result of scalar Higgs available in \cite{Anastasiou:2005qj} by using a formula which is equivalent to Eq. \eqref{eq:AH} for the NNLO case. We denote this approximate NNLO result by $\rm {NNLO_A}$. The analytic expression of this result is too big to be presented in this article, nevertheless, we reserve a section for the detailed numerical analysis of the results we computed. Further, in principle, one can extend the same ratio method discussed here to obtain the approximate results at N$^3$LO for the rapidity distribution of the pseudo-scalar Higgs. \textcolor{black}{However, since the complete N$^3$LO results for the Higgs rapidity distribution \cite{Dulat:2018bfe} are not yet available publicly, it is not possible to compute approximate N$^3$LO results for the rapidity distribution of the pseudo-scalar Higgs using the ratio method mentioned above.} Needless to say, computing the corrections beyond NNLO is not easy and the complexity level of the computation increases significantly which often prevents
us from achieving it. Hence, we resort to an alternate method based on soft-virtual and next-to-soft-virtual approximation \cite{Ajjath:2020lwb} which essentially captures the dominant contribution at the threshold to go beyond the NNLO accuracy, which will be the topic of the next section.     
\section{SV+NSV Formalism} \label{sec:SV+NSV}
The goal of this section is to study the rapidity distribution of pseudo-scalar Higgs in gluon fusion at the soft-virtual(SV) + next-to-soft-virtual(NSV) approximation. To be more precise, we consider the  
contributions to the partonic CF corresponding to the rapidity distribution of a pseudo-scalar Higgs boson in gluon fusion in the limit  $z_l \rightarrow 1$ by keeping only SV and NSV terms, hence we denote them by  $\Delta^{A,SV+NSV}_{d,g} $. Since we restrict ourselves to SV terms namely distributions of the kind $\delta(1-z_i)$ and ${\cal D}_k(z_i)$ and NSV terms
$\log^k(1-z_i)$ for the CF with gluon-gluon initiated channel, the expansion coefficients in Eq. \ref{expD} can be expressed as follows     

\begin{align*}
\label{ExpDel}
\Delta^{A,(i)}_{d,g}&=
\Delta^{A,(i)}_{d,g,\delta \delta} ~ \delta(\zo) \delta(\zt)
+ \sum \Delta^{A,(i)}_{d,g,\delta {\cal D}_j}~ \delta(\zo) {\cal D}_j(z_2)
 + \sum \Delta^{A,(i)}_{d,g,\delta {L}_j}~ \delta(\zo) {L}_j(z_2)
 \nonumber\\ &
+ \sum \Delta^{A,(i)}_{d,g,{\cal D}_j {\cal D}_k}
 {\cal D}_j(z_1)  {\cal D}_k(z_2) 
 +  \sum \Delta^{A,(i)}_{d,g,{\cal D}_j {L}_k}
 {\cal D}_j(z_1)  {L}_k(z_2) + \big( z_1 \leftrightarrow z_2 \big)\,, 
 \end{align*}
 \begin{align}
\text{with}~
{\cal D}_j(z_l)=\Bigg[{\ln^j(1-z_l) \over (1-z_l)}\Bigg]_+ , \delta(\bar z_l) = \delta(1-z_l) ~\text{and}~
{L}_j(z_l) = \log^j(1-z_l) ~ \text{for}~
l=1,2\,.
\end{align}

In \cite{Ajjath:2020lwb}, it has been already shown that, the SV+NSV contributions to the differential distributions arising from diagonal partonic channels which is the gluon-gluon channel in our case, can be factorised  in terms of the overall operator UV renormalization constant $Z_{g}^A$, the bare form factor $\hat{F}_g^A$ (FF), a function ${\cal S}^A_{d,g}$ that is sensitive to real emission contributions and the collinear singular AP kernels $\Gamma_{gg}$. This is always possible as $(Z_{g}^A)^2$ and $|\hat F^A_{g}|^{2}$ are simply proportional to 
$\delta(\zo)\delta(\zt)$ and can be factored out from these partonic channels. Hence, near threshold we obtain, for $\Delta_{d,g}^{A,SV+NSV}$
\begin{align}
\label{eq:normS}
\Delta^{A,SV+NSV}_{d,g} (z_1,z_2,m_A^2,\mu_F^2,\mu_R^2,\epsilon) &=\sigma^{A,(0)}(\mu_R^2) \left(Z^A_{g}(\hat{a}_s,\mu_R^2,\mu^2,\epsilon) \right)^{2} 
|\hat F^A_{g}(\hat a_s,\mu^2,m_A^2,\epsilon)|^{2} \nonumber \\& \times
\delta(\zt) \delta(\zo) 
\otimes {\cal S}^A_{d,g}\left(\hat a_s,\mu^2,m_A^2,z_1,z_2,\epsilon\right) 
\nonumber \\& 
\otimes \Gamma^{-1}_{gg}\left(\hat{a_s}, \mu^2, \mu^2_F, z_1,\epsilon \right)\delta(\zt) \nonumber \\& 
\otimes \Gamma^{-1}_{gg}\left(\hat{a_s}, \mu^2, \mu^2_F, z_2,\epsilon \right)\delta(\zo) \,.
\end{align}
The \textcolor{black}{symbol $\otimes$ refers to convolution, which is defined for functions, $f_i(x_i),i=1,2,\cdot \cdot \cdot,n$, as, 
\begin{eqnarray}
\label{conv}
	\left ( f_1 \otimes f_2 \otimes \cdot \cdot \cdot \otimes f_n\right)(z) 
= \prod_{i=1}^n \Bigg(\int dx_i f_i(x_i)\Bigg) \delta(z - x_1 x_2 \cdot\cdot \cdot x_n ) \,.
\end{eqnarray} }
As long as we are interested in computing the SV+NSV parts of the rapidity distribution, that is those resulting from the phase space region where $z_{1(2)}\rightarrow 1$, we keep only those terms that are proportional to distributions $\delta(\bar z_l)$,
${\cal D}_i(z_l)$ and NSV terms $\log^i(1-z_l)$ with $l=1,2$ and $i=0,1,\cdots$ and drop the rest of the terms resulting from the convolutions. Hence, we have kept only diagonal part of AP kernel $\Gamma_{ab}$ in Eq. \ref{eq:normS} and dropped the non-diagonal AP kernels. In addition, the diagonal kernels get contributions only from the diagonal splitting functions. The reason for the above simplification is due to the fact that the distributions and NSV logarithms can come only from convolutions of two or more distributions or a distribution with NSV logarithms. In summary, since our main focus here is on SV and NSV terms resulting from gluon initiated pseudo-scalar Higgs production, we have dropped contributions from non-diagonal partonic channels in the mass factorised result of $\Delta_{d,g}^A$. All the ingredients in Eq. \ref{eq:normS} that are required to get a finite CF, namely $Z_{g}^A$, $\hat{F}_g^A$, ${\cal S}^A_{d,g}$ and $\Gamma_{gg}$ are known to satisfy certain differential equations with respect to some mass scales \cite{Ajjath:2020lwb,Ravindran:2005vv,Ajjath:2020ulr,Ajjath:2020sjk}. The form of solutions to the respective differential equations which will be discussed in the subsequent sections along with the  well-established ideas of collinear factorisation lead to an all order formula for computing  $\Delta_{d,g}^{A,SV+NSV}$ in $z$-space 

\begin{align}
\label{eq:MasterF}
 \Delta_{d,g}^{A,SV+NSV}(q^2,\mu_R^2,\mu_F^2,z_1,z_2) = \mathcal{C}\exp \bigg( \Psi_{d,g}^A\big(q^2,\mu_R^2,\mu_F^2,z_1,z_2,\epsilon\big)\bigg)\bigg |_{\epsilon=0} \,,
\end{align}
where the function $\Psi_{d,g}^A$ is given by
\begin{align}
\label{eq:Psi}
    \Psi_{d,g}^A\big(q^2,\mu_R^2,\mu_F^2,z_1,z_2,\epsilon\big) &= \bigg( \ln ( Z_{g}^A (\hat{a}_s,\mu^2,\mu_R^2,\epsilon ) )^2   
    + \ln \big| \hat{F}_{g}^A\big(\hat{a}_s,\mu^2,Q^2,\epsilon)\big|^2\bigg) \delta(\zo)\delta(\zt)  \nonumber \\&
 +\mathcal{C} \ln {\cal S}^A_{d,g} \big(\hat{a}_s,\mu^2,q^2,z_1,z_2,\epsilon\big) 
    - \mathcal{C} \ln  \Gamma_{gg}\big(\hat{a}_s,\mu^2,\mu_F^2,z_1,\epsilon\big)\delta(\zt)  \nonumber \\&
    - \mathcal{C}\ln \Gamma_{gg}\big(\hat{a}_s,\mu^2,\mu_F^2,z_2,\epsilon\big)\delta(\zo) \,.
\end{align}
The symbol ``${\cal C}$'' stands for the convolution whose actions on a distribution $g(z_1,z_2)$ is defined as
\begin{align}
\label{eq:Cordered-expoen}
    {\cal C}e^{g(z_1,z_2)} = \delta(1-z_1)\delta(1-z_2)+ \frac{1}{1!} g(z_1,z_2)+ \frac{1}{2!} \left(g \otimes g\right)(z_1,z_2) +\cdots\,,
\end{align}
where $\otimes$ denotes the Mellin convolution. Though the constituents of $\Psi_{d,g}^A$ contain UV and IR divergent terms, the sum of all these terms is finite and is regular in the variable $\epsilon$. It contains the distributions such as $\delta(1-z_l)$, ${\cal D}_i(z_l)$ and the logarithms of the form $\log^i(1-z_l), l=1,2, i=0,1,\cdots$.
\subsection{Operator Renormalisation Constant}
\label{ORC}
Besides coupling constant renormalisation, the form factor also requires the renormalisation
of the effective operators in the effective Lagrangian, Eq.(\ref{eq:L}).
This additional
renormalisation is called the overall operator renormalisation which
is performed through the constant $Z^{A}_{g}$.
In Eq. \ref{eq:normS},  the overall operator renormalisation $Z^{A}_{g}$ is determined by
solving the underlying renormalisation group (RG) equation: 
\begin{align}
  \label{eq:ZRGE}
  \mu_{R}^{2} \frac{d}{d\mu_{R}^{2}} \ln Z^{A}_{g} \left( {\hat a}_{s},
  \mu_{R}^{2}, \mu^{2}, \epsilon \right) = \sum_{i=1}^{\infty}
  a_{s}^{i} \gamma^{A}_{g,i}\,.
\end{align}

Using the results of $\gamma^{A}_{g,i}$ given in Appendix \ref{App} and solving the above RG equation, we obtain the overall renormalisation
constant up to three loop level as
\begin{align}
  \label{eq:ZGG}
  Z^{A}_{g} &= 1 +  a_s \Bigg[ \frac{22}{3\epsilon}
              C_{A}  -
              \frac{4}{3\epsilon} n_{f} \Bigg] 
              + 
              a_s^2 \Bigg[ \frac{1}{\epsilon^2}
              \Bigg\{ \frac{484}{9} C_{A}^2 - \frac{176}{9} C_{A}
              n_{f} + \frac{16}{9} n_{f}^2 \Bigg\}
              + \frac{1}{\epsilon} \Bigg\{ \frac{34}{3} C_{A}^2  
              \nonumber\\
            &-
              \frac{10}{3} C_{A} n_{f}  - 2 C_{F} n_{f} \Bigg\} \Bigg] 
              + 
              a_s^3 \Bigg[   \frac{1}{\epsilon^3} 
              \Bigg\{ \frac{10648}{27} C_{A}^3 - \frac{1936}{9}
              C_{A}^2 n_{f}  + \frac{352}{9} C_{A} n_{f}^2  -
              \frac{64}{27} n_{f}^3 \Bigg\}  
              \nonumber\\
            &+   \frac{1}{\epsilon^2}
              \Bigg\{ \frac{5236}{27} C_{A}^3 - \frac{2492}{27}
              C_{A}^2 n_{f}  - \frac{308}{9} C_{A} C_{F} n_{f}  + 
              \frac{280}{27} C_{A} n_{f}^2  + \frac{56}{9} C_{F}
              n_{f}^2 \Bigg\}
              \nonumber\\
            &  
              +  \frac{1}{\epsilon} \Bigg\{ \frac{2857}{81} C_{A}^3  -
              \frac{1415}{81} C_{A}^2 n_{f}  - \frac{205}{27} C_{A} C_{F} n_{f} + 
              \frac{2}{3} C_{F}^2 n_{f} + \frac{79}{81} C_{A}
              n_{f}^2  + \frac{22}{27} C_{F} n_{f}^2 \Bigg\}
              \Bigg] \, ,
\end{align}
with the SU($N$) QCD color factors
\begin{equation}
  C_A=N,\quad \quad \quad C_F={N^2-1 \over 2 N}\,.
\end{equation}
Here, $n_f$ is the number of active light quark flavors.
It is to be noted that $Z_{g}^{A}=Z_{GG}$ which is given in
Eq.(3.49) of \cite{Ahmed:2015qpa} has been discussed extensively
in~\cite{Ahmed:2015qpa}. 


\subsection{Form Factor}
\label{FF}
The unrenormalised form factor
${F}^{A}_{g}(\hat{a}_{s}, Q^{2}, \mu^{2}, \epsilon)$
satisfies the so-called K + G differential equation \cite{Sudakov:1954sw,
  Mueller:1979ih, Collins:1980ih, Sen:1981sd} which is dictated by the
factorization property, gauge and renormalisation group (RG)
invariances:
\begin{align}\label{KGFF}
Q^2\frac{d}{dQ^2}\ln \hat{F}_g^A \big( \hat{a}_s, Q^2,\mu^2,\epsilon\big)= \frac{1}{2} \Big[ K_g^A \Big(& \hat{a}_s, \frac{\mu_R^2}{\mu^2},\epsilon\Big) 
+ G_g^A \Big( \hat{a}_s,\frac{Q^2}{\mu_R^2},\frac{\mu_R^2}{\mu^2},\epsilon \Big) \Big] \,,
\end{align}
where all poles in the dimensional regulator $\ep$ are contained in
the $Q^{2}$ independent function $K^{A}_{g}$ and the finite
terms in $\epsilon \rightarrow 0$ are encapsulated in
$G^{A}_{g}$. RG invariance of the form factor implies
\begin{align}
\label{RGKG}
	\mu_R^2\frac{d}{d\mu_R^2}K^A_g\Big(\hat{a}_s,\frac{\mu_R^2}{\mu^2},\epsilon \Big) =-
	\mu_R^2\frac{d}{d\mu_R^2}G^A_g\Big(\hat{a}_s,\frac{Q^2}{\mu_R^2},\frac{\mu_R^2}{\mu^2},\epsilon \Big) =  - \sum_{i=1}^{\infty}  a_s^i (\mu_R^2) A^A_{g,i} \,.
\end{align}
The cusp anomalous dimensions $A^{A}_{g,i}$\cite{Moch:2004pa, Vogt:2004mw, Catani:1989ne, Catani:1990rp, Vogt:2000ci} are given in Appendix \ref{App}. Solving the above renormalisation group equation (RGE) satisfied by $K^{A}_g$  we get
\begin{eqnarray}
K^A_g(\hat a_s,\mu^2,\mu_R^2,\epsilon) = \sum_{i=1}^\infty \hat a_s^i
\left({\mu_R^2 \over \mu^2}\right)^{i {\epsilon\over 2}} S_\epsilon^i
K^{A,(i)}_g(\epsilon)
\label{Kbexp}
\end{eqnarray}
with
\begin{eqnarray}
K^{A,(1)}_g(\epsilon)&=& {1 \over \epsilon} \Bigg\{ -2 A_{g,1}^{A}\Bigg\},\quad 
K^{A,(2)}_g(\epsilon)= {1 \over \epsilon^2} \Bigg\{ 2 \beta_0 A_{g,1}^{A} \Bigg\}
+{1 \over \epsilon} \Bigg\{ - A_{g,2}^{A} \Bigg\} \,,
\nonumber\\
K^{A,(3)}_g(\epsilon)&=& { 1 \over \epsilon^3} \Bigg\{ -{8 \over 3} \beta_0^2 A_{g,1}^{A} \Bigg\}
+{1 \over \epsilon^2} \Bigg\{ {2 \over 3} \beta_1 A_{g,1}^{A} 
+{8 \over 3} \beta_0 A_{g,2}^{A}\Bigg\} + {1 \over \epsilon} \Bigg\{ -{2 \over 3} A_{g,3}^{A} \Bigg\} \, .
\end{eqnarray}
Similarly upon solving the RGE in \eqref{RGKG} for $G^A_g$, we obtain
\begin{align}
\label{Gbexp}
 G^A_g (\hat{a}_s, \frac{Q^2}{\mu_R^2}, \frac{\mu_R^2}{\mu^2}, \epsilon ) 
&   = G^A_g (a_s (\mu_R^2), \frac{Q^2}{\mu_R^2}, \epsilon ) 
\nonumber\\
&   = G^A_g (a_s (Q^2), 1, \epsilon ) + \int_{Q^2/\mu_R^2}^{1} \frac{d \lambda^2}{\lambda^2} A^{A}_g (a_s(\lambda^2 \mu_R^2))
\nonumber\\
&   = G^A_g (a_s (Q^2), 1, \epsilon ) + \sum_{i=1}^{\infty} S_{\epsilon}^i \hat{a}_s^i \Big( \frac{\mu_R^2}{\mu^2} \Big)^{i \frac{\epsilon}{2}} 
                                                 \Big[ \Big( \frac{Q^2}{\mu_R^2} \Big)^{i \frac{\epsilon}{2}} - 1 \Big] K^{A,(i)}_g (\epsilon)\, .
\end{align}
We expand the finite function $G^A_g (a_s (Q^2), 1, \epsilon )$ in powers of $a_s(Q^2)$ as
\begin{equation}
 G^A_g (a_s (Q^2), 1, \epsilon ) = \sum_{i=1}^{\infty} a_s^i (Q^2) G_{g,i}^A (\epsilon) \, .
\end{equation}
After substituting these solutions in (\ref{KGFF}) and performing the final
integration,  we obtain the following solution for the form factor 
\begin{eqnarray}
\ln \hat F^A_g(\hat a_s,Q^2,\mu^2,\ep)
=\sum_{i=1}^\infty \hat a_s^i 
\left({Q^2 \over \mu^2}\right)^{i {\ep \over 2}}S^i_{\ep}~ \hat {\cal L}_{g,F}^{A,(i)}(\ep)\,,
\end{eqnarray}
where
\begin{eqnarray}
\hat {\cal L}_{g,F}^{A,(1)}&=&{1\over \ep^2} \Bigg(-2 A_{g,1}^A\Bigg) 
              +{1 \over \ep} \Bigg(G_{g,1}^A(\ep)\Bigg)
\nonumber\\
\hat {\cal L}_{g,F}^{A,(2)}&=&{1\over \ep^3} \Bigg(\beta_0 A_{g,1}^A\Bigg) 
                  +{1\over \ep^2} \Bigg(-{1 \over 2} A_{g,2}^A 
                  - \beta_0  G_{g,1}^A(\ep)\Bigg)
                  +{1 \over 2 \ep} G_{g,2}^A(\ep)
\nonumber\\
\hat {\cal L}_{g,F}^{A,(3)}&=& {1\over \ep^4} \Bigg(-{8 \over 9}\beta_0^2 A_{g,1}^A\Bigg) 
                  + {1\over \ep^3} \Bigg({2 \over 9} \beta_1 A_{g,1}^A 
                    +{8 \over 9} \beta_0 A_{g,2}^A +{4 \over 3} 
                     \beta_0^2 G_{g,1}^A(\ep)\Bigg) 
\nonumber\\
&&                  +{1\over \ep^2} \Bigg(-{2 \over 9} A_{g,3}^A 
                   -{1 \over 3} \beta_1 G_{g,1}^A(\ep) 
                   -{4 \over 3}\beta_0 G_{g,2}^A(\ep)\Bigg)
                  +{1 \over \ep}\Bigg({1 \over 3} G_{g,3}^A(\ep)\Bigg)\,,
\end{eqnarray}
One finds that $G^A_{g,i}$ can be expressed in terms of collinear $B^q_i$ and soft $f^q_i$ anomalous dimensions through the relation \cite{Ravindran:2004mb, Becher:2009cu, Gardi:2009qi}
\begin{eqnarray}
 G^A_{g,i} (\epsilon) = 2 (B^A_{g,i} - \gamma^A_{g,i}) + f^A_{g,i} + \sum_{k=0}^{\infty} \epsilon^k g_{g,i}^{A,k} \,.
\label{Gbexp}
\end{eqnarray}
Note that the single pole term of the form factor depends on three different anomalous dimensions, namely the collinear anomalous dimension $B^A_{g,i}$,
anomalous dimension of the coupling constant $\gamma^A_{g,i}$ and the soft anomalous dimension $f_{g,i}^A$.  $B^A_{g,i}$ can be obtained from the $\delta(1-z)$ part of the diagonal splitting function known up to three loop level \cite{Moch:2004pa, Vogt:2004mw} which are given in Appendix \ref{App}. The $f_{g,i}^A$ for $i=1,2$ can be found in \cite{Ravindran:2004mb} and in \cite{Moch:2004pa} for $i=3$. We list them in Appendix \ref{App}.
The constants $g^{A,0}_{g,i}$ are controlled by the beta function of the strong coupling constant through renormalization group invariance of the bare form factor as 
\begin{equation}
g^{A,0}_{g,1} = 0,\quad \quad  g^{A,0}_{g,2} = - 2 \beta_0 g_{g,1}^{A,1}, \quad \quad
g^{A,0}_{g,3} = - 2 \beta_1 g_{g,1}^{A,1} - 2 \beta_0 ( g_{g,2}^{A,1} + 2 \beta_0  g_{g,1}^{A,2}). 
\end{equation}
Below, we give the expressions of $g^{A,i}_{g,j}$ which are required to calculate the form factor up to $a_s^3$.   
\begin{align}
  \label{eq:g31}
  g^{A,1}_{g,1} &= C_A\Bigg\{ 4 + \zeta_2 \Bigg\}\,,
                  \nonumber\\
  g^{A,2}_{g,1} &= C_A\Bigg\{ - 6 - \frac{7}{3} \zeta_3 \Bigg\}\,,
                  \nonumber\\
  g^{A,3}_{g,1} &= C_A\Bigg\{ 7 - \frac{1}{2} \zeta_2 + \frac{47}{80} \zeta_2^2 \Bigg\}\,,
                  \nonumber\\
  g^{A,1}_{g,2} &= C_A^2 \Bigg\{ \frac{11882}{81} + \frac{67}{3}
                  \zeta_2 - \frac{44}{3} \zeta_3 \Bigg\} 
                  + 
                  C_A n_f \Bigg\{ - \frac{2534}{81} - \frac{10}{3} \zeta_2 -
                  \frac{40}{3} \zeta_3 \Bigg\} 
                  + C_F n_f \Bigg\{ - \frac{160}{3} 
                  \nonumber\\&
                + 12 \ln \left(\frac{\mu_R^2}{m_t^2}\right) + 16 \zeta_3 \Bigg\}\,,
                  \nonumber\\
  g^{A,2}_{g,2} &= C_F n_f \Bigg\{ \frac{2827}{18} - 18
                  \ln \left(\frac{\mu_R^2}{m_t^2}\right)  - \frac{19}{3} \zeta_2 - \frac{16}{3}
                  \zeta_2^2  - 
                  \frac{128}{3} \zeta_3 \Bigg\} 
                  + C_A n_f \Bigg\{
                  \frac{21839}{243}  - \frac{17}{9} \zeta_2 
                  \nonumber\\&
                  +
                  \frac{259}{60} \zeta_2^2  + 
                  \frac{766}{27} \zeta_3 \Bigg\} 
                  + {{C_{A}^2}} \Bigg\{ -
                  \frac{223861}{486}  + \frac{80}{9} \zeta_2 +
                  \frac{671}{120} \zeta_2^2  + 
                  \frac{2111}{27} \zeta_3 + \frac{5}{3} \zeta_2
                  \zeta_3  - 39 \zeta_5 \Bigg\}\,,
                  \nonumber\\ 
  g^{A,1}_{g,3} &=  n_f C_J^{(2)}  \Bigg\{ - 6 \Bigg\}  
                  + {{C_{F}
                  n_{f}^2}} \Bigg\{ \frac{12395}{27}  -
                  \frac{136}{9} \zeta_2  - 
                  \frac{368}{45} \zeta_2^2 - \frac{1520}{9} \zeta_3  -
                  24  \ln \left(\frac{\mu_R^2}{m_t^2}\right) 
                  \Bigg\}  
                  \nonumber\\
                &+ 
                  {{C_{F}^2 n_{f}}} \Bigg\{ \frac{457}{2} + 312 \zeta_3 -
                  480 \zeta_5 \Bigg\}  
                  + 
                  {{C_{A}^2 n_{f}}} \Bigg\{ - \frac{12480497}{4374} -
                  \frac{2075}{243} \zeta_2  - \frac{128}{45} \zeta_2^2 
                  \nonumber\\
                &- 
                  \frac{12992}{81} \zeta_3 - \frac{88}{9} \zeta_2
                  \zeta_3 + \frac{272}{3} \zeta_5 \Bigg\}
                  + 
                  {{C_{A}^3}} \Bigg\{ \frac{62867783}{8748} +
                  \frac{146677}{486} \zeta_2  - \frac{5744}{45}
                  \zeta_2^2  - 
                  \frac{12352}{315} \zeta_2^3 
                  \nonumber\\
                &- \frac{67766}{27}
                  \zeta_3 - \frac{1496}{9} \zeta_2 \zeta_3  - 
                  \frac{104}{3} \zeta_3^2 + \frac{3080}{3} \zeta_5 \Bigg\}
                  + 
                  {{C_{A} n_{f}^2}} \Bigg\{ \frac{514997}{2187} -
                  \frac{8}{27} \zeta_2  + \frac{232}{45} \zeta_2^2 
                  \nonumber\\
                &+ 
                  \frac{7640}{81} \zeta_3 \Bigg\} 
                  + {{C_{A} C_{F} n_{f}}} \Bigg\{
                  - \frac{1004195}{324}  + \frac{1031}{18} \zeta_2 + 
                  \frac{1568}{45} \zeta_2^2 + \frac{25784}{27} \zeta_3
                  + 40 \zeta_2 \zeta_3  + \frac{608}{3} \zeta_5 
                  \nonumber\\
                &+ 132 \ln \left(\frac{\mu_R^2}{m_t^2}\right) \Bigg\}\,.
\end{align}

 After substituting the above expressions in \eqref{KGFF} and performing the final
integration, we obtain the UV renormalised form factor up to ${\cal O}(a_s^3)$ as 
\begin{align}

   \ln |\hat F^A_g|^2 (-q^2, \epsilon)&=
       a_s(q^2) \bigg\{ - \frac{1}{\epsilon^2} 16 C_A  

       +  C_A   \big( 8 + 14 \zeta_2  \big) \bigg\}

       + a_s(q^2)^2 \bigg\{ \frac{1}{\epsilon^3} \bigg[ 16 C_A n_f   

       -  88 C_A^2   \bigg]
\nonumber\\&
       + \frac{1}{\epsilon^2}  \bigg[ C_A n_f   \bigg( \frac{40}{9} \bigg)

       +   C_A^2   \bigg(  - \frac{268}{9} + 8 \zeta_2 \bigg)  \bigg]

       +  \frac{1}{\epsilon}  \bigg[ C_A n_f   \bigg( - \frac{76}{27} + \frac{4}{3} \zeta_2 \bigg)
\nonumber\\&
       + C_A^2   \bigg( \frac{772}{27} - 4 \zeta_3 - \frac{22}{3} \zeta_2  \bigg)  \bigg]

       +  C_F n_f   \bigg(  - \frac{160}{3} + 12 \ln \left(\frac{\mu_R^2}{m_t^2}\right)  + 16 \zeta_3 \bigg)
\nonumber\\&
       +  C_A n_f   \bigg(  - \frac{1886}{81} - \frac{92}{9} \zeta_3 - \frac{50}{3} \zeta_2 \bigg)

       +  C_A^2   \bigg( \frac{8318}{81} - \frac{286}{9} \zeta_3 + \frac{335}{3} \zeta_2 - 24 \zeta_2^2 \bigg) \bigg\}
\nonumber\\&
       + a_s(q^2)^3 \bigg\{ \frac{1}{\epsilon^4}  \bigg[ C_A n_f^2   \bigg(  - \frac{1408}{81} \bigg)

       +  C_A^2 n_f   \bigg(\frac{15488}{81} \bigg)

       +   C_A^3   \bigg( - \frac{42592}{81}  \bigg)  \bigg] \nonumber\\&
       +  \frac{1}{\epsilon^3}  \bigg[ C_A n_f^2   \bigg( - \frac{1600}{243} \bigg) 
       +  C_A C_F n_f   \bigg(\frac{256}{9} \bigg)

       +  C_A^2 n_f   \bigg(\frac{31040}{243} - \frac{320}{27} \zeta_2 \bigg)
\nonumber\\&
       +   C_A^3   \bigg( - \frac{98128}{243} + \frac{1760}{27} \zeta_2 \bigg)  \bigg]

       +  \frac{1}{\epsilon^2}  \bigg[ C_A n_f^2   \bigg(\frac{224}{81} - \frac{32}{27} \zeta_2 \bigg)

 \nonumber\\&      +  C_A C_F n_f   \bigg(\frac{440}{27} - \frac{128}{9} \zeta_3 \bigg)

       -  C_A^2 n_f   \bigg( \frac{6176}{243} - \frac{544}{27} \zeta_3 - \frac{416}{81} \zeta_2 \bigg)

       +   C_A^3   \bigg(\frac{16328}{243} 
       \nonumber\\&
       - \frac{880}{27} \zeta_3 + \frac{1384}{81} \zeta_2 - \frac{704}{45}
          \zeta_2^2 \bigg)  \bigg]

       +  \frac{1}{\epsilon}  \bigg[ C_A n_f^2   \bigg( - \frac{3728}{2187} + \frac{224}{81} \zeta_3 - \frac{80}{81} \zeta_2 \bigg)
\nonumber\\&
       -  C_A C_F n_f   \bigg( \frac{3638}{81} - \frac{608}{27} \zeta_3 - \frac{8}{3} \zeta_2 - \frac{64}{15}
          \zeta_2^2 \bigg)

       +  C_A^2 n_f   \bigg(\frac{14980}{2187} - \frac{1424}{81} \zeta_3 \nonumber\\&
       + \frac{4792}{243} \zeta_2
          - \frac{656}{45} \zeta_2^2 \bigg)

       +   C_A^3   \bigg(\frac{234466}{2187} + \frac{64}{3} \zeta_5 - \frac{488}{9} \zeta_3 - \frac{24436}{243}
          \zeta_2 + \frac{160}{9} \zeta_2 \zeta_3 \nonumber\\&
         + \frac{2552}{45} \zeta_2^2 \bigg)  \bigg]

       -  n_f  C_J^{(2)} 

       +  C_F n_f^2   \bigg(\frac{1498}{9} - \frac{224}{3} \zeta_3 - \frac{40}{9} \zeta_2 - \frac{32}{45} \zeta_2^2 \bigg)

       + C_F^2 n_f   \bigg(\frac{457}{3} \nonumber\\& - 320 \zeta_5 + 208 \zeta_3 \bigg)

       + C_A n_f^2   \bigg(\frac{560290}{6561} + \frac{9152}{243} \zeta_3 - \frac{296}{27} \zeta_2 - \frac{424}{27}
          \zeta_2^2 \bigg)
\nonumber\\&
       +  C_A C_F n_f   \bigg( - \frac{623255}{486} + \frac{1216}{9} \zeta_5 + \frac{35176}{81} \zeta_3 - \frac{925}{9}
          \zeta_2 + \frac{368}{3} \zeta_2 \zeta_3 - \frac{128}{45} \zeta_2^2 \bigg)
\nonumber\\&
       +  C_A^2 n_f   \bigg( - \frac{7335209}{6561} + \frac{856}{9} \zeta_5 - \frac{2216}{81} \zeta_3 - 
         \frac{37054}{729} \zeta_2 - 104 \zeta_2 \zeta_3 + \frac{10616}{45} \zeta_2^2 \bigg)
\nonumber\\&
       +  C_A^3   \bigg(\frac{35421539}{13122} + \frac{4444}{9} \zeta_5 - \frac{322280}{243} \zeta_3 - \frac{208}{9}
          \zeta_3^2 + \frac{510619}{729} \zeta_2 - \frac{308}{3} \zeta_2 \zeta_3 
          \nonumber\\&
          - \frac{118534}{135} \zeta_2^2
          + \frac{75088}{945} \zeta_2^3 \bigg) \bigg\} \,.

\end{align}

\subsection{Mass factorisation kernel}
\label{MFK}
The mass factorisation kernels are the solutions to the AP evolution equation which is controlled by the AP splitting functions $P_{a a'} \big(z_l,\mu_F^2\big)$ as given below 
\begin{equation}\label{RGGam}
   \mu_F^2\frac{d}{d\mu_F^2}\Gamma_{ab}\big(z_l,\mu_F^2,\epsilon\big) = 
	\frac{1}{2}\sum_{a'=q,\overline q,g} P_{a a'} \big(z_l,\mu_F^2\big)\otimes \Gamma_{a'b}\big(z_l,\mu_F^2,\epsilon\big) \,,
\quad \quad a,b = q,\overline q,g \,,
\end{equation}
where the perturabative expansion of the AP splitting functions read as,  
\begin{eqnarray}
P\big(z_l,\mu_F^2\big)=
\sum_{i=1}^{\infty}a_s^i(\mu_F^2) P^{(i-1)}(z_l)\,.
\end{eqnarray}
As discussed in the previous section, only the diagonal parts of splitting functions $P_{ab}(z, \mu_F^2)$ in $\Gamma_{ab}(z,\mu_F^2, \varepsilon)$ 
need to be kept since the convolutions of two or more non-diagonal splitting functions give rise to terms which are of beyond NSV type. The diagonal  $P_{g g}\big(z_l,\mu_F^2\big)$ are expanded
around $z_l=1$ and all those terms which do not contribute to SV+NSV are eliminated.  The diagonal AP splitting functions
near $z_l=1$ take the following form:
\begin{eqnarray}
	P_{gg}\big(z_l,a_s(\mu_F^2)\big) &=& 2  \Bigg[ B^A_g(a_s(\mu_F^2)) \delta(1-z_l) + A^A_g(a_s(\mu_F^2)) {\cal D}_0(z_l)
\nonumber\\&&
                      + C^A_g(a_s(\mu_F^2)) \log(1-z_l) + D^A_g(a_s(\mu_F^2)) \Bigg] + {\cal O}((1-z_l)) \,,
\end{eqnarray}
where
\begin{eqnarray}
C^A_g(a_s(\mu_F^2)) = \sum_{i=1}^\infty a_s^i(\mu_F^2) C_{g,i}^A,
\quad \quad 
D^A_g(a_s(\mu_F^2)) = \sum_{i=1}^\infty a_s^i(\mu_F^2) D_{g,i}^A \,.
\end{eqnarray}
The constants $C^A_{g,i}$ and $D^A_{g,i}$ can be obtained from the
the splitting functions $P_{gg}$ which are known to three loops in QCD \cite{Moch:2004pa,Vogt:2004mw}
(see \cite{GonzalezArroyo:1979df,Curci:1980uw,Furmanski:1980cm,Hamberg:1991qt,Ellis:1996nn,Moch:2004pa,Vogt:2004mw,Soar:2009yh,Ablinger:2017tan,Moch:2017uml} for the lower order ones). We list  $C^A_{g,i}$ and $D^A_{g,i}$ below

\begin{align}
  C^A_{g,1} &= 0\,,

\nonumber\\
   C_{g,2}^A &=
        16 C_A^2   \,,

\nonumber\\
   C_{g,3}^A &=
        C_A^2 n_f   \bigg\{  - \frac{320}{9} \bigg\}

       + C_A^3   \bigg\{ \frac{2144}{9} - 64 \zeta_2 \bigg\}\,,
\nonumber\\
    D_{g,1}^A &=
        -4 C_A   \,,
\nonumber\\
   D_{g,2}^A &=
        C_A n_f   \bigg\{ \frac{40}{9} \bigg\}

       + C_A^2   \bigg\{  - \frac{268}{9} + 8 \zeta_2 \bigg\}\,,
\nonumber\\
    D_{g,3}^A &=
        C_A n_f^2   \bigg\{ \frac{16}{27} \bigg\}

       + C_A C_F n_f   \bigg\{ \frac{110}{3} - 32 \zeta_3 \bigg\}

       + C_A^2 n_f   \bigg\{ \frac{908}{27} + \frac{112}{3} \zeta_3 - \frac{160}{9} \zeta_2 \bigg\}
\nonumber\\&
       + C_A^3   \bigg\{  - 166 + \frac{56}{3} \zeta_3 + \frac{1072}{9} \zeta_2 - \frac{176}{5} \zeta_2^2 \bigg\}\,.

\end{align}
The RG equation in \eqref{RGGam} can be solved by employing the perturbative expansion of the AP kernels
\begin{eqnarray}
\Gamma_{gg}(z_l,\mu_F^2,\ep)=\delta(1-z_l)+\sum_{i=1}^\infty \hat a_s^i 
\left({\mu_F^2 \over \mu^2}\right)^{i {\ep \over 2}}S^i_{\ep} 
\Gamma_{gg}^{(i)}(z_l,\ep)\,.
\end{eqnarray}
 The solutions of $\Gamma_{gg}^{(i)}$ in the  
$\overline{MS}$ scheme are given by
\begin{eqnarray}
\Gamma_{gg}^{(1)}(z_l,\ep)&=&{1 \over \ep} \Pzz 
\nonumber\\
\Gamma_{gg}^{(2)}(z_l,\ep)&=&
                   {1 \over \ep^2}\Bigg({1 \over 2} \Pzz \otimes \Pzz 
                       -\beta_0 \Pzz\Bigg)
                 +{1 \over \ep} \Bigg({1 \over 2} \Poz\Bigg)
\nonumber\\
\Gamma_{gg}^{(3)}(z_l,\ep)&=&
               {1 \over \ep^3}\Bigg(
              {4 \over 3} \beta_0^2 \Pzz -\beta_0 \Pzz \otimes \Pzz
\nonumber\\
&&              +{1 \over 6} \Pzz \otimes \Pzz \otimes \Pzz
              \Bigg) +
              {1 \over \ep^2} \Bigg( {1 \over 2} \Pzz \otimes \Poz
\nonumber\\
&&                -{1 \over 3} \beta_1 \Pzz -{4 \over 3} \beta_0 \Poz\Bigg)
              +{1 \over \ep} \Bigg({1 \over 3} \Ptz\Bigg)\,.
\end{eqnarray}
The most remarkable fact is that these quantities are
universal, independent of the insertion of operators. Hence, for the process under consideration, we
make use of the existing process independent results of the AP kernels and splitting functions.

 \subsection{Soft-collinear function}
\label{SF}
Exploiting the fact that the CF $\Delta_{d,g}^{A,SV+NSV}$ is finite, the infrared structure of $ {\cal S}_{d,g}^A$ can be studied using the AP evolution equations of $\Gamma_{gg}$ and the K+G differential equation of
$\hat {F}^{P}_g$ provided with the renormalisation group equation of $Z^A_{g}$.
This is possible as we find that  $ {\cal S}_{d,g}^A$ also satisfies a K+G type differential equation:
\begin{eqnarray}
\label{KGSC}
q^2 {d \over dq^2}  {\cal S}_{d,g}^A =  {1 \over 2}  \Big[ \overline{K}^A_{d,g} \Big(\hat{a}_s, \frac{\mu_R^2}{\mu^2},\epsilon,z_1,z_2\Big) 
+ \overline{G}^A_{d,g} \Big( \hat{a}_s,\frac{q^2}{\mu_R^2},\frac{\mu_R^2}{\mu^2},\epsilon,z_1,z_2\Big) \Big]\otimes {\cal S}^A_{d,g}\,,
\label{Sc}
\end{eqnarray}
where the infrared singular part is contained in $\overline K^A_{d,g} $
in terms of universal anomalous dimensions while the finite $\overline G^A_{d,g}$ is controlled by certain process independent but initial state dependent functions and also certain process dependent pieces. Since the K+G equation \eqref{KGSC} corresponding to $ {\cal S}_{d,g}^A$ admits a solution of convoluted exponential form, we write
\begin{eqnarray}
\label{calS}
{\cal S}^A_{d,g} = {\cal C}\exp\left(2 {\mathrm \Phi}_{d,g}^A(\hat a_s,\mu^2,q^2,z_1,z_2,\epsilon)\right) \,,
\end{eqnarray} 
where the real emission contributions are encapsulated in the function $\mathrm \Phi_{d,g}^A$ which is termed as the soft-collinear function. Furthermore, ${\mathrm \Phi}^A_{d,g}$ being independent of $\mu_R^2$ satisfies the RG equation, $\mu_R^2\dfrac{ d }{d\mu_R^2}{\mathrm{\Phi}}^A_{d,g} = 0$ 
and consequently 
\begin{equation}
\mu_R^2 \frac{d}{d\mu_R^2} \overline{K}^{A}_{d,g} = - \mu_R^2 \frac{d}{d\mu_R^2} \overline{G}^{A}_{d,g} = - \delta(1-z_1)\delta(1-z_2) a_s(\mu_R^2) \overline{A}^{A}_g \,.
\label{rgesKG}
\end{equation}
The right hand side of the above equation is proportion to $\delta(1-z_1) \delta(1-z_2)$ as the most singular terms resulting from $\overline K^A_{d,g} $ should cancel with those
from the form factor contribution which is proportional to only pure delta functions. To make the CF $\Delta_{d,g}^{A,SV+NSV}$ finite, the poles from $\mathrm \Phi^{A}_{d,g}$ have to cancel with those coming from $\hat{F}^A_g$ and $\Gamma_{gg}$. Hence the constants $\overline{A}^{A}_g$ should satisfy $\overline{A}^{A}_g = - A^{A}_g$.
The RGE \eqref{rgesKG} for $\overline{G}^{A}_{d,g}$ can be solved using the above mentioned relation to get 
\begin{align}
& \overline G^{A}_{d,g} \left(\hat a_s, \frac{q^2}{\mu_R^2}, \frac{\mu_R^2}{\mu^2},z_1, z_2,\ep\right) 
\nonumber\\
& \qquad = \overline G^{A}_{d,g} \left( a_s(\mu_R^2), \frac{q^2}{\mu_R^2},z_1, z_2,\ep\right)
\nonumber\\
& \qquad = \overline G^{A}_{d,g} \left( a_s(q^2),1,z_1, z_2,\ep \right) - \delta(1-z_1) \delta(1-z_2)  \int_{ \frac{q^2}{\mu_R^2}}^1 \frac{d\lambda^2}{\lambda^2} A^A_g\left(a_s(\lambda^2 \mu_R^2)\right) \, .
\end{align}
With these solutions, it is now straightforward to solve the differential equation \eqref{KGSC} for obtaining the form of $\mathrm \Phi^{A}_{d,g}$. For convenience, we decompose the soft-collinear function as $\mathrm{\Phi}^A_{d,g} =  \mathrm{\Phi}^A_{d,g,SV}  + \mathrm{\Phi}^A_{d,g,NSV}$ in such a way that $\mathrm{\Phi}^{A}_{d,g,SV}$ contains only the SV terms $i.e$ all the distributions ${\cal D}_k(z_l)$ and $\delta(1-z_l)$ and $\mathrm{\Phi}^{A}_{d,g,NSV}$ contains the NSV terms namely $\log^k(1-z_l), l=1,2, k=0,\cdots$ in the limit $z_{1(2)} \rightarrow 1$. An all order solution for $\Phi^A_{d,g,SV}$ in powers of  $\hat a_s$ in dimensional regularisation is given in \cite{Ravindran:2006bu} and we reproduce it here for completeness:   
\begin{align}
\label{eq:PhiSV}
\mathrm{\Phi}^A_{d,g,SV} =& \sum_{i=1}^\infty \hat a_s^i \left(q^2 \zo \zt \over \mu^2\right)^{i{\epsilon \over 2}} S_\epsilon^i
\Bigg[ {(i \epsilon)^2 \over 4 \zo \zt } \hat \phi_{d,g}^{A,(i)}(\epsilon)\Bigg] \,,
\end{align}
with
\begin{eqnarray}
\label{Defplus}
\hat{\phi}_{d,g}^{A,(i)}(\epsilon) = \frac{1}{i\epsilon}\Big[ \overline{K}_{d,g}^{A(i)}(\epsilon) + \overline{G}_{d,g,SV}^{A(i)}(\epsilon) \Big]\,.
\end{eqnarray} 
The constants $\overline K^{A,(i)}_{d,g}(\ep)$ are determined by expanding $\overline K^A_{d,g} $ in powers of $\hat a_s$ as follows
\begin{equation}
\overline K^A_{d,g}  \left(\hat a_s, \frac{\mu_R^2}{\mu^2}, z_1, z_2, \ep \right) = \delta(1-z_1)\delta(1-z_2) \sum_{i=1}^\infty \hat a_s^i
   \left( \frac{\mu_R^2}{\mu^2} \right)^{i \frac{\ep}{2}}S^i_{\ep}~ \overline K^{A,(i)}_{d,g}(\ep) 
\label{Kbeps}
\end{equation}
and solving the RGE \eqref{rgesKG} for $\overline K^A_{d,g} $.   The constants $\overline K^{A,(i)}_{d,g}(\ep)$ are related to the constants $K^{A,(i)}_{d,g}(\ep)$ which appear in the form factor by  $\overline K^{A,(i)}_{d,g}(\ep) = K^{A,(i)}_{d,g}(\ep)|_{A^A_g -> -A^A_g} $, due to the IR pole cancellation.  $\overline {G}^{A,(i)}_{d,g,SV}(\ep)$ are related to the finite functions $\overline G^A_{d,g,SV}(a_s(q^2),1,z_1,z_2,\ep)$. In terms of renormalized coupling constant, we find \begin{align}
\sum_{i=1}^\infty \hat{a}_s^i \left( \frac{q^2 (1-z_1)(1-z_2)}{\mu^2} \right)^{i \frac{\ep}{2}} S^i_{\ep}~ \overline G_{d,g,SV}^{A,(i)}(\ep)
 = \sum_{i=1}^\infty a_s^i \left( q^2 (1-z_1)(1-z_2) \right) \overline{{\cal G}}^{A}_{d,g,i}(\ep) \,.
\label{Gbar1}
\end{align}
Using $\overline K^{A,(i)}_{d,g}$ from Eq.~\ref{Kbeps} after putiting the explicit values of $A^A_g$ and  $\overline G^{A,(i)}_{d,g,SV}$ from  Eq.~\ref{Gbar1} , we find that $  \hat \phi_{d,g}^{A,(i)}(\epsilon)$ in $\mathrm{\Phi}^A_{d,g,SV}$
up to third order in $\hat a_s$ takes the following form

\begin{align}
\hat \phi_{d,g}^{A ,(1)}(\epsilon) &=
        \frac{1}{\epsilon^2} 8 C_A  

       + \frac{1}{\epsilon}   \overline {\cal G}^A_{d,g,1}(\epsilon) \,,

\nonumber\\
  \hat \phi_{d,g}^{A ,(2)}(\epsilon) &=
        \frac{1}{\epsilon^3} \bigg\{ C_A n_f   \bigg( \frac{8}{3} \bigg)

       +  C_A^2   \bigg(  - \frac{44}{3} \bigg) \bigg\}

       + \frac{1}{\epsilon^2} \bigg\{ n_f   \bigg( \frac{2}{3}\overline {\cal  G}^A_{d,g,1}(\epsilon)\bigg)

       +  C_A   \bigg(  - \frac{11}{3}\overline {\cal  G}^A_{d,g,1}(\epsilon)\bigg)
\nonumber\\&
       +  C_A n_f   \bigg(  - \frac{20}{9} \bigg)

       +  C_A^2   \bigg( \frac{134}{9} - 4 \zeta_2 \bigg) \bigg\}

       +  \frac{1}{2 \ep}\overline {\cal  G}^A_{d,g,2}(\epsilon)  \,,
\nonumber\\
    \hat \phi_{d,g}^{A ,(3)}(\epsilon) &=
        \frac{1}{\epsilon^4} \bigg\{C_A n_f^2   \bigg( \frac{128}{81} \bigg)

       +  C_A^2 n_f   \bigg(  - \frac{1408}{81} \bigg)

       +  C_A^3   \bigg( \frac{3872}{81} \bigg) \bigg\}

       + \frac{1}{\epsilon^3}  \bigg\{ n_f^2   \bigg( \frac{16}{27}\overline {\cal  G}^A_{d,g,1}(\epsilon)\bigg)

\nonumber\\&       +  C_A n_f   \bigg(  - \frac{176}{27}\overline {\cal  G}^A_{d,g,1}(\epsilon)\bigg)

       + C_A n_f^2   \bigg(  - \frac{640}{243} \bigg)

       + C_A C_F n_f   \bigg( \frac{16}{9} \bigg)

       + C_A^2   \bigg( \frac{484}{27}\overline {\cal  G}^A_{d,g,1}(\epsilon)\bigg)
\nonumber\\&
       + C_A^2 n_f   \bigg( \frac{8528}{243} - \frac{128}{27} \zeta_2 \bigg)

       +  C_A^3   \bigg(  - \frac{26032}{243} + \frac{704}{27} \zeta_2 \bigg) \bigg\}

       + \frac{1}{\epsilon^2}  \bigg\{ n_f   \bigg( \frac{8}{9}\overline {\cal  G}^A_{d,g,2}(\epsilon) \bigg)

\nonumber\\&       + C_F n_f   \bigg( \frac{2}{3}\overline {\cal  G}^A_{d,g,1}(\epsilon)\bigg)

       +  C_A   \bigg(  - \frac{44}{9}\overline {\cal  G}^A_{d,g,2}(\epsilon) \bigg)

       +  C_A n_f   \bigg( \frac{10}{9}\overline {\cal  G}^A_{d,g,1}(\epsilon)\bigg)

       +  C_A n_f^2   \bigg(  - \frac{32}{243} \bigg)
\nonumber\\&
       +  C_A C_F n_f   \bigg(  - \frac{220}{27} + \frac{64}{9} \zeta_3 \bigg)

       + C_A^2   \bigg(  - \frac{34}{9}\overline {\cal  G}^A_{d,g,1}(\epsilon)\bigg)

       +  C_A^2 n_f   \bigg(  - \frac{1672}{243} - \frac{224}{27} \zeta_3 
       \nonumber\\& + \frac{320}{81} \zeta_2 \bigg)

       + C_A^3   \bigg( \frac{980}{27} + \frac{176}{27} \zeta_3 - \frac{2144}{81} \zeta_2 + \frac{352}{45} 
         \zeta_2^2 \bigg) \bigg\}

       + \frac{1}{3 \ep}\overline {\cal  G}^A_{d,g,3}(\epsilon) \,.
\end{align} 

In the above equations, $\overline {\cal G}^A_{d,g,i}(\ep) $ are parametrised as follows
\begin{align}
\overline {\cal G}^A_{d,g,i}(\ep) = -f^A_{g,i} + \sum_{k=0}^{\infty} \ep^k~ \overline {\cal G}^{A,(k)}_{d,g,i}\,,  \end{align}
where
\begin{equation}
\gt^{A,(0)}_{d,g,1} = 0,\quad \quad  \gt^{A,(0)}_{d,g,2} = - 2 \beta_0 \gt_{d,g,1}^{A,(1)}, \quad \quad
\gt^{A,(0)}_{d,g,3} = - 2 \beta_1 \gt_{d,g,1}^{A,(1)} - 2 \beta_0 ( \gt_{d,g,2}^{A,(1)} + 2 \beta_0  \gt_{d,g,1}^{A,(2)}). 
\end{equation}
The unknown constants $\gt^{A,(k)}_{d,g,i}$ will be determined in the next section.

Let us now study in detail the structure of $\mathrm{\Phi}_{d,g,NSV}^{A}$ using the Eq.(\ref{KGSC}).
Subtracting out the K+G equation for the SV part $\mathrm{\Phi}_{d,g,SV}^{A}$ from Eq.(\ref{KGSC}), we find 
that $\mathrm{\Phi}_{d,g,NSV}^{A}$ satisfies
\begin{align}\label{KGphiNSV}
q^2\frac{d}{dq^2}\mathrm{\Phi}_{d,g,NSV}^{A}(q^2,z_j,\epsilon) = \frac{1}{2} \Big[ G_{d,L}^{A,g} \Big( \hat{a}_s,\frac{q^2}{\mu_R^2},\frac{\mu_R^2}{\mu^2},\epsilon,z_j \Big) \Big] \,,
\end{align}
where $G_{d,L}^{A,g} = \overline G^A_{d,g} - \overline G^A_{d,g,SV}$,
\begin{align}
G_{d,L}^{A,g} \Big( \hat{a}_s,\frac{q^2}{\mu_R^2},\frac{\mu_R^2}{\mu^2},z_j,\epsilon \Big) &= 
\sum_{i=1}^\infty a_s^i\big(q^2 (1-z_j)^2\big)  \mathcal{G}_{d,L,i}^{A,g}(\overline z,\epsilon).
\end{align}
Now integrating \eqref{KGphiNSV}, we obtain the following structure for  $\mathrm{\Phi}_{d,g,NSV}^{A}$
\begin{align}
\label{eq:PhiNSV}
\mathrm{\Phi}^A_{d,g,NSV} =& \sum_{i=1}^\infty \hat a_s^i \left(q^2 \zo \zt \over \mu^2\right)^{i{\epsilon \over 2}} S_\epsilon^i
\Bigg[ {i \epsilon \over 4 \zo } \varphi_{d,g}^{A,(i)} (\overline z_2,\epsilon)
+ {i \epsilon \over 4 \zt } \varphi_{d,g}^{A,(i)} (\overline z_1,\epsilon)\Bigg] \,,
\end{align}
where $ \varphi_{d,g}^{A,(i)} (\overline z_1,\epsilon)$ are found to be as follows
\begin{align}

   \varphi_{d,G}^{A,(1)} (\overline z_j,\epsilon) &=
        \frac{1}{\epsilon}    {\cal G}^{A,g}_{d,L,1}(\overline z_j,\epsilon) 
\nonumber\\
 \varphi_{d,G}^{A,(2)} (\overline z_j,\epsilon) &=
        \frac{1}{\epsilon^2} \bigg\{ n_f   \bigg( \frac{2}{3}  {\cal G}^{A,g}_{d,L,1}(\overline z_j,\epsilon)  \bigg)

       + C_A   \bigg(  - \frac{11}{3}  {\cal G}^{A,g}_{d,L,1}(\overline z_j,\epsilon)  \bigg) \bigg\}

       + \frac{1}{2 \ep} {\cal G}^{A,g}_{d,L,2}(\overline z_j,\epsilon) 
       \nonumber\\
   \varphi_{d,G}^{A,(3)} (\overline z_j,\epsilon) &=
        \frac{1}{\epsilon^3} \bigg\{ n_f^2   \bigg( \frac{16}{27}  {\cal G}^{A,g}_{d,L,1}(\overline z_j,\epsilon)  \bigg)

       +  C_A n_f   \bigg(  - \frac{176}{27}  {\cal G}^{A,g}_{d,L,1}(\overline z_j,\epsilon)  \bigg)
\nonumber\\&
      + C_A^2   \bigg( \frac{484}{27}  {\cal G}^{A,g}_{d,L,1}(\overline z_j,\epsilon)  \bigg) \bigg\}

       + \frac{1}{\epsilon^2} \bigg\{ n_f   \bigg( \frac{8}{9} {\cal G}^{A,g}_{d,L,2}(\overline z_j,\epsilon) \bigg)

       +  C_F n_f   \bigg( \frac{2}{3}  {\cal G}^{A,g}_{d,L,1}(\overline z_j,\epsilon)  \bigg)

 \nonumber\\&      +  C_A   \bigg(  - \frac{44}{9} {\cal G}^{A,g}_{d,L,2}(\overline z_j,\epsilon) \bigg)

       +  C_A n_f   \bigg( \frac{10}{9}  {\cal G}^{A,g}_{d,L,1}(\overline z_j,\epsilon)  \bigg)

      +  C_A^2   \bigg(  - \frac{34}{9}  {\cal G}^{A,g}_{d,L,1}(\overline z_j,\epsilon)  \bigg) \bigg\}
\nonumber\\&
       +   \frac{1}{3 \ep}  {\cal G}^{A,g}_{d,L,3}(\overline z_j,\epsilon)  \,.
\end{align}
Here, $ \mathcal{G}_{d,L,i}^{A,g}(\overline z,\epsilon)$ are expanded in powers of $\ep$ as
\begin{align}
{\cal G}^{A,g}_{d,L,i}(\overline z_j, \ep) = 2 L^A_{g,i}(a_s,\overline z_l) + \sum_{k=0}^{\infty} \ep^k~  {\cal G}^{A,g,(k)}_{d,L,i} (\overline z_j)   \,, 
\end{align}
with
\begin{align}
{\cal G}^{A,g,(0)}_{d,L,1}(\overline z_j) = 0,\quad \quad {\cal G}^{A,g,(0)}_{d,L,2}(\overline z_j) = - 2 \beta_0 {\cal G}_{d,L,1}^{A,g,(1)}(\overline z_j)\,,  \nonumber\\
{\cal G}^{A,g,(0)}_{d,L,3}(\overline z_j) = - 2 \beta_1 {\cal G}_{d,L,1}^{A,g,(1)}(\overline z_j) - 2 \beta_0 ( {\cal G}_{d,L,2}^{A,g,(1)}(\overline z_j) + 2 \beta_0  {\cal G}_{d,L,1}^{A,g,(2)}(\overline z_j))\,.
\end{align}
The anomalous dimensions $L^A_{g,i}$ can be determined by demanding finiteness
of $\Delta^A_{d,g}$ and it turns out that it is half of NSV part of the AP splitting functions (see 
\cite{Ajjath:2020lwb}), that is 
\begin{eqnarray}
L^A_{g,i}(a_s,\overline z_l) = C^A_{g,i}(a_s) \log(1-z_l) + D^A_{g,i} (a_s) \,.
\end{eqnarray} 
The coefficients $\mathcal{G}_{d,L,i}^{A,g,(j)}(\overline z_j)$ in the above equations are parametrised in terms of
$\log^k(1-z_j), k=0,1,\cdots$ and all the terms that vanish as $z_j\rightarrow 1$ are dropped
\begin{eqnarray}
\label{GikLj}
\mathcal{G}_{d,L,i}^{A,g,(j)}(\overline z_j) = \sum_{k=0}^{i+j-1} \mathcal{G}_{d,L,i}^{A,g,(j,k)} \log^k(1-z_j) \,.
\end{eqnarray}
The highest power of the $\log(1-z_j)$ at every order depends on the order of the perturbation, namely
the power of $a_s$ and also the power of $\epsilon$ at each order in $a_s$.  Hence 
the summation runs from 0 to $i+j-1$. 

Next, we obtain an the integral representation of $\Phi^A_{d,g,SV}$ which is given by 
\begin{eqnarray}
\label{IntSV}
\mathrm{\Phi}^A_{d,g,SV}&=&
{1 \over 2}\delta(\zt) \Bigg( {1 \over \zo} \Bigg\{
\int_{\mu_F^2}^{q^2 \zo} {d \lambda^2 \over \lambda^2}
A_g^A \left(a_s(\lambda^2)\right) 
+ \overline G^{A}_{d,g,SV} \left(
a_s\left(q_1^2\right),\ep\right)\Bigg\} \Bigg)_+
\nonumber\\[2ex]
&&+q^2 {d \over dq^2} \Bigg[
\Bigg({1 \over 4 \zo \zt}
\Bigg\{\int_{\mu_R^2}^{q^2\zo\zt} {d \lambda^2 \over \lambda^2} 
A^A_g\left(a_s(\lambda^2)\right)
+\overline G^A_{d,g,SV}\left(a_s\left(q_{12}^2\right),\ep\right)\Bigg\}\Bigg)_+\Bigg]
\nonumber\\
&&+{1 \over 2} \delta(\zo)~\delta(\zt)~ \sum_{i=1}^\infty \hat a_s^i
\left({q^2 \over \mu^2}\right)^{i {\ep \over 2}}
S_{\ep}^i~
\hat \phi^{A,(i)}_{d,g}(\ep)
\nonumber\\
&&+{1 \over 2} \delta(\zt)
\left({1 \over \zo}\right)_+ ~\sum_{i=1}^\infty
\hat a_s^i \left({\mu_F^2 \over \mu^2}\right)^{i {\ep \over 2}}
S_{\ep}^i~
 K^{A,(i)}_g(\ep)
+ (z_1 \leftrightarrow z_2)\,,
\end{eqnarray}
where  $q_l^2 = q^2~(1-z_l) $ and $q^2_{12}=q^2 \overline z_1 \overline z_2$. The subscript $+$ indicates the standard plus distribution. 
 Similarly, we find an integral representation of the NSV part, $\Phi^A_{d,g,NSV}$ which reads as
\begin{align}
\label{IntNSV}
\Phi^{A}_{d,g,NSV} =&
{1 \over 2}\delta(\overline z_2) \Bigg( \Bigg\{
\int_{\mu_F^2}^{q^2 \overline z_1} {d \lambda^2 \over \lambda^2}
L^A_g \left(a_s(\lambda^2), \overline z_1\right) 
+ {\varphi}_{d,f,g}^{A} \left(
a_s\left(q_2^2\right),\overline z_1,\ep\right)\Bigg\} \Bigg)
\nonumber\\&
+q^2 {d \over dq^2} \Bigg[
\Bigg({1 \over 2 (\overline z_1)}
\Bigg\{\int_{\mu_F^2}^{q^2 \overline z_1 \overline z_2} {d \lambda^2 \over \lambda^2} 
L^A_g \left(a_s(\lambda^2), \overline z_1\right) 
+{\varphi}^{A}_{d,f,g} \left(a_s\left(q_{12}^2\right),\overline z_1,\ep\right)\Bigg\}\Bigg)_+\Bigg]
\nonumber\\&
+{1 \over 2} \delta(\overline z_2)~
\varphi^A_{d,s,g}(\overline z_1,\ep)  
+  (z_1 \leftrightarrow z_2) \,,
\end{align}
where,
\begin{equation}
   \varphi^A_{d,a,g}\big(a_s(\lambda^2),\overline z_l\big) = \sum_{i=1}^\infty
    \hat{a}_s^i \bigg(\frac{\lambda^2}{\mu^2}\bigg)^{i\frac{\varepsilon}{2}} S^i_{\varepsilon}  \varphi_{d,a,g}^{A,(i)}\big(z_l,\varepsilon\big) .
\quad \quad \quad a = f,s
\end{equation}
$\varphi_{d,s,g}^A$ is the singular part of the NSV solution. The finite part $\varphi_{d,f,g}^A$ is parametrized in the following way,
\begin{align}
\label{eq:Phidf}
\varphi_{d,f,g}^{A}(a_s(\lambda^2),\overline z_l) &= \sum_{i=1}^\infty \sum_{k=0}^{\infty} \hat  a_s^i \left({\lambda^2 \over \mu^2}\right)^{i\frac{\epsilon}{2}}
S_\epsilon^i 
\varphi^{A,(i,k)}_{d,g}(\epsilon)\ln^k \overline z_l\,,
\nonumber\\
&= 
\sum_{i=1}^\infty \sum_{k=0}^i a_s^i(\lambda^2) \varphi^{A,(k)}_{d,g,i} \ln^k \overline z_l\,.
\end{align} 
The upper limit on the sum over $k$ is controlled by the dimensionally regularised Feynman integrals that contribute to order $a_s^i$. 
\section{Matching with the inclusive}
\label{MI}
The unknown coefficients of both SV and NSV solutions of $\mathrm \Phi^A_{d,g}$, namely $\gt^{A,(k)}_{d,g,i}$ and $\varphi^{A,(k)}_{d,g,i}$ can be determined using the fixed order predictions of $\Delta^A_{d,g}$, at every order in perturbation theory. However, it can also be determined alternatively from corresponding inclusive cross section using the relation,
\begin{align}
\int_0^1 
dx_1^0 \int_0^1 
dx_2^0 \left(x_1^0 x_2^0\right)^{N-1}
{d \sigma^A \over d y}
=\int_0^1 d\tau~ \tau^{N-1} ~\sigma^A\,,
\label{iden}
\end{align}
where $\sigma^A$ is the inclusive cross section. This relation in the large $N$ limit gives
\begin{align}
\label{eq:RapCoeff}
\sum_{i=1}^\infty \hat a_s^i \left({q^2  \over \mu^2} \right)^{i \epsilon \over 2} S_\epsilon^i 
\Big[t_1^{i}(\epsilon) \hat \phi_{d,g}^{A,(i)}(\epsilon)
-t_2^{i}(\epsilon) \hat \phi^{A,(i)}_g(\epsilon) 
+ \sum_{k=0}^{\infty} \Big( t_3^{(i,k)}(\epsilon) \varphi^{A,(i,k)}_{d,g}(\epsilon) \nonumber \\
-t_4^{(i,k)}(\epsilon) \varphi^{A,(i,k)}_g(\epsilon)\Big)\Big] = 0\,,
\end{align}
where
\begin{align}
&t_1^i~ = {i\epsilon (2 - i \epsilon) \over 4 N^{i \epsilon}}  \Gamma^2\left(1+i {\epsilon\over 2}\right)\,, \quad 
t_2^i={ i \epsilon  (1 - i \epsilon)  \over 2 N^{i \epsilon}} \Gamma(1+i \epsilon),
\nonumber \\
&t_3^{(i,k)} = \Gamma\left(1+i {\epsilon\over 2}\right){\partial^k \over \partial \alpha^k} \left({\Gamma (1+\alpha) \over N^{\alpha + i \epsilon/2}}\right)_{\alpha=i {\epsilon \over 2}}\,, \quad 
t_4^{(i,k)} = {\partial^k \over \partial {\hat \alpha }^k} \left({\Gamma (1+\hat \alpha )\over
N^{\hat \alpha}} \right)_{\hat \alpha =i \epsilon}. 
\end{align}
Here we keep $\ln^k N$ as well as  ${\cal O}(1/N)$ terms for the determination of the SV and NSV coefficients. The constants $\hat \phi^{A,(i)}_g$ and $\varphi^{A,(i,k)}_g$ are the inclusive counterparts to the SV and NSV coefficients $\hat \phi^{A,(i)}_{d,g}$ and $\varphi^{A,(i,k)}_{d,g}$ respectively which are known to third order in QCD for Drell-Yan, for Higgs production in gluon fusion and in bottom quark annihilation (for NSV see {\cite{Ajjath:2020ulr})}.

Using the above relations in Eq. \eqref{eq:RapCoeff}, the SV coefficients $\gt^{A,(k)}_{d,g,i}$ up to second order are found to be
\begin{eqnarray}
\gt^{A,(1)}_{d,g,1}  
&=&C_A~ \Big(- \zeta_2\Big) \,,
\nonumber\\
\gt^{A,(2)}_{d,g,1}
&=& C_A~ \Bigg({1 \over 3}  \zeta_3\Bigg)\,,
\nonumber\\
\gt^{A,(3)}_{d,g,1}
&=& C_A~ \Bigg({1 \over 80}  \zeta_2^2\Bigg)\,,
\nonumber\\
\gt^{A,(1)}_{d,g,2}
&=& C_A^2 ~ \Bigg({2428 \over 81} -{67 \over 3} \zeta_2
              -4 \zeta_2^2 -{44 \over 3} \zeta_3\Bigg)
\nonumber\\&&             +C_A n_f~ \Bigg(-{328 \over 81} + {10 \over 3} \zeta_2
                +{8 \over 3} \zeta_3 \Bigg)\,.
\end{eqnarray}
Note that the above SV coefficients are identical to the corresponding SV coefficients of scalar Higgs given in Eq.(35) of \cite{Ravindran:2006bu}. This universality nature of SV coefficients $\gt^{A,(k)}_{d,g,i}$ is expected to hold to all orders in perturbation theory because of the fact that it originates
entirely from the soft part of the differential cross section. Further, this property has been explicitly verified till third order in QCD perturbation theory \cite{Ravindran:2006cg} for the case of inclusive cross section. 
The explicit expressions for NSV coefficients $\varphi^{A(k)}_{d,g,i}$ for the pseudo-scalar production in gluon fusion can be obtained from the corresponding results for inclusive coefficients $\varphi^{A(k)}_{g,i}$ given in\cite{Bhattacharya:2021hae}. The results up to second order are provided below
\begin{align}
\label{eq:PhigA}
\varphi^{A,(0)}_{d,g,1} &= 
       2 C_A  \,,
\nonumber\\       
\varphi^{A,(1)}_{d,g,1} &= 0\,,
\nonumber\\ 
\varphi^{A,(0)}_{d,g,2} &=
        C_A n_f   \bigg(  - \frac{136}{27} + \frac{8}{3} \zeta_2 \bigg) + C_A^2   \bigg( \frac{904}{27} - 28 \zeta_3 - \frac{104}{3} \zeta_2 \bigg)\,,
\nonumber\\
\varphi^{A,(1)}_{d,g,2} &=
        C_A n_f   \bigg(  - \frac{2}{3} \bigg) + C_A^2   \bigg( \frac{2}{3} \bigg)\,,
\nonumber\\     
\varphi^{A,(2)}_{d,g,2} &= 
        -4 C_A^2  \,.
\end{align}
We notice that the NSV coefficients $\varphi^{A(k)}_{d,g,i}$ of the pseudo-scalar Higgs are also identical to the corresponding NSV coefficients of scalar Higgs up to second order in $a_s$ \cite{Ajjath:2020lwb,Ravindran:2022aqr}. 

\section{Results of the SV and NSV rapidity distributions}
\label{DNSV}
In this section we present the analytic results of the SV and NSV rapidity distributions, $\Delta^{A,SV}_{d,g}$ and $\Delta^{A,NSV}_{d,g}$ respectively, at the partonic level to $\rm{N^3LO}$ in QCD. 
By expanding the formula in Eq. \eqref{eq:MasterF} in powers of $a_s$ and substituting the explicit expressions for all the anomalous dimensions, also the SV and NSV coefficients, we find, at $a_s$ order,

\begin{align}
\label{Delta1NSV}    
\Delta_{d,g,1}^{A,NSV} &= 
{\color{blue} \pmb{ L_{z_1}}}
           \bigg\{C_A \big( - 4 \bar {\delta} \big)\bigg\}
 
       +\overline{{\cal D}}_0  \bigg\{  C_A  \big( - 4 \big) \bigg\}

       + C_A   \bigg\{ 2\bar {\delta} \bigg\}+ \big( z_1 \leftrightarrow z_2 \big) \,,
  \end{align}
  and at $a_s^2$ order
\begin{align}
\label{Delta2NSV}    
\Delta_{d,g,2}^{A,NSV} &= 

       {\color{blue} \pmb{ L^3_{z_1}}}  
       \bigg\{ C_A^2 \big( - 8\bar {\delta} \big) \bigg\}

    + {\color{blue} \pmb{ L^2_{z_1}}}  
        \Bigg\{C_A^2  \big(  - 24 \overline{{\cal D}}_0 \big)

   +    \bar {\delta} \bigg[ C_A n_f \bigg( - \frac{4}{3}  \bigg)

       +   C_A^2  \bigg( \frac{94}{3} \bigg) \bigg]  \Bigg\}
\nonumber\\&  
 + {\color{blue} \pmb{ L_{z_1}}}  \Bigg\{
       \overline{{\cal D}}_0 \bigg[ C_A n_f   \bigg(  - \frac{8}{3} \bigg)

       + C_A^2   \bigg( \frac{164}{3} \bigg) \bigg]

       + C_A^2 \big( - 48 \overline{{\cal D}}_1 \big)
\nonumber\\& 
       +  \bar {\delta} \bigg[ C_A n_f   \bigg( \frac{46}{9} \bigg)

       + C_A^2  \bigg( - \frac{616}{9}   - 8 \zeta_2 \bigg) \bigg] \Bigg\}

        +  \overline{{\cal D}}_0 
       \Bigg\{ C_A n_f   \bigg( \frac{52}{9} \bigg)
\nonumber\\& 
       + C_A^2   \bigg(  - \frac{622}{9} - 8 \zeta_2  \bigg) \Bigg\}

       +  \overline{{\cal D}}_1 
        \Bigg\{ C_A n_f   \bigg(  - \frac{8}{3} \bigg)

       + C_A^2   \bigg( \frac{116}{3} \bigg) \Bigg\}

       +  \overline{{\cal D}}_2
        \bigg\{  - 24 C_A^2   \bigg\}

\nonumber\\& 
      +  \bar {\delta} \Bigg\{ 
        C_A n_f   \bigg(  - \frac{136}{27} + \frac{8}{3} \zeta_2 \bigg)

       + C_A^2   \bigg( \frac{1336}{27} - 60 \zeta_3 - \frac{80}{3} \zeta_2 \bigg)
       
       \Bigg\}+ \big( z_1 \leftrightarrow z_2 \big) \,.
\end{align}
In the above expressions, $L_{z_1}=\ln(\overline z_1)$, $\overline \delta=\delta(\overline z_2)$, $\overline {\cal D}_j = \bigg(\frac{\ln^j(\overline z_2)}{(\overline z_2)}\bigg)_+$ and $\zeta_2 = 1.6449\cdots$ and $\zeta_3 = 1.20205\cdots$. 

Next at $a_s^3$, the CF requires the third order NSV coefficients $\varphi^{A,(k)}_{d,g,3}$ with $k=0,1,2,3$. Since the full N$^3$LO results for both inclusive and rapidity distribution of pseudo-scalar Higgs are not available, we could not extract these NSV coefficients either directly from the rapidity results or by matching with the inclusive results. However, we apply the same ratio method discussed in sub-section \ref{sec:NLO} on the SV+NSV results of scalar Higgs rapidity distribution computed in \cite{Ajjath:2020lwb} to obtain the third order CF of pseudo-scalar Higgs at SV+NSV accuracy. At $a_s^3$, we find 
\begin{align}
\label{Delta3NSV}    
\Delta_{d,g,3}^{A,NSV} &= 
  
       {\color{blue} \pmb{ L^5_{z_1}}}   \bigg\{ C_A^3   \big(  - 8 \bar {\delta} \big) \bigg\}

      +{\color{blue} \pmb{ L^4_{z_1}}}   \Bigg\{ C_A^3 \big(  - 40  \overline{{\cal D}}_0 \big)

      + \bar {\delta} \bigg[ C_A^3   \bigg( \frac{616}{9}  \bigg)

       + n_f C_A^2   \bigg(  - \frac{40}{9}  \bigg) \bigg] \Bigg\}
\nonumber\\&       
  
       +{\color{blue} \pmb{ L^3_{z_1}}}  \Bigg\{  \overline{{\cal D}}_0 \bigg[ C_A^3   \bigg( \frac{2320}{9}  \bigg)

       + n_f C_A^2   \bigg(  - \frac{160}{9}  \bigg) \bigg]

      - C_A^3 \big( 160  \overline{{\cal D}}_1 \big)

       + \bar {\delta} \bigg[ C_A^3   \bigg(  - \frac{2560}{9} 
       \nonumber\\& 
       + 64 \zeta_2  \bigg)

       + n_f C_A^2   \bigg( \frac{1036}{27}  \bigg)

       + n_f^2 C_A   \bigg(  - \frac{16}{27}  \bigg) \bigg]  \Bigg\}

      +{\color{blue} \pmb{ L^2_{z_1}}}    \Bigg\{ \overline{{\cal D}}_0 \bigg[ C_A^3   \bigg(  - \frac{7516}{9} + 192 \zeta_2  \bigg)   \nonumber\\& 

       + n_f C_A^2   \bigg( \frac{1016}{9}  \bigg)

       + n_f^2 C_A   \bigg(  - \frac{16}{9}  \bigg) \bigg]

     + \overline{{\cal D}}_1  \bigg[ C_A^3   \bigg( \frac{2128}{3}  \bigg)

       + n_f C_A^2   \bigg(  - \frac{160}{3}  \bigg) \bigg]
 \nonumber\\& 
  
       - C_A^3 \big( 240   \overline{{\cal D}}_2 \big)

      + \bar {\delta}  \bigg[ C_A^3   \bigg( \frac{24982}{27} - 488 \zeta_3 - 400 \zeta_2  \bigg)

       + n_f C_A^2   \bigg(  - \frac{3668}{27} + 38 \zeta_2  \bigg) \nonumber\\& 
  
       -  4 ~C_A C_F n_f 

       + n_f^2 C_A   \bigg( \frac{88}{27}  \bigg)  \bigg] \Bigg\}

      + {\color{blue} \pmb{ L_{z_1}}}  \Bigg\{ \overline{{\cal D}}_0 \bigg[ C_A^3   \bigg( \frac{44800}{27} - 976 \zeta_3 - \frac{2288}{3} \zeta_2  \bigg) \nonumber\\&

       + n_f C_A^2   \bigg(  - \frac{6860}{27} + \frac{224}{3} \zeta_2  \bigg)

       - 8 ~C_A C_F n_f  

       + n_f^2 C_A   \bigg( \frac{184}{27}  \bigg) \bigg]

 \nonumber\\& 
      + \overline{{\cal D}}_1  \bigg[ C_A^3   \bigg(  - \frac{14528}{9} + 384 \zeta_2  \bigg)

       + n_f C_A^2   \bigg( \frac{1960}{9}  \bigg)

       + n_f^2 C_A   \bigg(  - \frac{32}{9}  \bigg) \bigg]

 \nonumber\\& 
     +  \overline{{\cal D}}_2  \bigg[ C_A^3   \bigg( \frac{1888}{3}  \bigg)

       + n_f C_A^2   \bigg(  - \frac{160}{3}  \bigg) \bigg]

       + C_A^3   \bigg(  - 160  \overline{{\cal D}}_3  \bigg)

     + \bar {\delta}  \bigg[ C_A^3   \bigg(  - \frac{145670}{81} 
     \nonumber\\&  + \frac{4936}{3} \zeta_3 + 336 \zeta_2 + \frac{64}{5} \zeta_2^2 
    \bigg)

       + n_f C_A^2   \bigg( \frac{8528}{27} - 40 \zeta_3 - \frac{920}{9} \zeta_2  \bigg)
  \nonumber\\& 
       + C_A C_F n_f   \bigg( 258 
     
       - 48   \ln \Big( \frac{\mu_R^2}{m_t^2}\Big) - 96 \zeta_3 - \frac{8}{3} \zeta_2  \bigg)

       + n_f^2 C_A   \bigg(  - \frac{328}{81} + \frac{32}{9} \zeta_2  \bigg)

         \bigg] \Bigg\}

 \nonumber\\&
   
       + \overline{{\cal D}}_0 \bigg[ C_A^3   \bigg(  - \frac{127114}{81} + 1400 \zeta_3 + \frac{2200}{9} \zeta_2 + \frac{64}{5} \zeta_2^2  \bigg)

       + n_f C_A^2   \bigg( \frac{24488}{81} - 32 \zeta_3 
        \nonumber\\&
        - \frac{232}{3} \zeta_2  \bigg)

       + C_A C_F n_f   \bigg( 254 - 48   \ln \Big( \frac{\mu_R^2}{m_t^2}\Big) - 96 \zeta_3  \bigg)

       + n_f^2 C_A   \bigg(  - \frac{496}{81} + \frac{32}{9} \zeta_2  \bigg) \bigg]

  \nonumber\\&
     + \overline{{\cal D}}_1  \bigg[ C_A^3   \bigg( \frac{35044}{27} - 976 \zeta_3 - \frac{2384}{3} \zeta_2  \bigg)

       + n_f C_A^2   \bigg(  - \frac{6056}{27} + \frac{224}{3} \zeta_2  \bigg)

       - 8 C_A C_F n_f
 \nonumber\\&
       + n_f^2 C_A   \bigg( \frac{208}{27}  \bigg) \bigg]

      + \overline{{\cal D}}_2 \bigg[ C_A^3   \bigg(  - \frac{6748}{9} + 192 \zeta_2  \bigg)

       + n_f C_A^2   \bigg( \frac{896}{9}  \bigg)

       + n_f^2 C_A   \bigg(  - \frac{16}{9}  \bigg) \bigg]
 \nonumber\\&
     + \overline{{\cal D}}_3 \bigg[ C_A^3   \bigg( \frac{1600}{9}  \bigg)

       + n_f C_A^2   \bigg(  - \frac{160}{9}  \bigg)

       + C_A^3   \bigg(  - 40   \overline{{\cal D}}_4 \bigg) \bigg]

     + \bar {\delta}  \bigg[ C_A^3   \bigg( \frac{859052}{729} - 192 \zeta_5 
      \nonumber\\&
      - \frac{51068}{27} \zeta_3 - \frac{64400}{81} \zeta_2 + 
         \frac{608}{3} \zeta_2 \zeta_3 - 128 \zeta_2^2   
         \bigg)

       + n_f C_A^2   \bigg(  - \frac{150088}{729} + \frac{488}{3} \zeta_3 
        \nonumber\\&
        + \frac{12200}{81} \zeta_2 + \frac{88}{15} 
         \zeta_2^2  \bigg)

       + C_A C_F n_f   \bigg(  - \frac{5038}{27} + 24   \ln \Big( \frac{\mu_R^2}{m_t^2}\Big) + \frac{760}{9} \zeta_3 + \frac{16}{3} \zeta_2 + \frac{32}{5} 
         \zeta_2^2  \bigg)  \nonumber\\&

       + n_f^2 C_A   \bigg(  - \frac{232}{729} + \frac{32}{27} \zeta_3 - \frac{176}{27} \zeta_2  \bigg)

      \bigg] + \big( z_1 \leftrightarrow z_2 \big) \,.

\end{align}

Now, using the above result of $\Delta_{d,g,3}^{A,NSV}$ at $a_s^3$, we extract the NSV coefficients $\varphi_{d,g,3}^{A,(k)}$ with $k=0,1,2,3$ and they are given by
\begin{align}
\label{eq:PhigA3loop}
\varphi^{A,(0)}_{d,g,3} &=
        C_A n_f^2   \bigg(  - \frac{232}{729} + \frac{32}{27} \zeta_3 - \frac{176}{27} \zeta_2 \bigg) + C_A^2 n_f   \bigg(  - \frac{80860}{729} + \frac{704}{9} \zeta_3 + \frac{11960}{81} \zeta_2 - \frac{24}{5} 
         \zeta_2^2 \bigg) \nonumber\\&  + C_A^3   \bigg( \frac{423704}{729} + 192 \zeta_5 - \frac{18188}{27} \zeta_3 - \frac{55448}{81} \zeta_2 + 
         \frac{176}{3} \zeta_2 \zeta_3 + \frac{1384}{15} \zeta_2^2 
         \nonumber\\& 
         + C_F C_A n_f   \bigg(  - \frac{2158}{27} + \frac{472}{9} \zeta_3 + \frac{16}{3} \zeta_2 + \frac{32}{5} \zeta_2^2 \bigg)
          \,,
\nonumber\\
\varphi^{A,(1)}_{d,g,3} &=
       C_A n_f^2   \bigg( \frac{56}{27} \bigg) + C_A^2 n_f   \bigg( \frac{1528}{81} - 8 \zeta_3 - \frac{152}{9} \zeta_2 \bigg) + C_A^3   \bigg( - \frac{18988}{81} + \frac{448}{3} \zeta_3 + \frac{752}{9} \zeta_2 
       \nonumber\\& 
       \bigg)  + C_F C_A n_f   \bigg( 4 - \frac{8}{3} \zeta_2 \bigg) \,,
\nonumber\\
\varphi^{A,(2)}_{d,g,3} &=
       C_A n_f^2   \bigg( \frac{8}{27} \bigg) + C_A^2 n_f   \bigg( \frac{164}{27} + \frac{2}{3} \zeta_2 \bigg) + C_A^3   \bigg( - \frac{1432}{27} + \frac{40}{3} \zeta_2 \bigg)\,,
\nonumber\\
\varphi^{A,(3)}_{d,g,3} &= C_A^2 n_f   \bigg( \frac{32}{27} \bigg) + C_A^3   \bigg( - \frac{176}{27} \bigg)\,.
\end{align}
Here, we notice that the above NSV coefficients are same for both pseudo-scalar and scalar Higgs production via gluon fusion \cite{Ajjath:2020lwb}. However, the universality of $\varphi_{d,g,3}^{A,(k)}$ at third order can be checked only when the explicit N$^3$LO results are available for the
pseudo-scalar Higgs boson production in gluon fusion. The results of SV rapidity distributions to N$^3$LO can be found in Appendix \ref{app:SV}.    

\section{More on the $\vec z$ space solution ${\Phi}_{d,g}^A$} \label{MorePhi}
In the following, we discuss in detail the characteristic structure of SV and NSV solutions given in \ref{eq:PhiSV} and \ref{eq:PhiNSV}, respectively. Both the SV and NSV parts of $\mathrm \Phi_{d,g}^A$ satisfy the K+G equation and they contain singular as well as finite parts at every order. The pole part in the SV solution namely the soft and collinear divergences which are proportional to the distributions $\delta(1-z_l)$ and ${\cal D}_{0}(z_l)$ get cancelled against those resulting from the FFs entirely and the AP kernels partially. And the $z$ dependent finite part correctly reproduces all the distributions in the SV part of CFs $\Delta_{d,g}^A$. The NSV part, $\mathrm \Phi^A_{d,g,NSV}$, which comprises of terms like ${\cal D}_i(z_l)\ln^k(1-z_j)$ and $\delta(1-z_l) \ln^k(1-z_j)$ with ($l,j=1,2),~(i,k=0,1,\cdots$) removes the remaining collinear divergences of the AP kernels. The finite part of it along with SV counterpart give rises to next to SV terms to CFs $\Delta_{d,g}^A$. Note that the SV part $\mathrm \Phi^A_{d,g}$ plays a vital role in producing the next to SV terms for the CFs $\Delta^A_{d,g}$ at every order, when the exponential is expanded in powers of $a_s$. This is due to the fact that the convolutions of two or more distributions contribute to certain next to SV logarithms in addition to the distributions. 

Let us now focus on a peculiar feature that the NSV solution exhibits. Unlike in the case of SV solution, the NSV solution has the explicit $z$ dependency due to two pieces. One of them is from the ansatz $(1-z_l)^{i \epsilon/2} (1-z_j)^{i \epsilon/2}/(1-z_l)$ and the other one is from the coefficient $\varphi^{A,(i)}_{d,g}(z_j,\epsilon)$. This enables us to construct a class of solutions, a minimal class, to the K+G equation, satisfying the correct divergent structure as well as the dependence on $\ln^k(1-z_j)$ with ($l,j=1,2),~(i,k=0,1,\cdots$) \cite{Ajjath:2020ulr} as given below
\begin{align}
&\Phi^{A,j}_{d,g, NSV} = \sum_{i=1}^\infty \hat a_s^i \left(q^2 \zo^{\alpha_1} \zt^{\alpha_2} \over \mu^2\right)^{i{\epsilon \over 2}} S_\epsilon^i \Bigg[ {i \epsilon \over 4 \zo } \varphi_{d,g,\alpha_2}^{A,(i)} (\overline z_2,\epsilon)\Bigg]   
+ \big( \bar z_1 \leftrightarrow \bar z_2 \big) |_{(\alpha_1 \rightarrow \beta_1 , \alpha_2 \rightarrow \beta_2)} \,,
\end{align}
with $j = (\alpha_1,\alpha_2,\beta_1,\beta_2)$ and $\bar z_l = (1-z_l)$ for $l=1,2$. It is to be noted that $\alpha_1 = \beta_1 =1 $ for obtaining a finite $\Delta_{d,g}^A$, whereas $\alpha_2$ and $\beta_2$ can be arbitrary. The predictions from the soultions $\mathrm \Phi^{A,j}_{d,g,NSV}$ are found to be independent of the choice of $\alpha_2$ and $\beta_2$ owing to the explicit $z$-dependence of the coefficients $\varphi^{A,(i)}_{d,g,j}(z_l,\epsilon)$ with $j=\alpha_2$ for $l=2$ and $j=\beta_2$ for $l=1$ at every order in $\hat a_s$ and in $\epsilon$. It is straightforward to show that any variation of $\alpha_2$ and $\beta_2$ in the factors $(1-z_2)^{i \alpha_2 \epsilon/2}$ and $(1-z_1)^{i \beta_2 \epsilon/2}$ can always be compensated by suitably adjusting the $z$ independent coefficients of $\ln \left( 1 - z_1 \right)$ and $\ln \left( 1 - z_2 \right)$ terms in $\varphi^{A,(i)}_{d,g,\beta_2}(z_1,\epsilon)$ and $\varphi^{A,(i)}_{d,g,\alpha_2}(z_2,\epsilon)$, respectively at every order in $\hat a_s$. Here, the logarithmic structure of $\varphi_{d,f,g}^{A,j}$ plays a crucial role. Under this scale transformation, the expression given in \eqref{eq:Phidf} takes the following form
\begin{align}
\label{eq:Phidfj}
\varphi_{d,f,g}^{A, j}(a_s(q^2 {\bar z_l}^j),\overline z_l) &= 
\sum_{i=1}^\infty \sum_{k=0}^i a_s^i(q^2 {\bar z_l}^j) \varphi^{A,(k)}_{d,g,j,i} \ln^k \overline z_l\,.
\end{align} 
The fact that the predictions are insensitive to $j$ relate the coefficients $\varphi_{d,g,j,i}^{A,(k)}$ and $\varphi_{d,g,i}^{A,(k)}$, the solution corresponding to $j =1$, as given below  
\begin{eqnarray}
\varphi_{d,g,j,1}^{A,(0)}  &=& \varphi_{d,g,1}^{A,(0)},\quad 
\varphi_{d,g,j,1}^{A,(1)}  = -D^A_1 {\overline j} + \varphi_{d,g,1}^{A,(1)}, \quad
\varphi_{d,g,j,2}^{A,(0)}  = \varphi_{d,g,2}^{A,(0)} , \nonumber\\
\varphi_{d,g,j,2}^{A,(1)}  &=&  -{\overline j} \Big(D^A_2 - {\beta_0} \varphi_{d,g,1}^{A,(0)}\Big) + \varphi_{d,g,2}^{A,(1)} , \quad
\nonumber\\
\varphi_{d,g,j,2}^{A,(2)}  &=& -  \frac{1}{2}{\overline j}^2 \beta_0 D^A_1 -  {\overline j} \Big(C^A_2   - {\beta_0} \varphi_{d,g,1}^{A,(1)} \Big) + \varphi_{d,g,2}^{A,(2)}
\nonumber\\
\varphi_{d,g,j,3}^{A,(0)}  &=& \varphi_{d,g,3}^{A,(0)},
\quad
\varphi_{d,g,j,3}^{A,(1)}  = -{\overline j} \Big(D^A_3  - {\beta_1}  \varphi_{d,g,1}^{A,(0)} - 2 {\beta_0}  \varphi_{d,g,2}^{A,(0)}\Big) + \varphi_{d,g,3}^{A,(1)}
\nonumber\\
\varphi_{d,g,j,3}^{A,(2)}  &=&- {\overline j}^2 \Big( \frac{1}{2}{\beta_1} D^A_1 + {\beta_0} D^A_2  - {\beta_0}^2  \varphi_{d,g,1}^{A,(0)} \Big)-{\overline j} \Big(C^A_3 {\overline j}  - {\beta_1}  \varphi_{d,g,1}^{A,(1)} - 2 {\beta_0}  \varphi_{d,g,2}^{A,(1)}\Big) + \varphi_{d,g,3}^{A,(2)}
\nonumber\\
\varphi_{d,g,j,3}^{A,(3)}  &=& {\beta_0}^2 \bigg(-\frac{1}{3}D^A_1 {\overline j}^3 + {\overline j}^2 \varphi_{d,g,1}^{A,(1)}\bigg) +
     {\beta_0} {\overline j} \bigg(-C^A_2 {\overline j} + 2 \varphi_{d,g,2}^{A,(2)}\bigg) + \varphi_{d,g,3}^{A,(3)}
\nonumber\\
\varphi_{d,g,j,4}^{A,(0)}  &=& \varphi_{d,g,4}^{A,(0)}, \quad
\varphi_{d,g,j,4}^{A,(1)}  = - D^A_4 {\overline j} + {\beta_2} {\overline j} \varphi_{d,g,1}^{A,(0)} + 2 {\beta_1} {\overline j} \varphi_{d,g,2}^{A,(0)} +
     3 {\beta_0} {\overline j} \varphi_{d,g,3}^{A,(0)} + \varphi_{d,g,4}^{A,(1)}
\nonumber\\
\varphi_{d,g,j,4}^{A,(2)}  &=& -C^A_4 {\overline j} - \frac{1}{2}{\beta_2} D^A_1 {\overline j}^2 - {\beta_1} D^A_2 {\overline j}^2 -
     \frac{3}{2} {\beta_0} D^A_3 {\overline j}^2 + \frac{5}{2} {\beta_0} {\beta_1} {\overline j}^2 \varphi_{d,g,1}^{A,(0)}
     + {\beta_2} {\overline j} \varphi_{d,g,1}^{A,(1)} +
     3 {\beta_0}^2 {\overline j}^2 \varphi_{d,g,2}^{A,(0)} 
     \nonumber\\&&
     + 2 {\beta_1} {\overline j} \varphi_{d,g,2}^{A,(1)} 
     + 3 {\beta_0} {\overline j} \varphi_{d,g,3}^{A,(1)} + \varphi_{d,g,4}^{A,(2)}
\nonumber\\
\varphi_{d,g,j,4}^{A,(3)}  &=& {\beta_0}^3 {\overline j}^3 \varphi_{d,g,1}^{A,(0)} + {\beta_0}^2 {\overline j}^2 \bigg(-D^A_2 {\overline j} + 3 \varphi_{d,g,2}^{A,(1)}\bigg) -
     \frac{1}{6}{\beta_1} {\overline j} \bigg(6 C^A_2 {\overline j} + 5 {\beta_0} {\overline j} \bigg(D^A_1 {\overline j} - 3 \varphi_{d,g,1}^{A,(1)}\bigg) 
     \nonumber\\&&
     - 12 \varphi_{d,g,2}^{A,(2)}\bigg) 
      - \frac{3}{2} {\beta_0} {\overline j} \bigg(C^A_3 {\overline j} - 2 \varphi_{d,g,3}^{A,(2)}\bigg) + \varphi_{d,g,4}^{A,(3)}
\nonumber\\
\varphi_{d,g,j,4}^{A,(4)}  &=& {\beta_0}^3\bigg( - \frac{1}{4}D^A_1 {\overline j}^4 + {\overline j}^3 \varphi_{d,g,1}^{A,(1)}\bigg) +
     {\beta_0}^2 {\overline j}^2 \bigg(-C^A_2 {\overline j} + 3 \varphi_{d,g,2}^{A,(2)}\bigg) + 3 {\beta_0} {\overline j} \varphi_{d,g,3}^{A,(3)} + \varphi_{d,g,4}^{A,(4)}\,,
\end{eqnarray}

with $j$ being $\alpha_2$ and $\beta_2$ for the coefficients of $\ln \left( 1-z_2 \right)$
 and $\ln \left( 1-z_1 \right)$ respectively. In the above equations, $\bar j = (\alpha_2-1$) and ($\beta_2-1)$ for $j=\alpha_2$ and $\beta_2$ respectively. The above relations are the transformations for $\varphi_{d,g,j,i}^{A,(k)}$ that are required to compensate the contributions resulting from the change in the exponents of $(1-z_1)$ and $(1-z_2)$ from $(i \epsilon)/2$ to $(i j \epsilon)/2$. The function $\mathrm \Phi^{A, j}_{d,g,NSV}$ being insensitive to the choice of the scales $\alpha_2$ and $\beta_2$ indicates its invariance under certain gauge like transformations on both $(1-z_l)^{i \epsilon/2}$ and $\varphi^{A,j}_{d,f,g}(z_l,\epsilon)$. Due to this invariance, these transformations neither alter the divergent structure nor the finite parts of $\mathrm \Phi^A_{d,g,NSV}$. However, we choose to work with $\alpha_2 = \beta_2 = 1$ in the NSV solution to have more resemblance with its SV counterpart. In summary, we find a minimal class of solutions to the K+G equation without affecting neither the all order structure nor the predictions for $\Delta^A_{d,g}$.  We will show later that this choice will allow us to study resummation in two-dimensional Mellin space for SV as well as NSV parts with single ${\mathcal O}(1)$ term denoted by $\omega = a_s \beta_0 \ln \left(N_1 N_2\right)$.


\section{Resummation in the Mellin $\vec N$ space}\label{RES}
This section is devoted to the study of all order perturbative structure of $\Delta_{d,g}^A$ in the Mellin space. To find the structure of $\Delta_{d,g}^A$ in the Mellin space, we use the integral representations of both $\mathrm{\Phi}^{A}_{d,g,SV}$ and $\mathrm{\Phi}^{A}_{d,g,NSV}$ given in \eqref{IntSV} and \eqref{IntNSV} respectively. As a result, $\Psi^A_{d,g} $ in \eqref{eq:MasterF} takes the following form 
\begin{align}
\label{Psiint}
\Psi^A_{d,g} (q^2,\mu_F^2,z_1,z_2)   =& {\delta(\overline z_1) \over 2} \Bigg(\!\!\displaystyle {\int_{\mu_F^2}^{q^2 \overline z_2}
\!\!{d \lambda^2 \over \lambda^2}}\! {\cal P}_{gg}\left(a_s(\lambda^2),\zt\right) 
\!+\! {\cal Q}^A_{d,g}\left(a_s(q_2^2),\zt\right)
\!\!\Bigg)_+ 
\nonumber\\&
+ {1 \over 4} \Bigg( {1 \over \overline z_1 }
\Bigg\{{\cal P}_{gg}(a_s(q_{12}^2),\zt ) + {\color{black} 2 }L^A_{g}(a_s(q_{12}^2) ,\zt)
\nonumber \\
& + q^2{d \over dq^2} 
\left({\cal Q}^A_{d,g}(a_s(q_2^2 ),\zt) +  {\color{black} 2 }\varphi^A_{d,f,g}(a_s(q_2^2 ),\zt)\right)
\Bigg\}\Bigg)_+
\nonumber\\&
+ {1 \over 2}
\delta(\overline z_1) \delta(\overline z_2)
\ln \Big(g^A_{d,g,0}(a_s(\mu_F^2))\Big)
+ \overline z_1 \leftrightarrow \overline z_2,
\end{align}
where ${\cal P}_{gg} (a_s, \overline z_l)= P_{gg}(a_s,\overline z_l) - 2 B_g^A(a_s) \delta(\overline z_l)$, $q_l^2 = q^2~(1-z_l) $ and $q^2_{12}=q^2 \overline z_1 \overline z_2$. The subscript $+$ indicates standard plus distribution. The function ${\cal Q}^P_{d,g}$ in \eqref{Psiint} is given as
\begin{eqnarray}
\label{CalQ}
{\cal Q}^A_{d,g}(a_s,\overline z_l) = {2 \over \overline z_l}  \mathbf{D}^A_{d,g}(a_s) + 2 \varphi^A_{d,f,g} (a_s,\overline z_l)\,.
\end{eqnarray}
The SV coefficient $ \mathbf{D}_{d,g}^A$ are given in the Appendix \ref{App}. The constant $g^A_{d,g,0}$ in \eqref{Psiint} results from finite part of the virtual contributions and pure $\delta(\overline z_l)$ terms of $\Phi^A_{d,g}$. 

Now we take the double Mellin transform of $\Delta_{d,g}^A$ in $\vec N$ space as
\begin{align}
\label{DeltaN}
\Delta^A_{d,g,\vec N} (q^2,\mu_R^2,\mu_F^2) = \int_0^1 dz_1 z_1^{N_1-1}\int_0^1 dz_2 z_2^{N_2-1}\Delta^A_{d,g}(z_1,z_2) (q^2,\mu_R^2,\mu_F^2)
\nonumber\\ = \tilde g_{d,g,0}^A (q^2,\mu_R^2,\mu_F^2) \exp\left(
\Psi^A_{d,g,\vec N} (q^2,\mu_F^2) 
\right)\,.
\end{align} 
The $N$-independent constant $\tilde g_{d,g,0}^A$ is given in Appendix \ref{app:g0t}.
The resummed result for $\Psi^A_{d,g,\vec N}$ takes the following form
\begin{align}
\label{PsiN}
\Psi^A_{d,g,\vec N} &= 
  \bigg(g_{d,g,1}^A(\omega) 
+\frac{1}{N_1} \overline{g}_{d,g,1}^A(\omega)\bigg) \ln N_1
+ \sum\limits_{i=0}^\infty a_s^i \bigg( \frac{1}{2}  g_{d,g,i+2}^A(\omega) + \frac{1}{N_1} \overline{g}_{d,g,i+2}^A(\omega) \bigg)
\nonumber\\&
+\frac{1}{N_1} \left(h^A_{d,g,0}(\omega,N_1)  + 
\sum\limits_{i=1}^{\infty} a_s^i h^A_{d,g,i}(\omega,\omega_1,N_1)\right) 
+ (N_1 \leftrightarrow N_2,\omega_1 \leftrightarrow \omega_2) \,,
\end{align}
with 
\begin{align}
\label{hg}
h^A_{d,g,0}(\omega,N_l) &= h^A_{d,g,00}(\omega) + h^A_{d,g,01}(\omega) \ln N_l\,,
\nonumber \\ 
         h^A_{d,g,i}(\omega,\omega_l,N_l) &= \sum_{k=0}^{i-1} h^A_{d,g,ik}(\omega)~ \ln^k N_l 
         + \tilde{h}^A_{d,g,ii}(\omega,\omega_l)~ \ln^k N_l\,.
\end{align}
In the above expressions, $\omega = a_s \beta_0 \ln N_1 N_2$ and $\omega_l = a_s \beta_0 \ln N_l$ for $l=1,2$.
Here, $ g_{d,g,i}^A$ are the resummation constants resulting from the SV contributions and $\bar g^A_{d,g,i}$ result entirely from $A^A_g,B^A_g$ coefficients of $P_{gg}$ and from the function $\mathbf{D}_{d,g}^A$ in \eqref{CalQ}. The function $\bar g^A_{d,g,1}$ is found to be identically zero and we find that none of the coefficients $\bar g^A_{d,g,i}$ contains
explicit $\ln N_l$. The functions $h^A_{d,g,i}$ comprise of $C^A_g$ and $D^A_g$ which are present in $P_{gg}$ as well as the pure NSV coefficients present in  
$\varphi^A_{d,f,g}$.
We find that coefficient of $h^A_{d,g,01}$ is proportional to $C^A_{g,1}$ which is identically zero.
Hence, at order $a_s^0$, there is no ${\ln N_l \over N_l}$ term. The SV resummation constant $g_{d,g,i}^A$ has been discussed in great detail in references \cite{Banerjee:2017cfc,Banerjee:2017ewt,Ahmed:2020caw} and the NSV resummation coefficients $\bar g^A_{d,g,i}$, $h^A_{d,g,ij}$ and $\tilde{h}^A_{d,g,ii}$ are provided in Appendix \ref{app:gbdN} and \ref{app:hdN}. Our next aim is to include these resummed contributions consistently in the fixed order predictions to understand the phenomenological relevance of resumming the NSV contributions for the case of psuedo-scalar Higgs production in gluon fusion channel.

\section{Numerical analysis}\label{NUM}

\begin{figure}[H]
\centering
\includegraphics[scale=0.350]{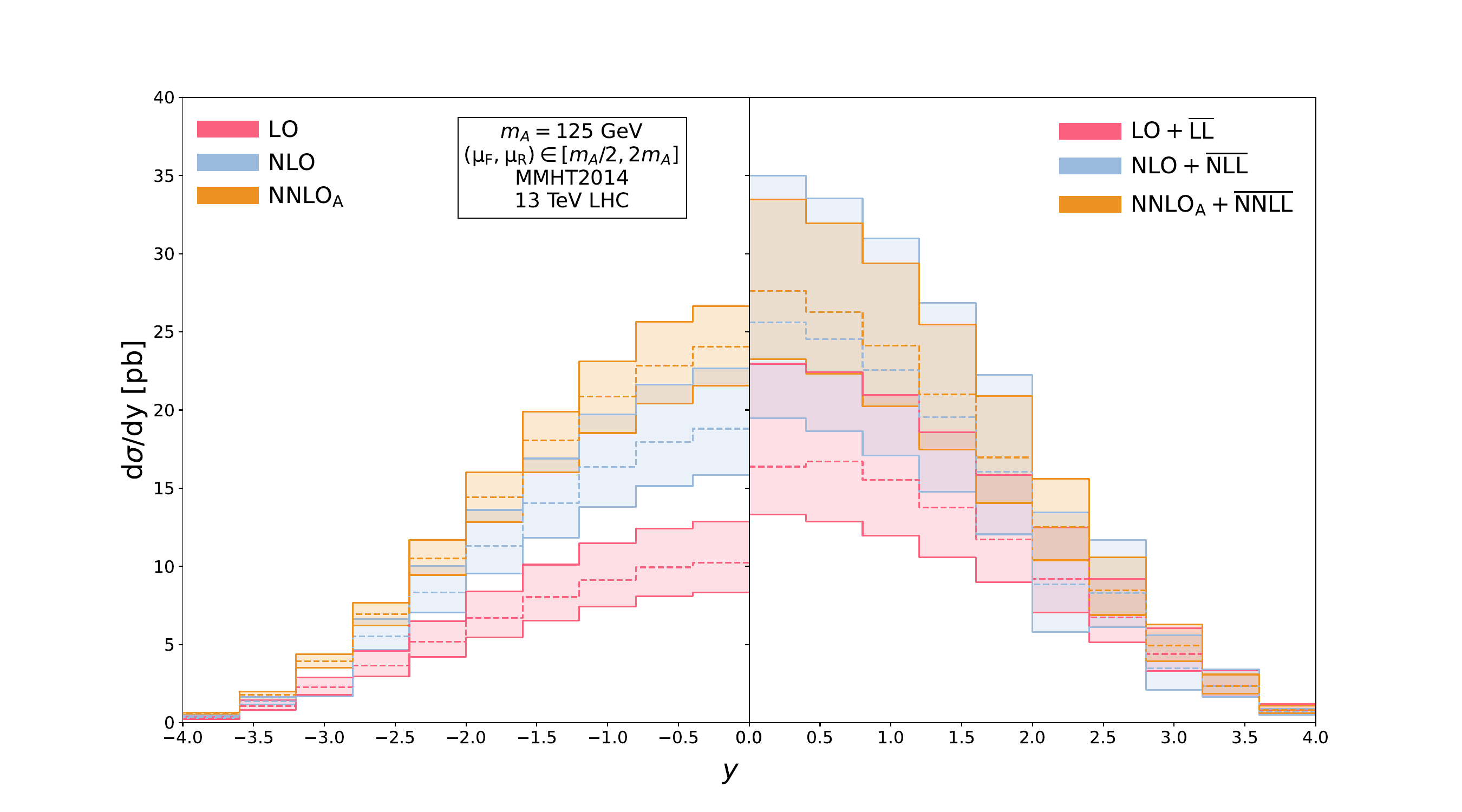}
\includegraphics[scale=0.350]{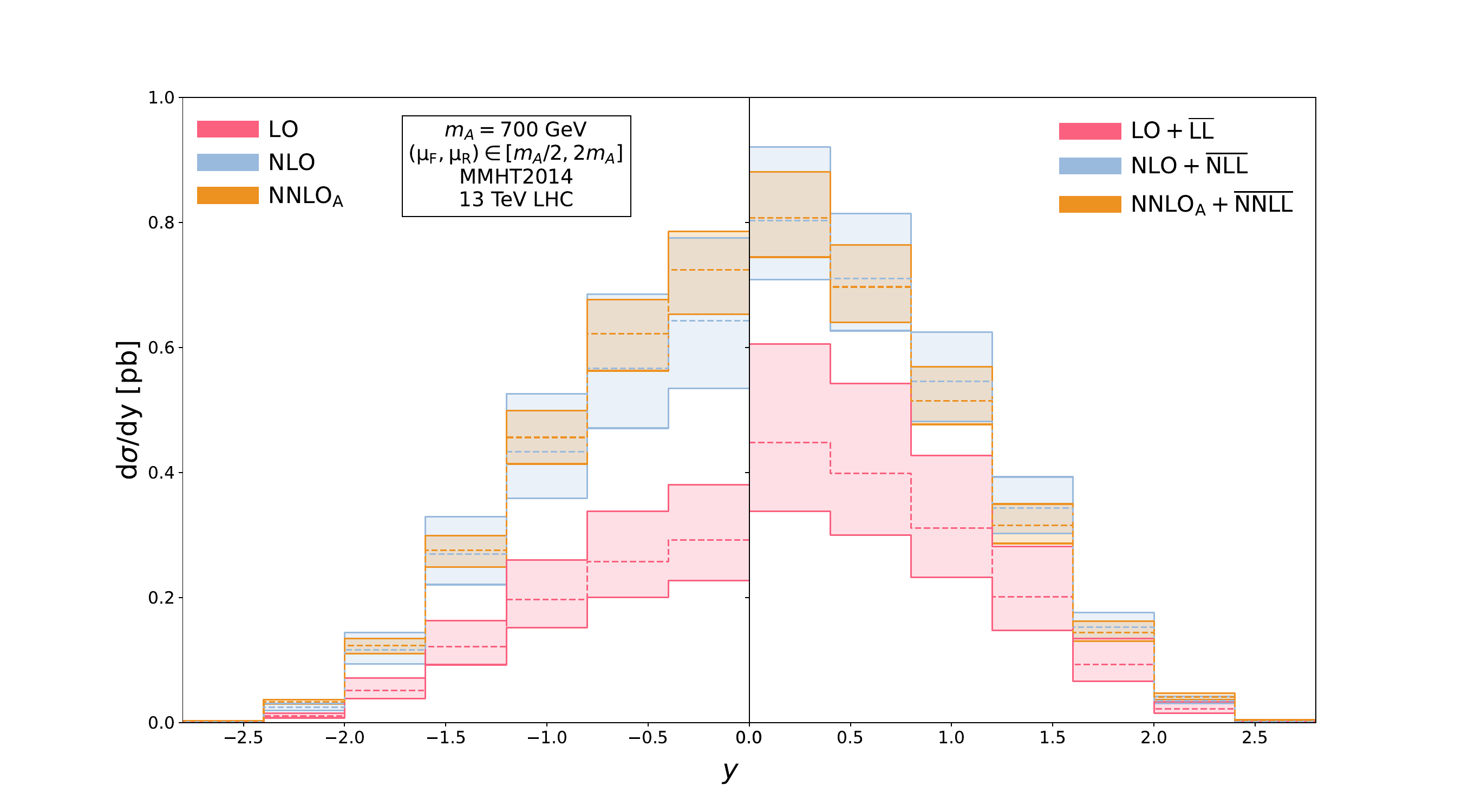}
\caption{Comparison of 7-point scale variation between fixed order and SV+NSV resummed results for $m_A=125$ (top) and $m_A=700$ (bottom) GeV.
The dashed lines refer to the corresponding central scale values at each order.}
\label{fig:7pt125}
\end{figure}

In this section, we study the impact of resummed soft-virtual and next-to-soft virtual (SV+NSV) contributions for the rapidity distribution of the pseudo-scalar Higgs production in gluon fusion channel at the LHC to $\rm{NNLO_A}$+$\rm \overline{NNLL}$ accuracy. We use the MMHT2014(68cl) PDF set \cite{Harland-Lang:2014zoa} 
and the corresponding strong coupling $a_s$ through the Les Houches Accord PDF (LHAPDF) interface \cite{Buckley:2014ana} 
at each order in perturbation theory with $n_f = 5$ active massless quark flavours throughout. Our predictions are based on Higgs effective field theory where the top quarks are
integrated out at higher orders. Nevertheless, we retain the top
quark mass dependence at LO. The term $C_J^{(2)}$
in the Wilson
coefficient $C_J$ in (\ref{eq:const}) is taken to be zero in our analysis
because it is not available in the literature yet. For
simplicity, we have set $\cot \beta =1$ in our numerical analysis.
Results for other values of $\cot \beta$ can be easily obtained by
rescaling the cross sections with $\cot^2 \beta$. For the fixed order rapidity distribution of the pseudo-scalar Higgs,
we use the publicly available code FEHiP \cite{Anastasiou:2005qj} of the scalar Higgs by taking into account the ratio factor discussed in section \ref{sec:NLO}. \textcolor{black}{The resummed contribution is obtained from $\Delta^A_{d,g,\vec N}$ 
in (\ref{DeltaN}) after performing Mellin inversion which
is done using an in-house FORTRAN based code.} The resummed results are matched to the fixed order result in order to avoid any double counting of threshold logarithms as

%

\begin{align}\label{match}
{d\sigma^{A,\rm match} \over dY } = {d\sigma^{A,\rm (SV+NSV)} \over dY }{\bigg|_{\rm resum}} - {d\sigma^{A,\rm (SV+NSV)}\over dY }{\bigg|_{\rm FO}}  + {d\sigma^{A,\rm FO} \over dY }\,.
\end{align}
We do the analysis for centre of mass energy $\sqrt{S} = 13$ TeV with the pseudo-scalar Higgs mass $m_A = 125$ GeV and $m_A= 700$ GeV, top quark pole mass $m_t = 173.3$ GeV and the Fermi constant $G_F = 4541.63$ pb. The numerical values for the aforementioned parameters are taken from the
Particle Data Group 2020 \cite{ParticleDataGroup:2020ssz}. To distinguish between the SV and SV+NSV resummed results, the NSV included resummed results have been denoted by $\rm \overline{N^n LL}$ for the $n^{th}$ level logarithmic accuracy. \\ \\
\textbf{K-factor analysis}: We begin our analysis by studying the higher order effects which are quantified through the K-factors as
\begin{equation}\label{eq:K125}
 \mathrm{K}  = \dfrac{\dfrac{d\sigma}{dY}\left(\mu_R=\mu_F=m_A\right)}{\dfrac{d\sigma^{\text{LO}}}{dY}(\mu_R=\mu_F=m_A)} \,.
 \end{equation}
 \begin{table*}[ht]
\small
\centering
 \renewcommand{\arraystretch}{2.0}
\begin{tabular}{ |P{1.5cm}||P{1.3cm}|P{1.1cm}|P{1.9cm} |P{1.2cm}|P{2.3cm}|}
 \hline
  y &$\rm{K_{LO+\rm{\overline{LL}}}}$&$\rm{K_{NLO}}$&$\rm{K_{NLO+\rm{\overline{NLL}}}}$&$\rm{K_{NNLO_{A}}}$&$\rm{K_{NNLO_A+\rm{\overline{NNLL}}}}$\\
 \hline
\hline
 0-0.4    & 1.602 &  1.839 & 2.505 &  2.352 & 2.699 \\
 0.4-0.8  & 1.681 &  1.806 & 2.469 & 2.297 &  2.644 \\
 0.8-1.2  & 1.703 & 1.792 & 2.472 & 2.285 &  2.643\\
 1.2-1.6  & 1.713 &  1.746 & 2.433 & 2.248 & 2.613 \\
 1.6-2.0  & 1.748 & 1.688 &  2.397 & 2.151 & 2.533 \\
\hline
\end{tabular}
\caption{K-factor values of fixed order and resummed results at the central scale $\mu_R=\mu_F= m_A$ for $m_A=125$ GeV.} \label{tab:K125}
\end{table*}
\begin{table*}[ht]
\small
\centering
 \renewcommand{\arraystretch}{2.0}
\begin{tabular}{ |P{1.5cm}||P{1.3cm}|P{1.1cm}|P{1.9cm} |P{1.2cm}|P{2.3cm}|}
 \hline
  y &$\rm{K_{LO+\rm{\overline{LL}}}}$&$\rm{K_{NLO}}$&$\rm{K_{NLO+\rm{\overline{NLL}}}}$&$\rm{K_{NNLO_{A}}}$&$\rm{K_{NNLO_A+\rm{\overline{NNLL}}}}$\\
 \hline
\hline
 0-0.4    & 1.533 &  2.200 & 2.749 & 2.478 & 2.763 \\
 0.4-0.8  & 1.547 &  2.199 & 2.755 & 2.414 & 2.703  \\
 0.8-1.2  & 1.579 & 2.200 & 2.769 & 2.315 & 2.613 \\
 1.2-1.6  & 1.653 &  2.212 & 2.819 & 2.266 & 2.592 \\
 1.6-2.0  & 1.797 & 2.238 & 2.947 & 2.370 & 2.781 \\
\hline
\end{tabular}
\caption{K-factor values of fixed order and resummed results at the central scale $\mu_R=\mu_F= m_A$ for $m_A=700$ GeV.} \label{tab:K700}
\end{table*}
We fix the central scale at $\mu_R=\mu_F=m_A$ throughout our analysis. 
In Table \ref{tab:K125}, we present the K-factor values of fixed order and resummed predictions at $m_A=125$ GeV for benchmark
rapidity values. We observe that the NLO result at the central scale is enhanced by 83.9\% with respect to the LO one around the central rapidity region. However, the enhancement of the approximate NNLO ($\rm{NNLO_A}$) result at the central scale is 27.9\% in comparison to the NLO result. For the SV+NSV resummed results, we notice an enhancement of 60\% and 36.2\% when $\rm \overline{LL}$ and $\rm \overline{NLL}$ are added to LO and NLO respectively at the central rapidity region. The rapidity distribution increases by 14.76\% when we include $\rm \overline{NNLL}$ to $\rm{NNLO_A}$. Further, at the central scale, the resummed rapidity distribution at NLO+$\rm \overline{NLL}$ ($25.6$ pb) mimics that at $\rm{NNLO_A}$ ($24$ pb) around the central rapidity region. We also study the K-factor values for the high mass region i.e $m_A=700$ GeV as given in Table \ref{tab:K700}. For the fixed order results, there is a large increment of 120\% when we go from LO to NLO. Interestingly the higher order effects at $\rm{NNLO_A}$ give rise to only 12.6\% correction to NLO around the central rapidity region. 
We find that there is an enhancement of 53.3\% and 24.97\% by the inclusion of $\rm \overline{LL}$ and $\rm \overline{NLL}$ resummed results at LO and NLO respectively around the central rapidity region. At $\rm{NNLO_A}$, the rapidity distribution increases by 11.48\% when we include $\rm \overline{NNLL}$. From Tables \ref{tab:K125} and \ref{tab:K700}, it can be observed that resummed predictions not only bring in considerable enhancement in the fixed order results, but also improve the perturbative convergence till $\rm {NNLO_A}+\rm {\overline{NNLL}}$ accuracy. \\ \\
\textbf{7-point scale variation}:  Next, we study the theoretical uncertainties due to the unphysical renormalization ($\mu_R$) and factorization ($\mu_F$) scales in our results using the standard canonical 7-point variation approach. Here, $\mu = \{ \mu_F, \mu_R \}$ is varied in the range $\frac{1}{2} \leq \frac{\mu}{m_A} \leq 2$, keeping the ratio $\mu_R/\mu_F$ not larger than 2 and smaller than 1/2. In figure \ref{fig:7pt125}, we depict the bin-integrated rapidity distribution of the pseudo-scalar Higgs boson for the fixed order results in the left panel and the resummed results in the right panel around the central scale $\mu_R = \mu_F = m_A$ for $m_A = 125$ GeV(top) and $m_A = 700$ GeV(bottom). We have provided the fixed order as well as SV+NSV resummed results for benchmark rapidity values at the central scale $\mu_R = \mu_F = m_A$ for $m_A=125$ GeV and  $m_A=700$ GeV in Table \ref{tab:7pt125} and \ref{tab:7pt700} respectively at various perturbative orders. These tables also contain the maximum increments and decrements from the corresponding central scale values obtained by varying $\{ \mu_R, \mu_F \}$ in the range $\{ 1/2, 2 \}m_A$.
\begin{table*}[ht]
\centering
\smaller
 \renewcommand{\arraystretch}{2.5}
\begin{tabular}{ |P{1.1cm}||P{1.99cm}|P{1.99cm}|P{1.99cm}||P{1.99cm}|P{1.99cm}|P{2.19cm}|}
 \hline
  y &LO&LO+$\rm{\overline{LL}}$&NLO&NLO+$\rm{\overline{NLL}}$&$\rm{NNLO_A}$&$\rm{NNLO_A}$+$\rm{\overline{NNLL}}$\\
 \hline
\hline

 0-0.4 & 10.225 $^{+ 2.645  }_{-1.902 }$ 
 & 16.379 $^{+  6.580 }_{- 3.064 }$ 
 & 18.805  $^{+ 3.862   }_{-2.953 }$ 
 & 25.623  $^{+9.360  }_{- 6.130 }$ 
 & 24.054 $^{+2.599  }_{-2.490  }$ 
 &27.605  $^{+ 5.873}_{-4.340}$  \\

  0.4-0.8 & 9.938 $^{+ 2.490 }_{- 1.860}$ 
 & 16.704 $^{+  5.7120 }_{-3.833  }$ 
 & 17.951  $^{+ 3.682  }_{- 2.815 }$ 
 & 24.543  $^{+8.994  }_{- 5.883 }$ 
 & 22.836 $^{+ 2.798 }_{- 2.428 }$ 
 & 26.281 $^{+ 5.675 }_{-3.959  }$  \\

 0.8-1.2 & 9.128 $^{+2.363  }_{-1.707 }$ 
 &  15.54 $^{+ 5.424 }_{- 3.581 }$ 
 &  16.362 $^{+ 3.350  }_{- 2.563 }$ 
 &  22.572  $^{+ 8.396 }_{-5.474  }$ 
 & 20.856 $^{+2.251  }_{-2.327  }$ 
 & 24.126 $^{+ 5.273 }_{-3.873  }$  \\

  1.2-1.6 &  8.033 $^{+ 2.080 }_{-1.495 }$ 
 &  13.763 $^{+4.820  }_{-3.171  }$ 
 & 14.034  $^{+ 2.863  }_{-2.193  }$ 
 & 19.546  $^{+ 7.315 }_{-4.765  }$ 
 & 18.067 $^{+1.825  }_{- 2.037 }$ 
 & 20.992 $^{+4.491  }_{- 3.508 }$  \\
 
  1.6-2.0 & 6.698 $^{+ 1.714 }_{-1.252 }$ 
 & 11.711 $^{+ 4.123 }_{-2.723  }$ 
 &  11.311 $^{+  2.299  }_{-1.762  }$ 
 & 16.061  $^{+ 6.203 }_{- 4.008 }$ 
 &14.412  $^{+  1.607 }_{-1.566  }$ 
 & 16.968 $^{+ 3.926 }_{-2.906  }$  \\

\hline
\end{tabular}
\caption{Values of resummed rapidity distribution at various orders in comparison to the fixed order results in pb at the central scale $\mu_R=\mu_F= m_A=125$ GeV for 13 TeV LHC.} \label{tab:7pt125}
\end{table*}

\begin{table*}[ht]
\centering
\smaller
 \renewcommand{\arraystretch}{2.5}
\begin{tabular}{ |P{1.1cm}||P{1.99cm}|P{1.99cm}|P{1.99cm}||P{1.99cm}|P{1.99cm}|P{2.19cm}|}
 \hline
  y &LO&LO+$\rm{\overline{LL}}$&NLO&NLO+$\rm{\overline{NLL}}$&$\rm{NNLO_A}$&$\rm{NNLO_A}$+$\rm{\overline{NNLL}}$\\
 \hline
\hline
 0-0.4 & 0.292 $^{+ 0.088 }_{- 0.065}$ 
 & 0.123 $^{+0.158  }_{-0.1098  }$ 
 &  0.643 $^{+ 0.0132  }_{- 0.1078 }$ 
 &  0.803 $^{+ 0.1178 }_{- 0.0943 }$ 
 & 0.724 $^{+ 0.0613 }_{- 0.07045 }$ 
 &  0.807$^{+ 0.0735 }_{- 0.0625 }$  \\
 
 0.4-0.8 & 0.257 $^{+ 0.0804 }_{- 0.0574}$ 
 &  0.399 $^{+ 0.14379 }_{-0.09845  }$ 
 & 0.567  $^{+ 0.1179 }_{- 0.0959 }$ 
 & 0.710 $^{+  0.1040 }_{-  0.0834}$ 
 &  0.622 $^{+ 0.0543  }_{- 0.0596 }$ 
 & 0.697 $^{+ 0.06733 }_{-  0.0566}$ \\
 
 0.8-1.2 & 0.197  $^{+ 0.0633 }_{- 0.0450 }$ 
 & 0.311 $^{+ 0.1158 }_{- 0.0788 }$ 
 & 0.434  $^{+0.0922  }_{- 0.0749 }$ 
 & 0.546 $^{+ 0.07919 }_{- 0.0641 }$ 
 & 0.456 $^{+ 0.0428 }_{- 0.0424 }$ 
 & 0.515 $^{+ 0.0543 }_{-  0.0379}$ \\

 1.2-1.6 & 0.122 $^{+ 0.0417 }_{- 0.0291 }$ 
 & 0.201 $^{+ 0.0803 }_{- 0.0535 }$ 
 & 0.434 $^{+ 0.06  }_{- 0.0486 }$ 
 & 0.343 $^{+ 0.0497}_{-0.0405 }$ 
 & 0.276 $^{+ 0.02322 }_{- 0.02688 }$ 
 & 0.316 $^{+ 0.03399 }_{- 0.0291 }$  \\

 1.6-2.0 & 0.052  $^{+ 0.0195 }_{- 0.0132 }$ 
 & 0.093 $^{+0.0416  }_{- 0.0267 }$ 
 & 0.116 $^{+ 0.0279 }_{- 0.0224 }$ 
 & 0.153 $^{+ 0.0232}_{- 0.0181 }$ 
 & 0.123  $^{+ 0.01131 }_{-  0.0125}$ 
 & 0.144 $^{+ 0.0177}_{-0.01365 }$ \\
 
\hline
\end{tabular}
\caption{Values of resummed rapidity distribution at various orders in comparison to the fixed order results in pb at the central scale $\mu_R=\mu_F= m_A=700$ GeV for 13 TeV LHC.} \label{tab:7pt700}
\end{table*}

\begin{figure}[H]
\begin{center}
\includegraphics[scale=0.350]{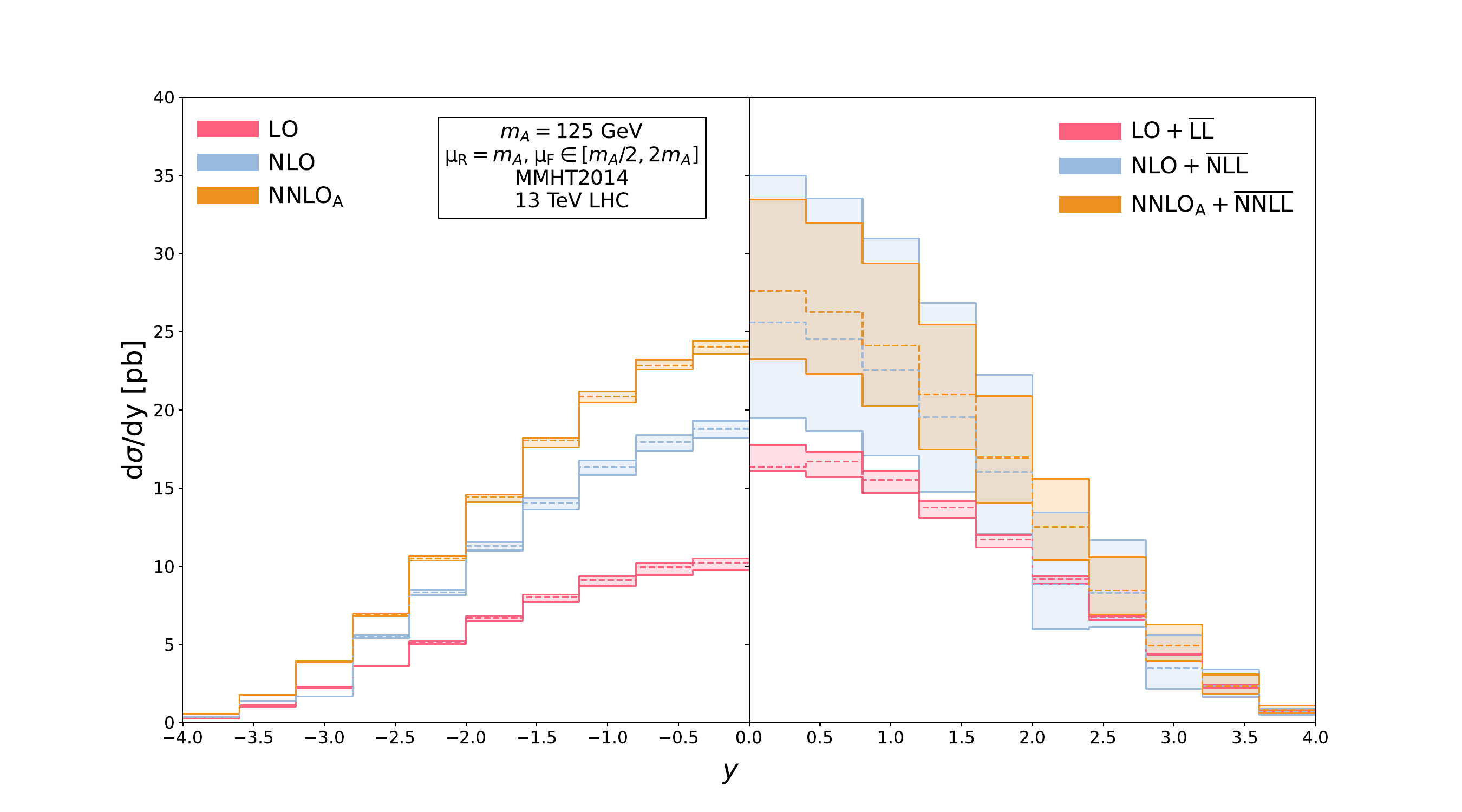}
\includegraphics[scale=0.350]{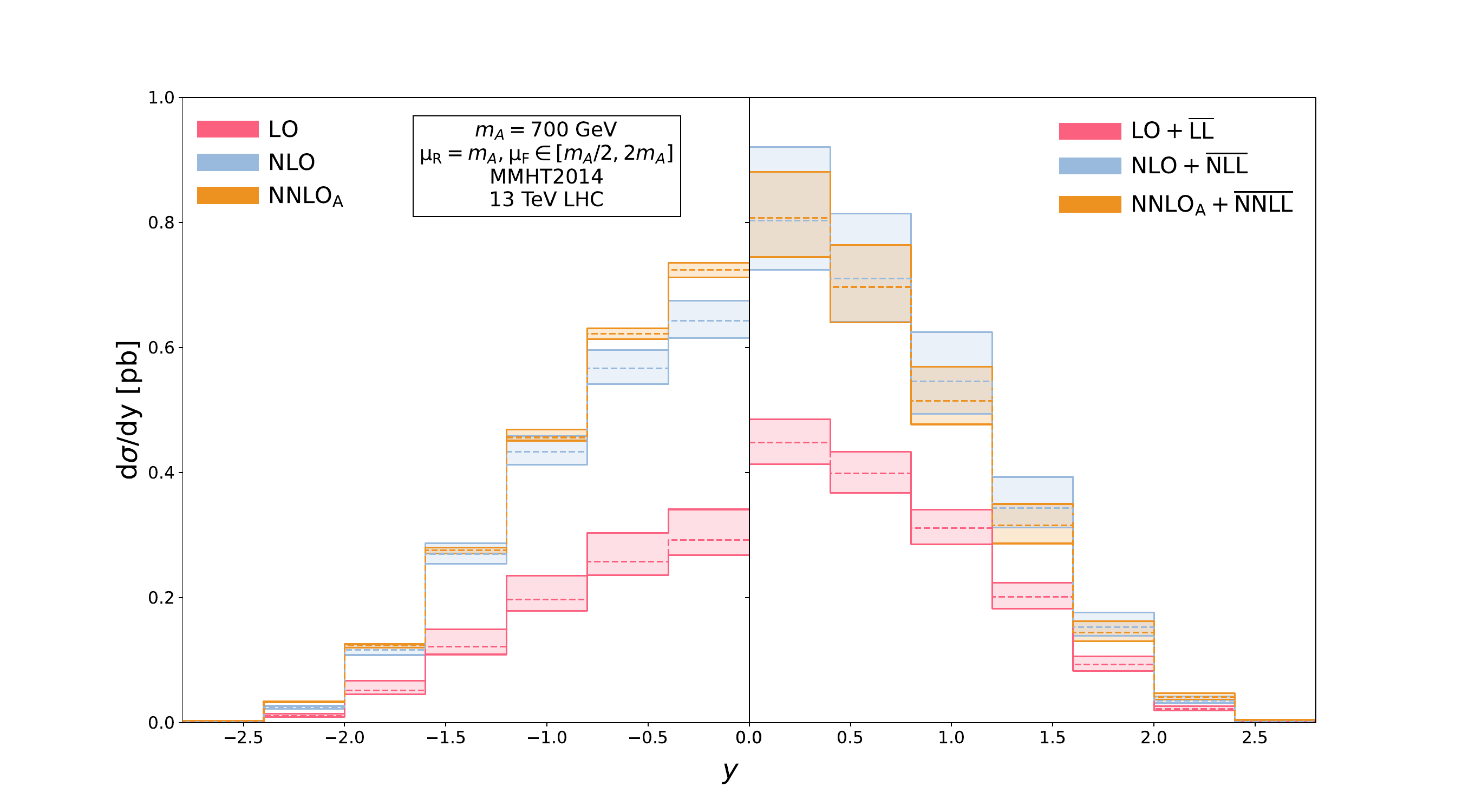}
\caption{Comparison of $\mu_F$ scale variation between fixed order and SV+NSV resummed results for $m_A=125$ (top) and $m_A=700$ (bottom) GeV.
The dashed lines refer to the corresponding central scale values at each order.}
\label{fig:MuF125}
\end{center}
\end{figure}

Let us first look at the plot with $m_A = 125$ GeV in figure \ref{fig:7pt125}. This plot shows that the addition of SV+NSV resummed results to the fixed order ones increases the rapidity distribution at each order up to $\rm {NNLO_A}$ in perturbation theory. However, the percentage enhancement in the rapidity distribution decreases from 60\% at LO to 14.76\% at ${\rm NNLO_A}$
by the inclusion of $\rm \overline{LL}$ and $\rm \rm {\overline{NNLL}}$ respectively at the central rapidity region. This indicates better perturbative convergence of the truncated series at higher orders due to the addition of the resummed predictions. We now compare the 7-point uncertainties of fixed order and SV+NSV resummed results due to $\mu_R$ and $\mu_F$ scales. We find that the combined uncertainty due to $\mu_R$ and $\mu_F$ scales lies in the range (+25.87\%,-18.60\%) at LO while at ${\rm NNLO_A}$, it gets substantially reduced to (+10.80\%,-10.35\%) around the central rapidity region. We see that the bands of resummed predictions till ${\rm NNLO_A}+\rm \rm {\overline{NNLL}}$ are wider than that of the corresponding fixed order results throughout the rapidity spectrum for $m_A = 125$ GeV. Numerically, the combined uncertainty due to these unphysical scales lies between (+40.17\%,-18.71\%) at $\rm LO + {\overline{LL}}$, (+36.53\%,-23.92\%) at $\rm NLO + {\overline{NLL}}$ and (+21.27\%, -15.72\%) at ${\rm NNLO_A} + {\rm \overline{NNLL}}$ order around $y=0$. This shows that there is a systematic decrease in the uncertainty when we go to higher logarithmic accuracy for SV+NSV resummed results. The plot for $m_A=700$ GeV in figure \ref{fig:7pt125} shows a similar trend of enhancement in the rapidity distribution by the addition of SV+NSV resummed results as was depicted above. The 7-point uncertainty values show that at lower orders, the resummed results show significantly more $\mu_R$ and $\mu_F$ variation as compared to the fixed order ones similar to the case of $m_A=125$ GeV. However, at ${\rm NNLO_A}+ \rm {\overline{NNLL}}$ accuracy, the combined uncertainty of the resummed result lies in the range (+9.11\%,-7.74\%) which is comparable to the uncertainty of (+8.47\%, -9.73\%) for the fixed order prediction at ${\rm NNLO_A}$ around central rapidity region. Thus, the SV+NSV resummed results become more relevant for higher values of pseudo-scalar Higgs boson mass. The above analysis suggests the need to understand the behavior of the resummed results w.r.t $\mu_R$ and $\mu_F$ scale variations in a better way. 
Hence, we study the impact of each scale individually by keeping the other fixed. \\ \\
\textbf{Uncertainties due to $\mu_R$ and $\mu_F$ scales individually}: We now, discuss the effect of the factorization scale $\mu_F$ individually by keeping the renormalization scale $\mu_R$ fixed. Figure \ref{fig:MuF125} shows the bin-integrated rapidity distributions for the fixed order(left panel) as well as the SV+NSV resummed results(right panel) at various perturbative orders for $m_A = 125$ GeV (top) and $m_A = 700$ GeV (bottom) keeping the renormalization scale fixed at $\mu_R = m_A$. The factorization scale is varied in the range $\{1/2, 1\}m_A$ around the central scale $\mu_F = \mu_R = m_A$ to get the uncertainty bands. The fixed order results show negligible dependence on the $\mu_F$ scale both at $m_A = 125$ GeV and $m_A = 700$ GeV. On the other hand, the resummed predictions show substantial dependence w.r.t the $\mu_F$ scale especially at $m_A = 125$ GeV. The uncertainty lies in the range (+36.53 \%, - 23.92\%) at $\rm NLO + {\overline{NLL}}$ accuracy which comes down to (+ 21.27\%, - 15.72\%) at  ${\rm NNLO_A}+ {\rm \overline{NNLL}}$ accuracy around central rapidity region for $m_A = 125$ GeV. When we compare these $\mu_F$ scale uncertainty values with those at $m_A = 700$ GeV, we find that it vary in the range (+ 14.66 \%, - 9.88\%) and (+9.11 \%, - 7.75\%) at $\rm NLO + {\overline{NLL}}$ and $\rm{NNLO}_A + {\overline{NNLL}}$ order respectively around $y=0$. Hence, as suggested by the 7-point scale variation analysis, the uncertainty decreases considerably at the higher value of the pseudo-scalar Higgs Boson mass. The uncertainty due to the factorisation scale decreases at higher orders for both the cases of $m_A$. Also, the higher order uncertainty bands lie within the lower order ones. These two observations hint towards improved reliability of the perturbative results and better perturbative convergence at higher orders. 
\begin{figure}[H]
\begin{center}
\includegraphics[scale=0.350]{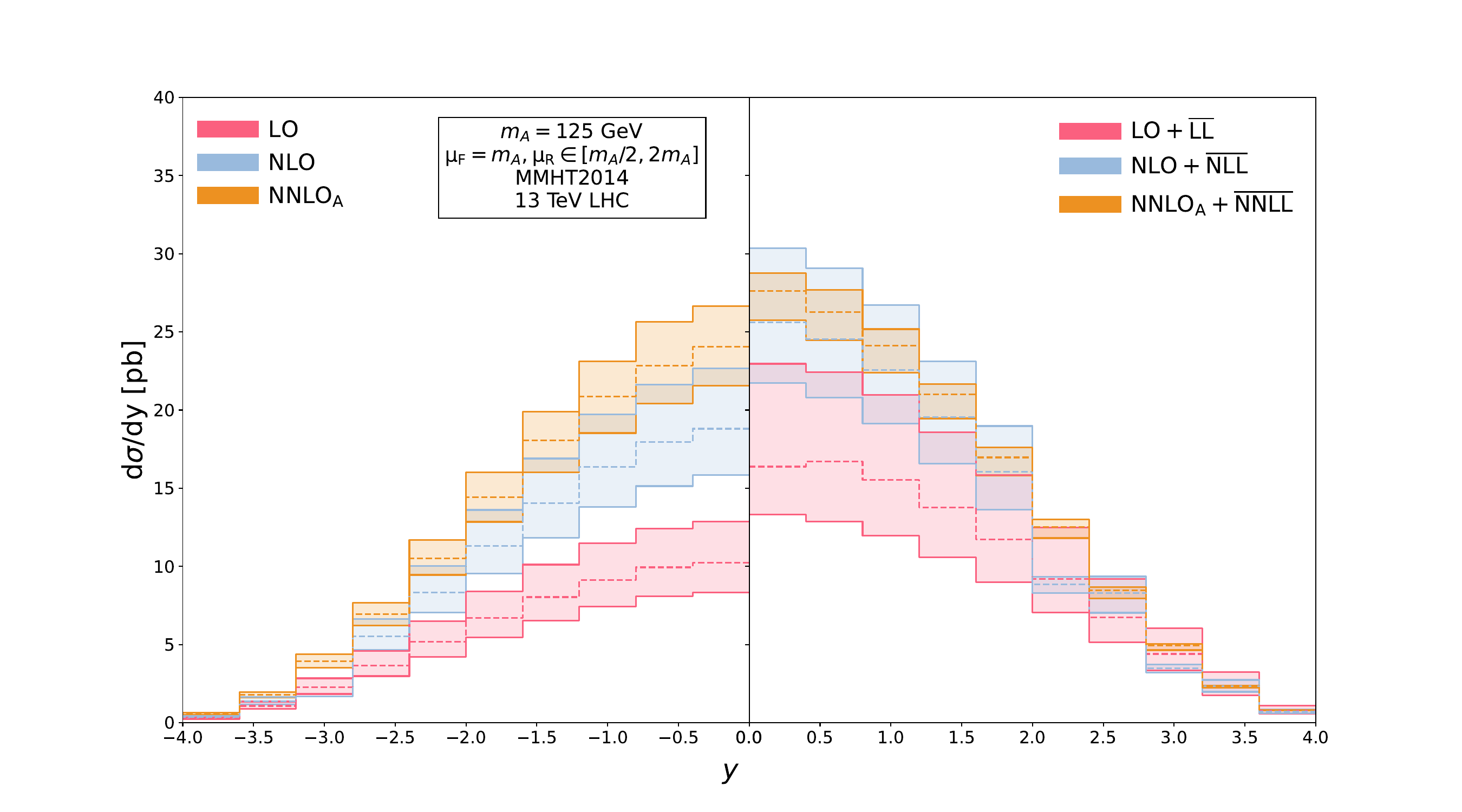}
\includegraphics[scale=0.350]{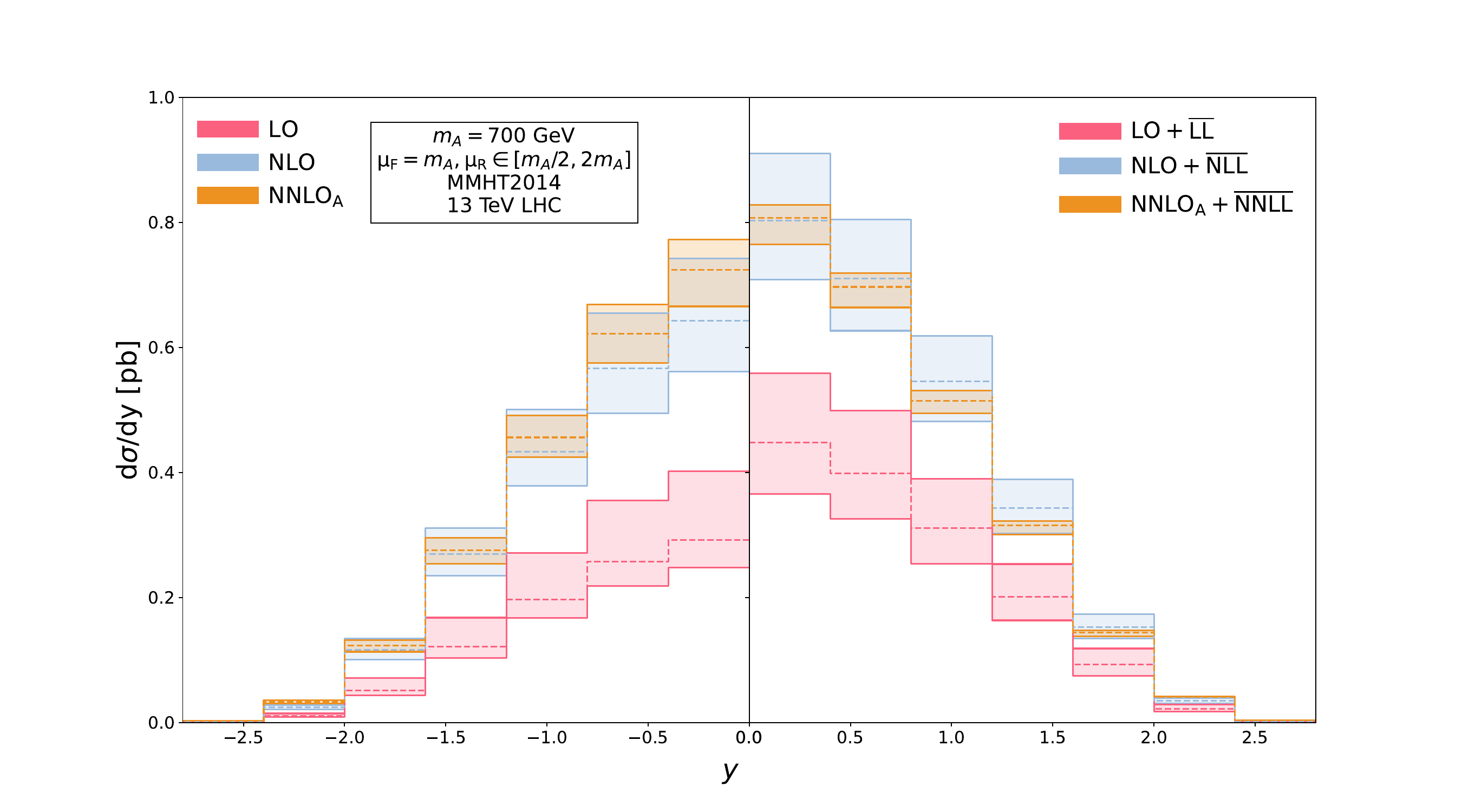}
\caption{Comparison of $\mu_R$ scale variation between fixed order and SV+NSV resummed results for $m_A=125$ (top) and $m_A=700$ (bottom) GeV.
The dashed lines refer to the corresponding central scale values at each order.}
\label{fig:MuR125}
\end{center}
\end{figure}

Next, we study the variation in the fixed order and SV+NSV resummed results w.r.t the renormalisation scale by keeping the factorisation scale fixed at $\mu_F = m_A$. The uncertainty bands are obtained by varying $\mu_R$ in the range $\{1/2, 1\}m_A$ around the central scale $\mu_F = \mu_R = m_A$. In figure \ref{fig:MuR125}, we observe that from the NLO level, the $\mu_R$ scale uncertainty of the fixed order results decreases by the addition of resummed predictions for $m_A = 125$ GeV as well as $m_A = 700$ GeV. The numerical values of the $\mu_R$ uncertainties lie between (+18.45\%, -15.24\%) and (+4.16\%, -6.69\%) at $\rm NLO + {\overline{NLL}}$ and $\rm{NNLO_A} + {\overline{NNLL}}$ order respectively which is a considerable reduction from (+20.54\%, -15.70\%) and (+10.80\%, -10.35\%) at NLO and $\rm{NNLO_A}$ accuracy respectively around $y=0$ for $m_A = 125$ GeV. Similarly, for the case of $m_A = 700$ GeV, they lie in the range (+13.33 \%, - 11.74\%) and (+ 2.62\%, - 5.19\%) at $\rm NLO + {\overline{NLL}}$ and $\rm{NNLO_A} + {\overline{NNLL}}$ respectively whereas they vary between (+15.48 \%, - 12.69\%) and  (+6.76 \%, - 8.04\%) for NLO and $\rm{NNLO_A}$ order respectively around central rapidity region. From the above percentages, we also find that the uncertainty decreases as we go to higher orders for both cases of pseudo-scalar Higgs boson masses. In addition, the uncertainty bands of resummed results at higher orders are well within the lower orders from NLO level onwards.

Here, we performed a comparative study between the fixed order results and the SV+NSV resummed predictions for the rapidity distribution of pseudo-scalar Higgs boson in gluon fusion process.
This has been done through the K-factor analysis, 7-point variation approach, and finally by studying the variation of $\mu_F$ and $\mu_R$ scales individually. We did the analysis for two different cases of pseudo-scalar Higgs boson masses $m_A = 125, 700$ GeV. The K-factor analysis showed that the inclusion of SV+NSV resummed predictions resulted in the enhancement of the fixed order results at every order in perturbation theory up to ${\rm NNLO_A}$ accuracy for both the cases of $m_A$. Also, we observed that the percentage enhancement by adding the resummed results decreases as we go from LO to ${\rm NNLO_A}$ accuracy. This shows that the resummed results are more reliable and have a better perturbative convergence. The study of factorisation scale variation showed that the addition of the resummed results especially at $m_A = 125$ GeV significantly increased the uncertainty of the fixed order results, which otherwise was almost independent of the $\mu_F$ scale variation. However, the dependence of the resummed results on the $\mu_F$ scale decreases considerably for the case of $m_A = 700$ GeV. The renormalisation scale dependence, on the other hand, gets improved by the inclusion of resummed predictions. In order to understand this behavior of SV+NSV resummed results in a better way, we compare them with the well established SV resummed results at various orders in the next section. \\ \\
\textbf{SV+NSV vs SV resummed predictions:} 
In previous sections, we presented the observations on the behavior of SV+NSV resummed corrections by comparing them with the fixed-order results at various perturbative orders. 
Here, we try to understand the reasons behind those observations by comparing the full SV+NSV resummed predictions with the well-established SV resummed results which would help us to infer the behavior of resummed NSV logarithms in particular.
\begin{table*}[ht]
\centering
 \renewcommand{\arraystretch}{1.9}
\begin{tabular}{ |P{1.2cm}||P{1.3cm}|P{1.3cm}|P{1.8cm}|P{1.8cm} |P{2.3cm}|P{2.3cm}|}
 \hline
  y &$\rm{K_{LO+\rm{LL}}}$&$\rm{K_{LO+\rm{\overline{LL}}}}$&$\rm{K_{NLO+\rm{NLL}}}$&$\rm{K_{NLO+\rm{\overline{NLL}}}}$&$\rm{K_{NNLO_A+\rm{NNLL}}}$&$\rm{K_{NNLO_A+\rm{\overline{NNLL}}}}$\\
 \hline
\hline
  0-0.4  &  1.411 & 1.602 &  2.159  & 2.505 & 2.502  & 2.699 \\
 0.4-0.8 &1.410 & 1.681 & 2.126  & 2.469 &  2.447   &  2.644 \\
 0.8-1.2 &1.428 & 1.703 & 2.125  & 2.472 &  2.441  &  2.643\\
 1.2-1.6 &1.437 & 1.713 & 2.086  & 2.433 &  2.409  & 2.613 \\
 1.6-2.0 & 1.466 & 1.748 &  2.047 &  2.397 & 2.324   & 2.533 \\
\hline
\end{tabular}
\caption{K-factor values of SV and SV+NSV resummed results at the central scale $\mu_R=\mu_F= m_A=125$ GeV.} \label{tab:SVNSV125}
\end{table*}
\begin{table*}[ht]
 \renewcommand{\arraystretch}{1.9}
\begin{tabular}{ |P{1.2cm}||P{1.3cm}|P{1.3cm}|P{1.8cm}|P{1.8cm} |P{2.3cm}|P{2.3cm}|}
 \hline
  y &$\rm{K_{LO+\rm{LL}}}$&$\rm{K_{LO+\rm{\overline{LL}}}}$&$\rm{K_{NLO+\rm{NLL}}}$&$\rm{K_{NLO+\rm{\overline{NLL}}}}$&$\rm{K_{NNLO_A+\rm{NNLL}}}$&$\rm{K_{NNLO_A+\rm{\overline{NNLL}}}}$\\
 \hline
\hline
 0-0.4   & 1.375 & 1.533& 2.530  &  2.749& 2.630& 2.763  \\
 0.4-0.8 & 1.388  & 1.547 &2.536  & 2.755 & 2.571 & 2.703  \\
 0.8-1.2 &  1.416 & 1.579 & 2.550 & 2.769 & 2.479 &2.613  \\
 1.2-1.6 & 1.480 &  1.653 &2.596  & 2.819 & 2.451 & 2.593 \\
 1.6-2.0 & 1.6035 & 1.797 &2.703  & 2.947  & 2.612 & 2.781 \\
\hline
\end{tabular}
\caption{K-factor values of SV and SV+NSV resummed results at the central scale $\mu_R=\mu_F= m_A=700$ GeV.} \label{tab:SVNSV700}
\end{table*}

We begin our analysis with the K-factor values of SV and SV+NSV resummed results. We provide the K-factor values of SV and SV+NSV resummed results at various perturbative orders for benchmark rapidity values for $m_A = 125, 700$ GeV in Tables \ref{tab:SVNSV125} and \ref{tab:SVNSV700}. Looking at the values in Tables \ref{tab:SVNSV125} and \ref{tab:SVNSV700}, we find that the addition of resummed NSV logarithms enhances the SV resummed predictions for the rapidity distribution at each order in perturbation theory for both the values of $m_A$. For instance, there is an enhancement of 16.03\% and 7.87\% by the inclusion of resummed NSV logarithms at NLO+${\rm \overline{NLL}}$ and $\rm{NNLO_A} + {\overline{NNLL}}$  respectively for $m_A = 125$ GeV. Similarly, for $m_A = 700$ GeV, the rapidity distribution increases by 8.66\% and 5.06\% when we go from NLO+NLL and $\rm{NNLO_A} + NNLL$  to $\rm NLO + {\overline{NLL}}$ and $\rm{NNLO_A} + {\overline{NNLL}}$ respectively. We also observe that the percentage enhancement in the rapidity distribution due to the resummed NSV logarithms decreases as we go from $\rm NLO + {\overline{NLL}}$ to $\rm{NNLO_A} + {\overline{NNLL}}$. This suggests better perturbative convergence of the SV+NSV resummed result which was already noticed while comparing it with the fixed order results.
\begin{figure*}[ht]
\includegraphics[scale=0.350]{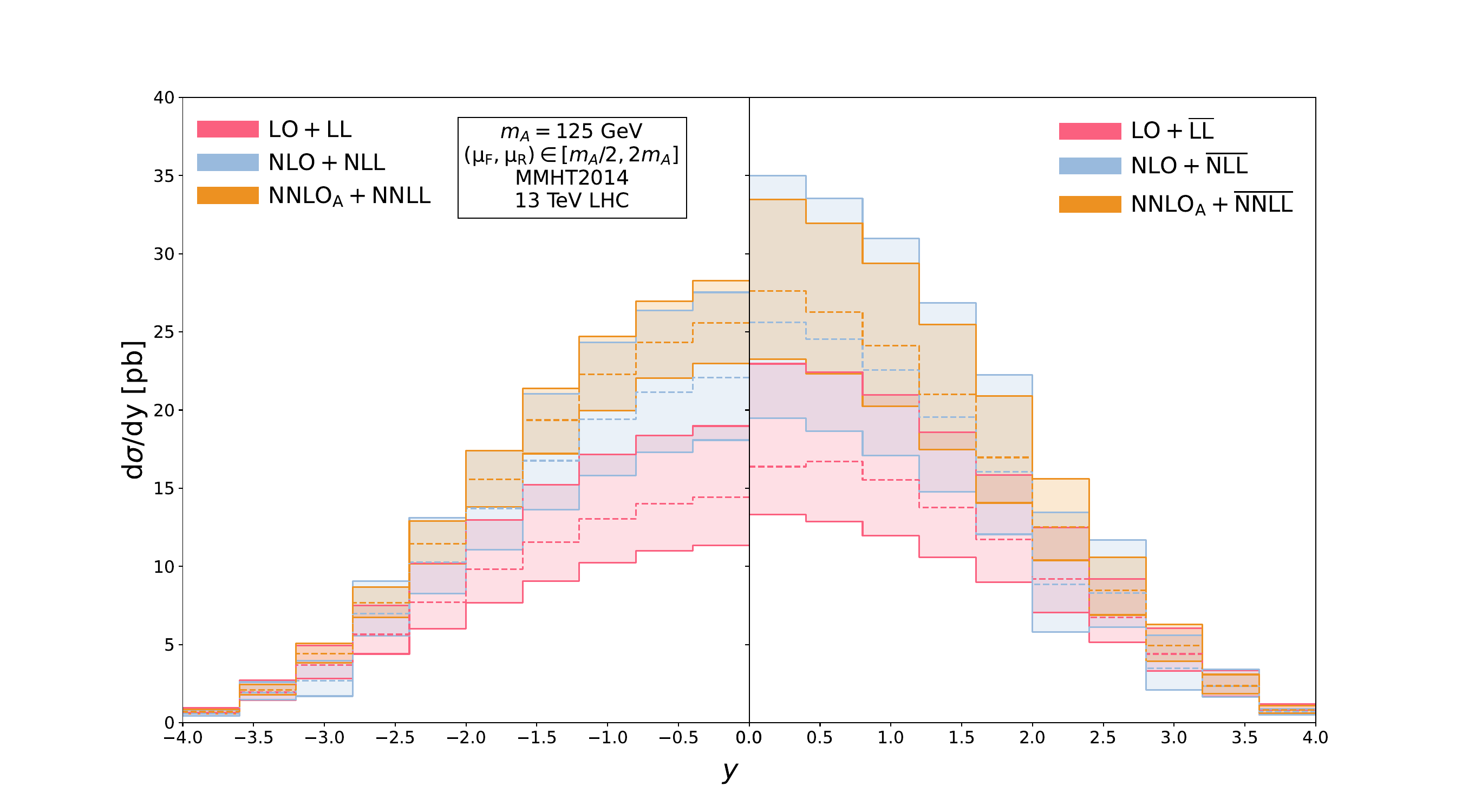}
\includegraphics[scale=0.350]{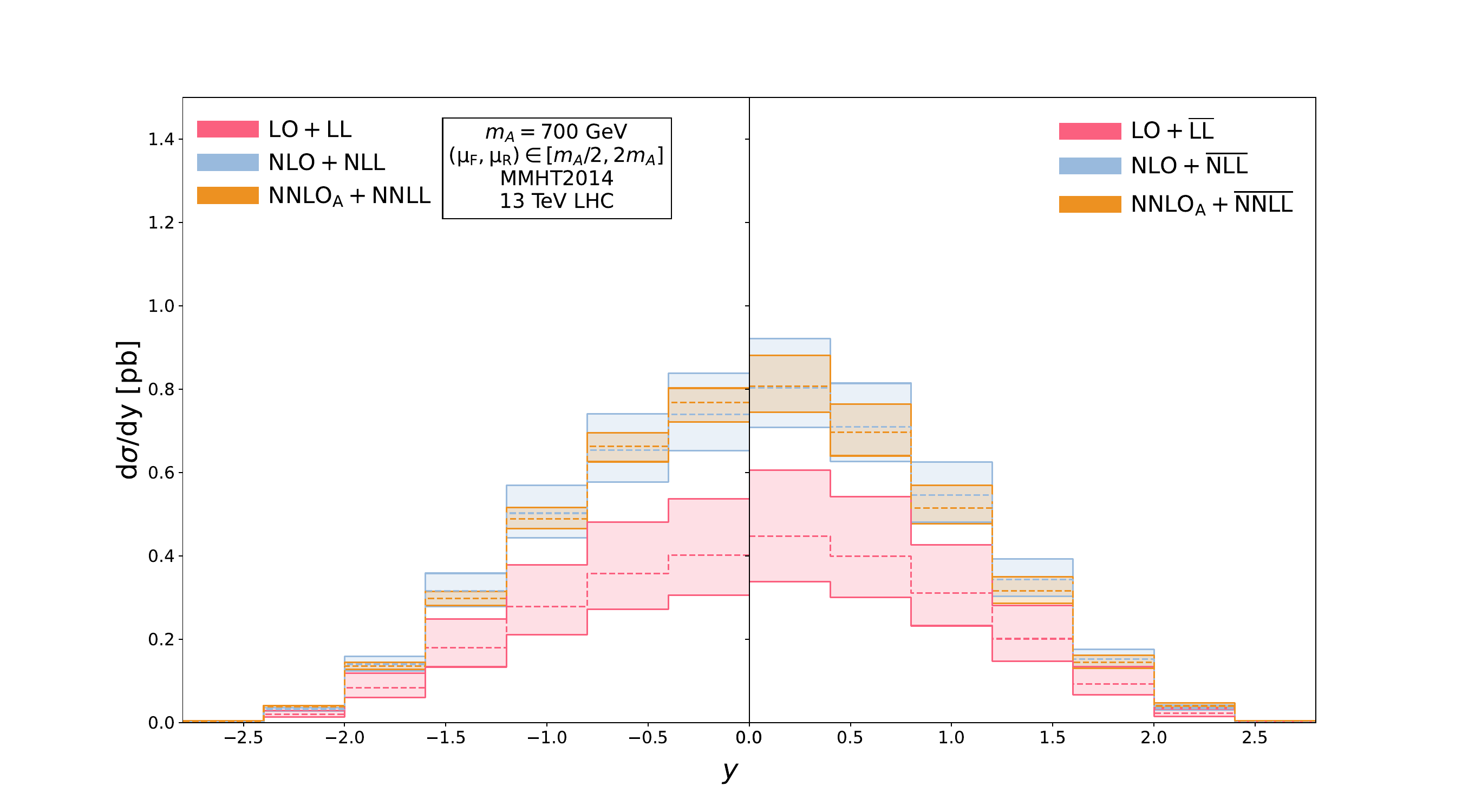}
\caption{Comparison of 7-point scale variation between SV and SV+NSV resummed results for $m_A=125$ (top) and 700 (bottom) GeV. The
dashed lines refer to the corresponding central scale values at each order.}
\label{fig:7ptsvnsv}
\end{figure*}

Next, we study the uncertainties of the resummed NSV logarithms w.r.t the $\mu_R$ and $\mu_F$ scale variations. We first present the plots for canonical 7-point scale variation of the bin-integrated rapidity distribution in figure \ref{fig:7ptsvnsv} for the pseudo-scalar Higgs boson mass, $m_A = 125, 700$ GeV in the top and bottom panels respectively. The scales $\mu = \{ \mu_F, \mu_R \}$ are varied in the range $\frac{1}{2} \leq \frac{\mu}{m_A} \leq 2$, keeping the ratio $\mu_R/\mu_F$ not larger than 2 and smaller than 1/2 around the central scale $\mu_R = \mu_F = m_A$. 

\begin{table*}[ht]
\smaller
 \renewcommand{\arraystretch}{2.8}
\begin{tabular}{ |P{1.1cm}||P{1.99cm}|P{1.99cm}|P{1.99cm}||P{1.99cm}|P{2.3cm}|P{2.3cm}|}
 \hline
  y &LO+$\rm{LL}$&LO+$\rm{\overline{LL}}$&NLO+$\rm{NLL}$&NLO+$\rm{\overline{NLL}}$&$\rm{NNLO}_A+\rm{NNLL}$&$\rm{NNLO}_A+\rm{\overline{NNLL}}$\\
 \hline
\hline
 0-0.4 & 14.430 $^{+ 4.547  }_{-3.090 }$ 
 & 16.379 $^{+  6.580 }_{- 3.064 }$ 
 &  22.086 $^{+ 5.446   }_{-4.008 }$ 
 & 25.623  $^{+9.360  }_{- 6.130 }$ 
 & 25.583 $^{+ 2.696 }_{- 2.609}$ 
 &27.605  $^{+ 5.873}_{-4.340}$  \\

  0.4-0.8 & 14.022 $^{+4.339  }_{-3.015 }$ 
 & 16.704 $^{+  5.712}_{-3.833  }$ 
 & 21.137  $^{+ 5.237  }_{- 3.846 }$ 
 & 24.543  $^{+8.994  }_{- 5.883 }$ 
 & 24.326 $^{+2.621  }_{- 2.292 }$ 
 & 26.281 $^{+ 5.675 }_{-3.959  }$  \\

 0.8-1.2 & 13.035 $^{+4.135  }_{-2.814 }$ 
 &  15.54 $^{+ 5.424 }_{- 3.581 }$ 
 & 19.402  $^{+ 4.919  }_{-3.587  }$ 
 &  22.572  $^{+ 8.396 }_{-5.474  }$ 
 &  22.287 $^{+ 2.434 }_{-2.322  }$ 
 & 24.126 $^{+ 5.273 }_{-3.873  }$  \\

  1.2-1.6 & 11.546  $^{+ 3.677 }_{-2.491 }$ 
 &  13.763 $^{+4.820  }_{-3.171  }$ 
 & 16.759  $^{+ 4.291  }_{-3.120 }$ 
 & 19.546  $^{+ 7.315 }_{-4.765  }$ 
 & 19.3604 $^{+ 2.012 }_{-2.148 }$ 
 & 20.992 $^{+4.491  }_{- 3.508 }$  \\
 
  1.6-2.0 & 9.823 $^{+ 3.140 }_{-2.140 }$ 
 & 11.711 $^{+ 4.123 }_{-2.723  }$ 
 & 13.715  $^{+ 3.681  }_{- 2.636 }$ 
 & 16.061  $^{+ 6.203 }_{- 4.008 }$ 
 &15.571 $^{+ 1.843  }_{-1.764  }$ 
 & 16.968 $^{+ 3.926 }_{-2.906  }$  \\
 
\hline
\end{tabular}
\caption{Values of SV+NSV resummed rapidity distribution at various orders in comparison to the SV resummed results in pb at the central scale $\mu_R=\mu_F= m_A=125$ GeV for 13 TeV LHC.} \label{tab:SVNSV7pt125}
\end{table*}

\begin{table*}[ht]
\smaller
 \renewcommand{\arraystretch}{2.7}
\begin{tabular}{ |P{1.1cm}||P{1.99cm}|P{1.99cm}|P{1.99cm}||P{1.99cm}|P{2.3cm}|P{2.3cm}|}
 \hline
  y &LO+$\rm{LL}$&LO+$\rm{\overline{LL}}$&NLO+$\rm{NLL}$&NLO+$\rm{\overline{NLL}}$&NNLO$_A$+$\rm{NNLL}$&NNLO$_A$+$\rm{\overline{NNLL}}$\\
 \hline
\hline
 0-0.4 & 0.4018 $^{+ 0.135 }_{- 0.094 }$ 
  & 0.123 $^{+0.158  }_{-0.1098  }$  
 &  0.7391$^{+ 0.0990 }_{- 0.0863}$ 
  &  0.803 $^{+ 0.1178 }_{- 0.0943 }$ 
 & 0.7685 $^{+ 0.0340 }_{- 0.0466 }$ 
  &  0.807$^{+ 0.0735 }_{- 0.0625 }$ \\
 
 0.4-0.8 & 0.3577 $^{+ 0.123 }_{- 0.0855 }$ 
  &  0.399 $^{+ 0.14379 }_{-0.09845  }$ 
 & 0.6538 $^{+ 0.0875 }_{- 0.0763 }$ 
  & 0.710 $^{+  0.1040 }_{-  0.0834}$ 
 & 0.6628 $^{+ 0.0325 }_{- 0.0368 }$ 
  & 0.697 $^{+ 0.06733 }_{-  0.0566}$ \\

 0.8-1.2 & 0.2791 $^{+ 0.099 }_{- 0.0685 }$ 
  & 0.311 $^{+ 0.1158 }_{- 0.0788 }$ 
 & 0.5026 $^{+ 0.0672}_{- 0.0586 }$ 
  & 0.546 $^{+ 0.07919 }_{- 0.0641 }$
 & 0.4886 $^{+ 0.0276 }_{- 0.0233 }$ 
 & 0.515 $^{+ 0.0543 }_{-  0.0379}$  \\

 1.2-1.6 & 0.1803 $^{+ 0.0686}_{- 0.0465}$ 
  & 0.201 $^{+ 0.0803 }_{- 0.0535 }$ 
 & 0.3162 $^{+ 0.0423 }_{- 0.0369 }$ 
 & 0.343 $^{+ 0.0497}_{-0.0405 }$ 
 & 0.2985 $^{+ 0.0167 }_{- 0.0178 }$ 
  & 0.316 $^{+ 0.03399 }_{- 0.0291 }$  \\
 
 1.6-2.0 & 0.0831 $^{+ 0.0355}_{- 0.0231 }$ 
  & 0.093 $^{+0.0416  }_{- 0.0267 }$ 
 & 0.1403 $^{+ 0.0186 }_{- 0.0164 }$ 
 & 0.153 $^{+ 0.0232}_{- 0.0181 }$ 
 & 0.1355 $^{+ 0.009 }_{- 0.008 }$ 
 & 0.144 $^{+ 0.0177}_{-0.01365 }$  \\
\hline
\end{tabular}
\caption{Values of SV+NSV resummed rapidity distribution at various orders in comparison to the SV resummed results in pb at the central scale $\mu_R=\mu_F= m_A=700$ GeV for 13 TeV LHC.} \label{tab:SVNSV7pt700}
\end{table*}

We also provide Tables \ref{tab:SVNSV7pt125} and \ref{tab:SVNSV7pt700} with the numerical values of the SV and SV+NSV resummed rapidity distributions at the central scale for benchmark rapidity values for $m_A = 125, 700$ GeV respectively. These tables also contain the corresponding maximum increments and decrements in the rapidity distribution from the central scale values. From figure \ref{fig:7ptsvnsv}, we observe that the inclusion of resummed NSV logarithms to the SV resummed predictions increases the 7-point scale uncertainty tremendously at $m_A = 125$ GeV for each perturbative order till $\rm{NNLO_A}$. Quantitatively, the uncertainty lies between (+24.66\%, -18.15\%) and (+10.54\%, -10.2\%) for NLO+NLL and $\rm{NNLO_A}$+NNLL respectively around $y=0$. When we include the resummed NSV logarithms to these predictions, the uncertainty increases to (+36.53\%, -23.92\%) and (+21.28\%, -15.72\%) at $\rm NLO + {\overline{NLL}}$ and $\rm{NNLO_A} + {\overline{NNLL}}$ respectively. However, for $m_A = 700$ GeV, the increase in the 7-point uncertainty due to the addition of resummed NSV logarithms is not very large. For instance, the uncertainty varies between (+14.67\%, -11.74\%) and (+9.11\%, -7.74\%) for $\rm NLO + {\overline{NLL}}$ and $\rm{NNLO_A} + {\overline{NNLL}}$ respectively which is not significantly higher than (+13.40\%, -11.68\%) and (+4.42\%, -6.06\%) at NLO+NLL and NNLO+NNLL respectively around central rapidity region. We also observe that the bands of SV+NSV resummed results at $\rm{NNLO_A} + {\overline{NNLL}}$ are completely within the bands of $\rm NLO + {\overline{NLL}}$ results for both the values of $m_A$. On the other hand, this is not the case with SV resummed results at $m_A = 125$ GeV. This suggests that the inclusion of resummed NSV logarithms improves the convergence of the perturbative result especially for $m_A = 125$ GeV.

Before moving forward to the comparison of SV and SV+NSV resummed predictions under the variation of $\mu_R$ and $\mu_F$ scales individually, we would like to make few comments which would help in the better understanding of our results. The resummed predictions that we compute numerically for the phenomenological analysis, when truncated to a particular logarithmic accuracy, contains not only the distributions and logarithms that we are resumming using the all-order structure but also certain spurious terms. These spurious terms arise from the “inexact” Mellin inversion of the $N$-space resummed result and are beyond the precision of the resummed quantity. For instance, the spurious terms developed in the SV resummation are at the NSV and beyond NSV accuracy and those developed through NSV resummation are beyond NSV accuracy in perturbative QCD. We have discussed the effect of these spurious terms in our numerical results in great detail for the case of inclusive cross-section and the rapidity distribution of the Higgs Boson production through gluon fusion in Refs.\cite{Ajjath:2021bbm,Ravindran:2022aqr}. The same behaviour is expected to be followed by the SV and SV+NSV resummed results of rapidity distribution of the pseudo-scalar Higgs Boson as well. 
\begin{figure*}[ht]
\includegraphics[scale=0.350]{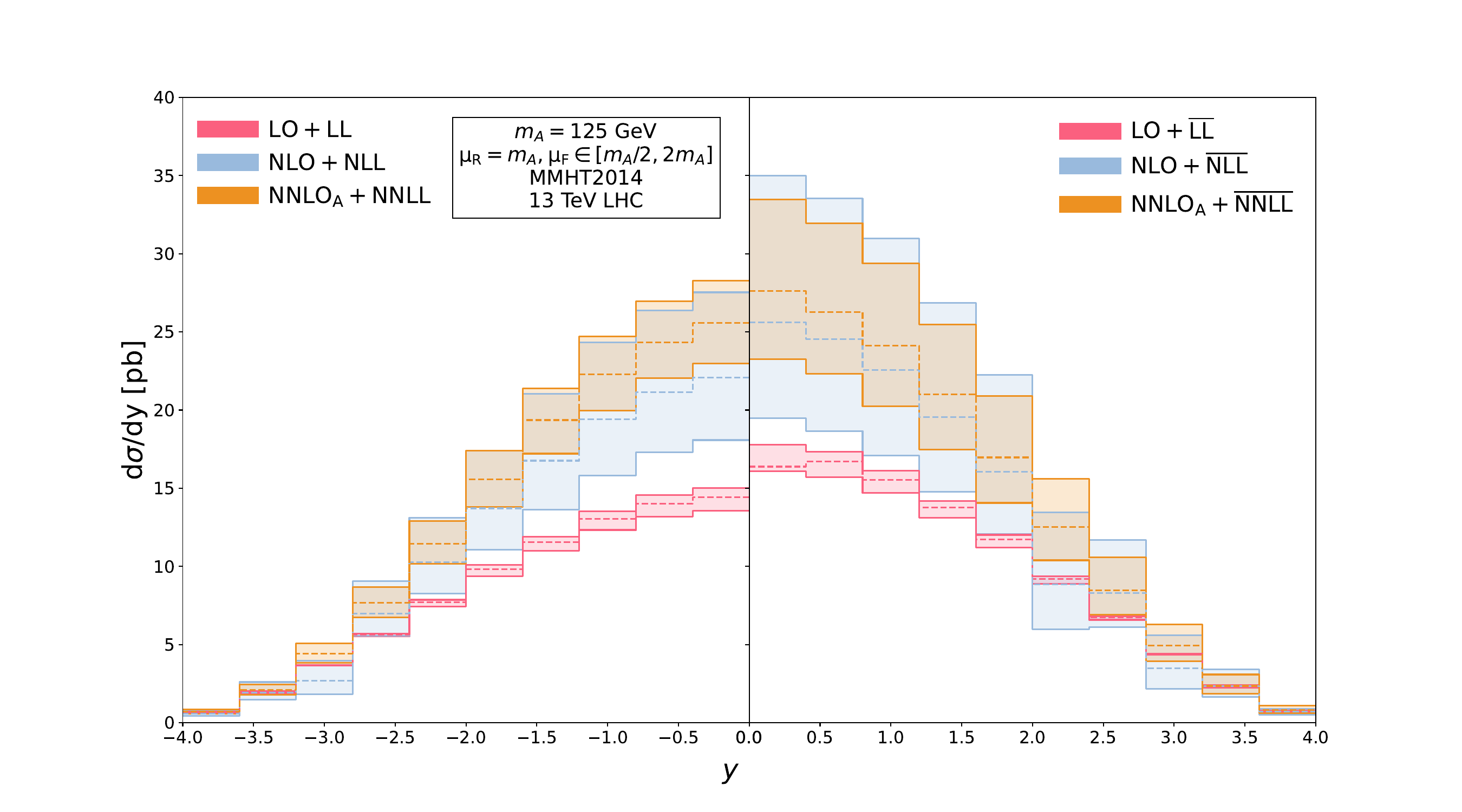}
\includegraphics[scale=0.350]{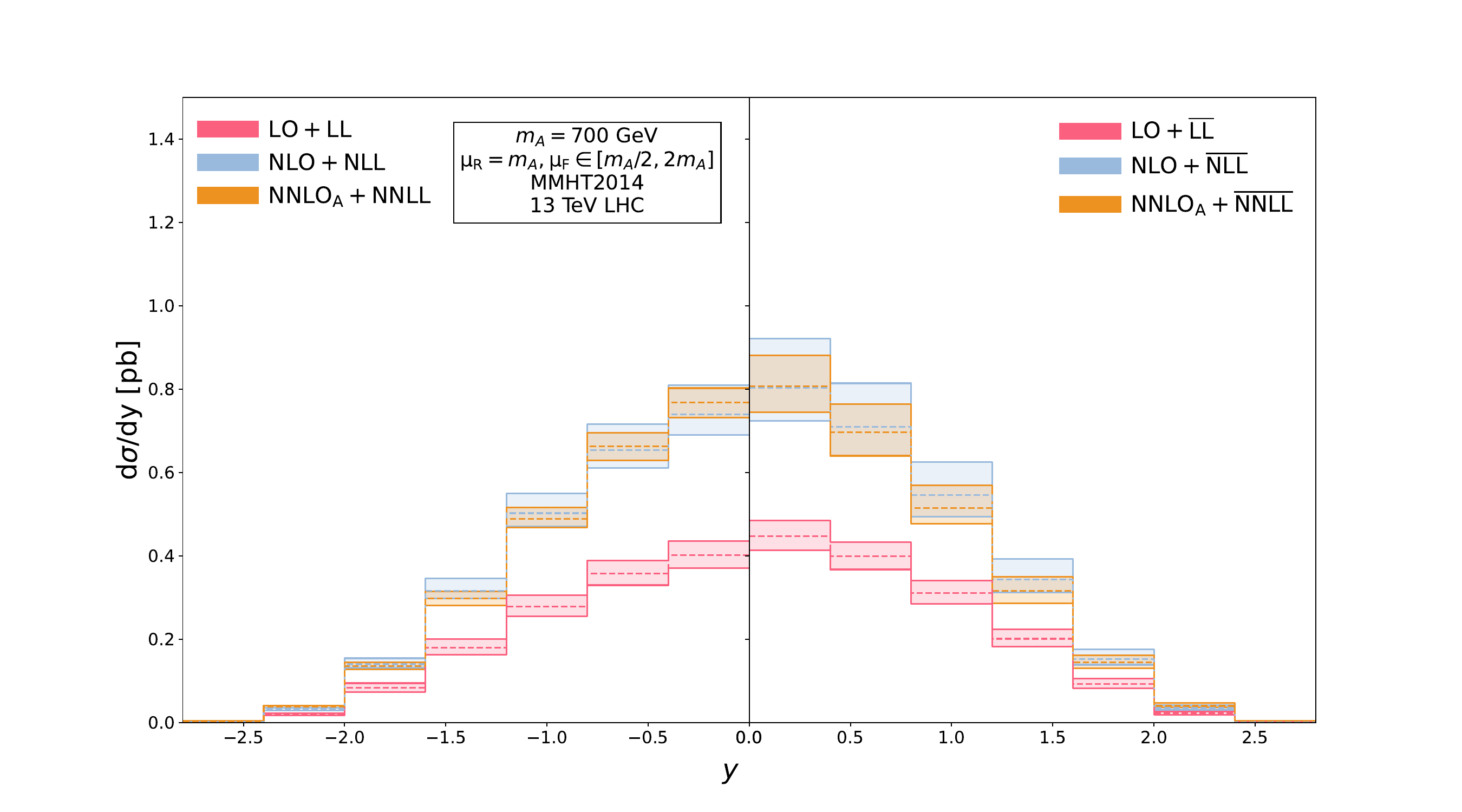}
\caption{Comparison of $\mu_F$ scale variation between SV and SV+NSV resummed results with the scale $\mu_R = m_A$. The dashed lines refer to the corresponding central scale
values at each order.}
\label{fig:mufsvnsv}
\end{figure*}
\begin{figure*}[ht]
\includegraphics[scale=0.350]{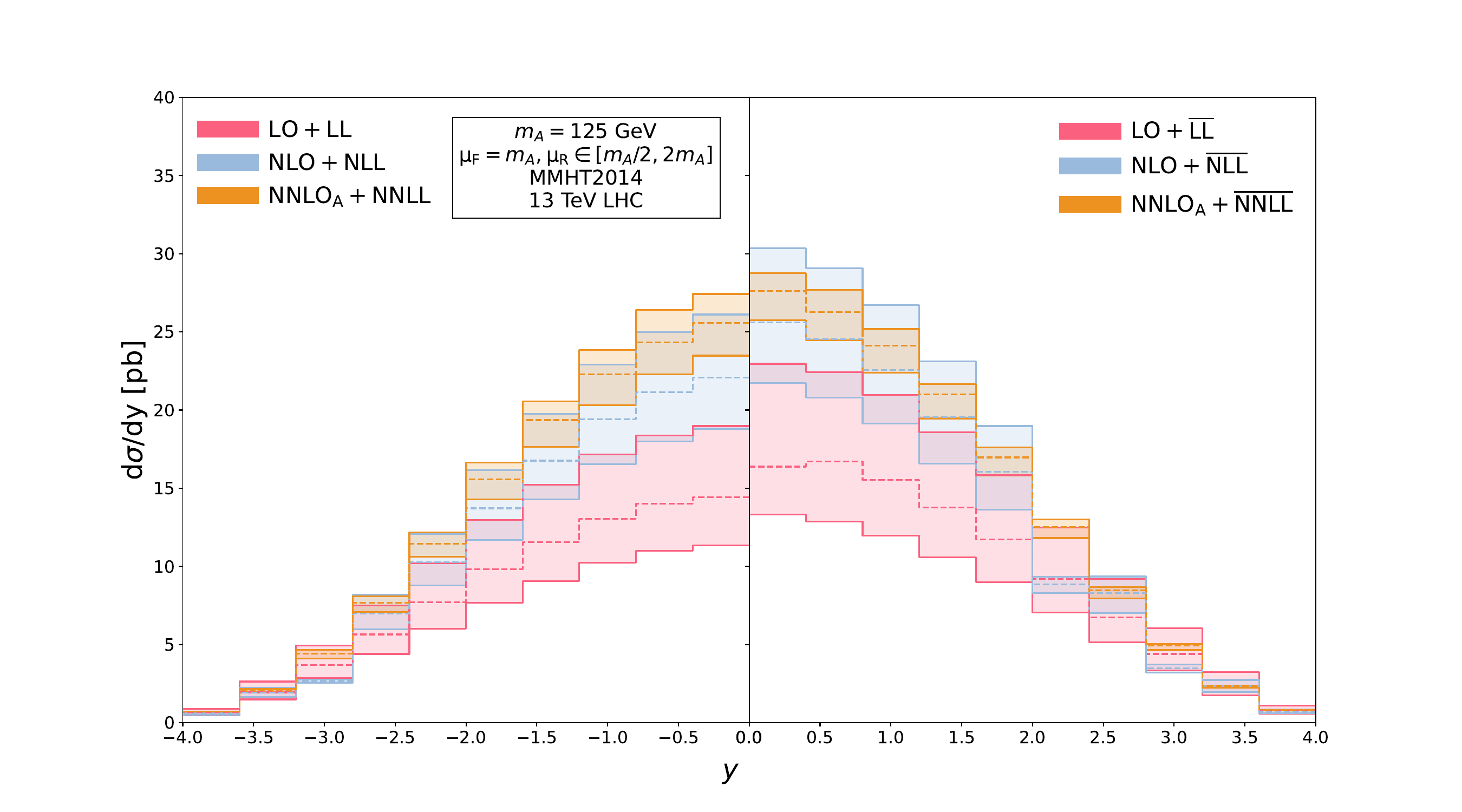}
\includegraphics[scale=0.350]{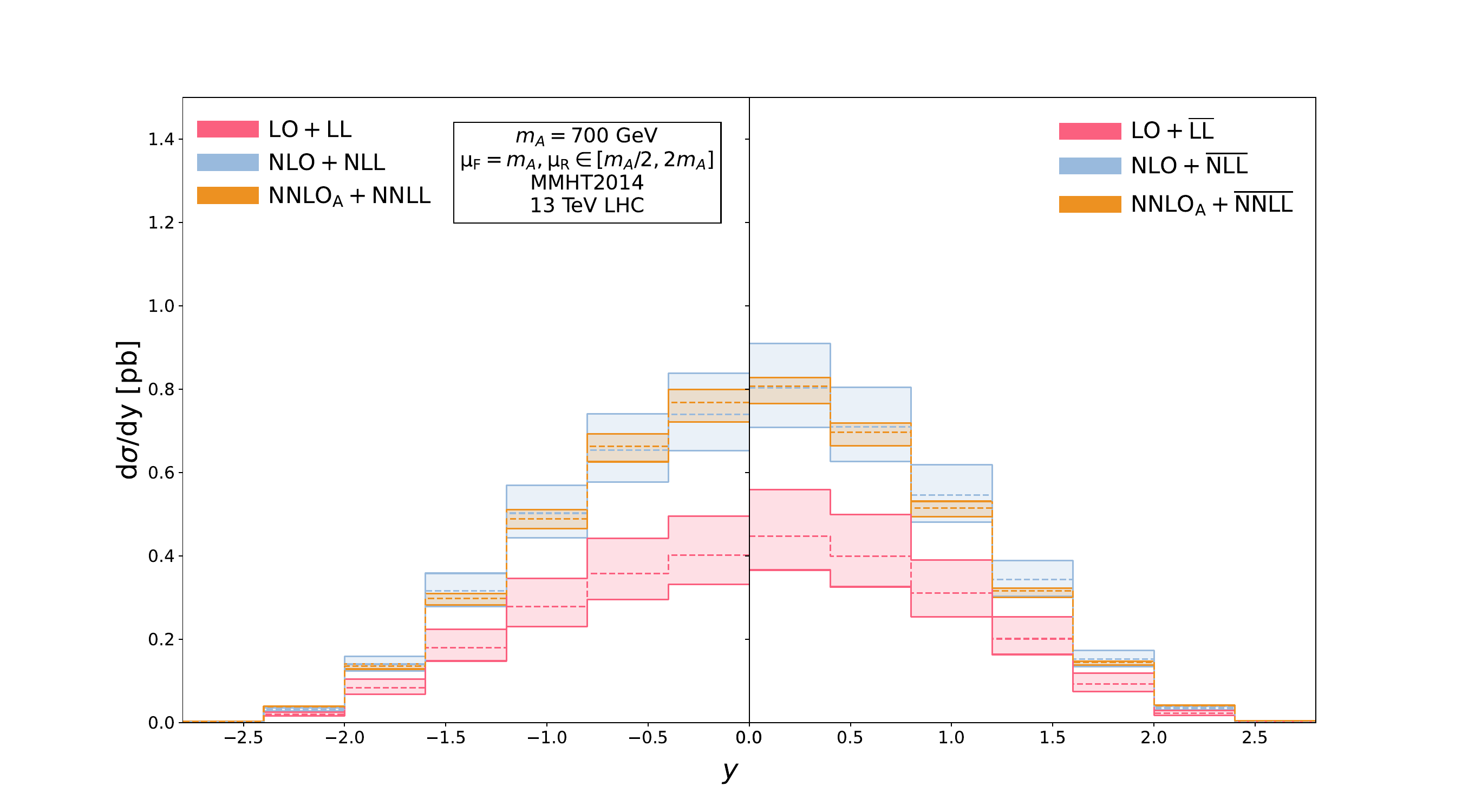}
\caption{Comparison of $\mu_R$ scale variation between SV and SV+NSV resummed results with the scale $\mu_F = m_A$. The dashed lines refer to the corresponding central scale
values at each order.}
\label{fig:mursvnsv}
\end{figure*}

Now, let us do the comparison of SV and SV+NSV resummed predictions by varying the factorisation scale $\mu_F$ keeping $\mu_R$ fixed. In figure \ref{fig:mufsvnsv}, we provide plots for bin-integrated rapidity distributions for the resummed SV(left panel) and resummed SV+NSV(right panel) corrections for $m_A = 125, 700$ GeV keeping $\mu_R = m_A$ in the top and bottom panels respectively. The uncertainty bands are obtained by varying the factorization scale in the range $\{1/2, 1\}m_A$ around the central scale $\mu_F = \mu_R = m_A$. The plots given in figure \ref{fig:mufsvnsv} show that the inclusion of resummed NSV corrections to the SV resummed results worsens the variation of the result w.r.t the $\mu_F$ scale for both $m_A = 125, 700$ GeV. This can be seen directly from the numerical value of the uncertainty which lies between (+36.53\%, -23.92\%) and (+21.27\%, - 15.72\%) at $\rm NLO + {\overline{NLL}}$ and $\rm{NNLO_A} + {\overline{NNLL}}$ respectively around central rapidity region for $m_A = 125$ GeV. These values are significantly larger than the corresponding SV resummed uncertainties of (+24.66\%, -18.15\%) and (+10.54\%, -10.19\%) at NLO+NLL and $\rm NNLO_A$+NNLL respectively. Likewise, for $m_A = 700$ GeV, the uncertainty lies between (+14.66\%, -9.88\%) and (+9.11\%, -7.75\%) for SV+NSV resummed results at $\rm NLO + {\overline{NLL}}$ and $\rm{NNLO_A} + {\overline{NNLL}}$ respectively whereas it varies between (+9.58\%, -6.63\%) and (+4.43\%, -4.71\%) for SV resummed results at NLO+NLL and $\rm NNLO_A$+NNLL respectively around $y=0$. These values also suggest that the variation w.r.t the factorization scale decreases when we go from $m_A = 125$ GeV to $m_A = 700$ GeV for both SV and SV+NSV resummed predcitions. We need to understand the reason behind this considerable $\mu_F$ scale variation of the resummed predictions. We stated in the paragraph above that our resummed results contain the spurious terms existing due to the "inexact" Mellin inversion. The detailed analysis done in the references \cite{Ajjath:2021bbm} and \cite{Ravindran:2022aqr} showed us that these spurious terms play an important role in the $\mu_F$ scale variation in our results. The study demonstrated that the $\mu_F$ scale uncertainty arising due to the NSV logarithms gets compensated by the variation coming from the beyond NSV logarithms and this compensation increases with the increase in the order of perturbation theory. 

First, we try to understand the behaviour of SV resummed results under $\mu_F$ variation. The $\mu_F$ scale variation seen in the SV resummed results comes mainly from the spurious beyond SV terms arising from the inexact Mellin inversion of the $N$-space SV resummed results. The plots in figure \ref{fig:mufsvnsv} also show us that the uncertainty decreases when we go from NLO+NLL to $\rm NNLO_A$+NNLL. This confirms our analysis mentioned above that the compensation between $\mu_F$ uncertainty coming from spurious NSV and beyond NSV terms increases at higher orders thereby decreasing the overall scale dependency for the SV resummed result. Now, let us explore the reason for the huge dependency of SV+NSV resummed results on the factorization scale. For the case of SV resummed results, the spurious terms were the main source of $\mu_F$ uncertainty, however, for the SV+NSV resummed results, the NSV logarithms contribute significantly towards $\mu_F$ variation as well. This uncertainty due to the resummed NSV terms can be compensated by adding the resummed beyond NSV terms which is missing in our calculation. In this case, as well, we have the spurious beyond NSV terms, although now it acts as a compensating factor and cancels the uncertainty due to the resummed NSV logarithms. As a result, we observe that the $\mu_F$ variation decreases when we go from $\rm NLO + {\overline{NLL}}$ to $\rm{NNLO_A} + {\overline{NNLL}}$ accuracy. However, it can not compensate much and we need to resumm the beyond NSV logarithms in order to completely cancel the uncertainty arising from the resummed NSV logarithms.  

We next move on to compare the $\mu_R$ scale uncertainties of SV and SV+NSV resummed predictions. The plots given in figure \ref{fig:mursvnsv} illustrate the variation of $\mu_R$ scale in the range $\{1/2, 1\}m_A$ around the central scale $\mu_F = \mu_R = m_A$ for the bin-integrated rapidity distributions of the resummed SV(left panel) and resummed SV+NSV(right panel) predictions keeping $\mu_F = m_A$ for $m_A = 125, 700$ GeV. The plots show that the inclusion of resummed NSV logarithms reduces the uncertainty due to the $\mu_R$ scale. The uncertainty varies between (+18.22\%, -14.87\%) and (+7.22\%, -8.21\%) for the SV resummed predictions at NLO+NLL and  $\rm{NNLO_A}$+NNLL around the central rapidity region for $m_A = 125$ GeV. The corresponding uncertainty bands for the SV+NSV resummed results lie in the range (+18.45\%, -15.24\%) and (+4.16\%, -6.69\%) for $\rm NLO + {\overline{NLL}}$ and $\rm{NNLO_A} + {\overline{NNLL}}$ respectively. Similar trends are observed for the case of $m_A = 700$ GeV where the $\mu_R$ scale variation lies in the range (+13.40\%, -11.67\%) and (+4.01\%, -6.06\%) for the SV resummed results at NLO+NLL and  $\rm{NNLO_A}$+NNLL respectively whereas for the SV+NSV resummed results, it lies between (+13.33\%, -11.74\%) and (+2.62\%, -5.19\%) at $\rm NLO + {\overline{NLL}}$ and $\rm{NNLO_A} + {\overline{NNLL}}$ level respectively around the central rapidity region. We see from these numerical values that the uncertainty remains almost the same at the next-to-leading level for both SV and SV+NSV resummed results, but at the next-to-next-to-leading order, the $\mu_R$ scale uncertainty decreases by the addition of the resummed NSV logarithms to the SV resummed results. We know that the inclusion of higher order logarithmic corrections within a particular channel leads to a decrease in the sensitivity of the rapidity distribution w.r.t the renormalization scale. This suggests that the percentage contribution of the resummed NSV logarithms is higher at the $\rm{NNLO_A} + {\overline{NNLL}}$ as compared to the $\rm NLO + {\overline{NLL}}$ which results in the significant reduction in the $\mu_R$ uncertainty at this order.

To summarize the findings of this section, we observed that the resummed SV+NSV results are significantly dependent on the factorization scale and have large uncertainties related to this scale. In order to understand this, we compared our results with the fixed order as well as the SV resummed predictions. We found that the fixed order corrections have negligible dependence on $\mu_F$ scale whereas for the case of SV resummed results, the main source of $\mu_F$ variation is the spurious beyond SV terms arising from the "inexact" Mellin inversion of the $N$-space resummed results. When we add the resummed NSV logarithms, the $\mu_F$ uncertainty increases further. Our analysis showed that the reason for this large dependency of the SV+NSV resummed results is the absence of resummed beyond NSV terms which are supposed to cancel the $\mu_F$ variation of the NSV logarithms. This suggests that it is important to include resummed beyond NSV terms to get a more accurate and reliable prediction for the rapidity distribution of pseudo-scalar Higgs boson in gluon fusion process. We would also like to mention that we have used the same PDF set for both fixed order and resummed predictions. In order to understand the $\mu_F$ variation in a better way, resummed PDFs should be used if they are available. For the renormalization scale, we found that the SV+NSV resummed predictions are the least sensitive when we vary $\mu_R$ around the central scale value keeping $\mu_F$ fixed. Also, the $\mu_R$ scale uncertainty decreases when we go to higher orders in perturbation theory. Thus, as expected the $\mu_R$ scale variation decreases by the addition of higher order logarithmic contributions.



%



\section{Discussions and Conclusions}\label{CL}
We present the resummed rapidity distribution of pseudo-scalar Higgs Boson production via gluon fusion at LHC up to next-to-next-to-leading-logarithmic($\rm \overline{NNLL}$) accuracy containing both resummed threshold SV contributions as well as next-to-SV ones. It has been matched to the fixed order predictions up to next-to-next-to-leading order($\rm{NNLO_A}$) accuracy. Beyond NLO, the fixed order rapidity distribution of the pseudo-scalar Higgs boson has been computed using the corresponding result for the scalar Higgs case by appropriately multiplying it with the ratio factor $R_{AH}$. This ratio method was first established in ref(37) by one of the authors for obtaining the inclusive cross-section of pseudo-scalar Higgs from that of the scalar Higgs boson. In \cite{Ahmed:2016otz}, it was shown that the approximate result for the inclusive cross-section obtained in this way has an excellent agreement with the exact result and the difference is found only in terms of next-to-next-to soft distributions which are eventually suppressed in the threshold limit $z \rightarrow 1$. The same trend is expected to follow for the rapidity distribution as well. The resummed corrections have been obtained by using our formalism described in \cite{Ajjath:2020lwb} where we restrict ourselves to the diagonal channel for the production of the pseudo-scalar Higgs.  \\
We have performed a detailed numerical analysis of our computed results around the central scale values $\mu_R = \mu_F = m_A$ for benchmark rapidity values for two different cases of pseudo-scalar mass $m_A = 125,700$ GeV. The K-factor values showed that there is a significant enhancement in the rapidity distribution by the addition of resummed SV+NSV corrections up to next-to-leading order. At $\rm{NNLO_A}$, the inclusion of $\rm \overline{NNLL}$ resummed results increases the rapidity distribution however, the percentage enhancement drops substantially compared to that of lower order results. For instance, there is an enhancement of 53.3\% and 24.97\% by the inclusion of $\rm \overline{LL}$ and $\rm \overline{NLL}$ resummed results to LO and NLO respectively around the central rapidity region which comes down to an 11.48\% increase when we include $\rm \overline{NNLL}$ to $\rm NNLO_A$ accuracy at $m_A = 700$ GeV. This shows that the addition of resummed corrections improves the perturbative convergence of the result thereby making it more reliable. We further used canonical 7-point variation approach to show that the combined uncertainty due to $\mu_F$ and $\mu_R$ scales increases by the inclusion of SV+NSV resummed corrections to the fixed order results throughout the rapidity spectrum and for both the cases of pseudo-scalar Higgs masses. Although, the increase in the sensitivity to the unphysical scales decreases when we go to higher values of pseudo-scalar Higgs mass. For example, for $m_A = 700$ GeV, the 7-point scale uncertainty of the resummed result at $\rm NNLO_A + \overline{NNLL}$ becomes comparable to that of the fixed order rapidity distribution at $\rm NNLO_A$.\\
We studied the impact of the renormalisation and the factorisation scales individually on our result for the better understanding of their behaviour. We found that at higher orders, the uncertainty of our result is mainly driven by the factorisation scale. The inclusion of the resummed NSV logarithms to the well-established threshold SV resummed rapidity distribution increases the sensitivity of our result w.r.t the $\mu_F$ scale. The main reason behind this is the absence of resummed beyond NSV terms which is responsible for the cancellation of the uncertainty arising due to the resummed NSV logarithms. On the other hand, the uncertainty due to the $\mu_R$ scale decreases by the addition of the resummed NSV logarithms. This is expected because the addition of more corrections within the same partonic channel
improves the $\mu_R$ scale uncertainties.

\section{Acknowledgments}
We would like to thank the computer administrative unit of IMSc for their help and support. S.T. is funded by the Deutsche Forschungsgemeinschaft(DFG, German Research Foundation) under grant 396021762 — TRR 257 “Particle Physics Phenomenology after the Higgs Discovery”.
\appendix
\section{QCD $\beta$ functions} \label{App}

\begin{align}
  \beta_0&={11 \over 3 } C_A - {2 \over 3 } n_f \, ,
           \nonumber \\[0.5ex]
  \beta_1&={34 \over 3 } C_A^2- 2 n_f C_F -{10 \over 3} n_f C_A \, ,
           \nonumber \\[0.5ex]
  \beta_2&={2857 \over 54} C_A^3 
           -{1415 \over 54} C_A^2 n_f
           +{79 \over 54} C_A n_f^2
           +{11 \over 9} C_F n_f^2
           -{205 \over 18} C_F C_A n_f
           + C_F^2 n_f 
\end{align}
\section{NLO Results}\label{App:NLORes}
In this section, we present the analytical results of the NLO hadronic rapidity distribution for the production of the pseudo-scalar Higgs boson via gluon fusion as follows:
\begin{align}
\begin{autobreak}

   \frac{d\sigma_{gg}^{A,\rm (1)}}{dY} =

     C_A \Bigg[ H_{gg}(x_1^0, x_2^0 , \mu_F^2)  \Bigg\{ \big(
           4
          + 6 \zeta_2
          \big)

       +    \big(
           2 L_{QF} 
          \big) ~{\cal K}(x_1^0, x_2^0)

       +    {\cal K}^2(x_1^0, x_2^0)  \Bigg\}


       + \int dx_1  ~ \frac{H_{gg,1}(x_1, x_2^0 , \mu_F^2)}{(x_1-x_1^0)} \big(
           4 {\cal K}_b^1
         \big)
          

       + \int dx_1 ~H_{gg}(x_1, x_2^0 , \mu_F^2) \Bigg\{
        \bigg(  - 4 \frac{(x_1^0)^2}{x_1^3}
          + 4 \frac{x_1^0}{x_1^2}
          -  \frac{8}{x_1}
          +  \frac{4}{x_1^0} \bigg)~{\cal K}_a^1 +  \frac{4 {\cal K}_c^1}{(x_1-x_1^0)}
         \Bigg\}

      + \int dx_1 \int dx_2  \frac{H_{gg,1}(x_1, x_2 , \mu_F^2)}{(x_1-x_1^0)} \bigg(
          - 4 \frac{(x_1^0)^2}{x_1^3}
          + 4 \frac{x_1^0}{x_1^2}
          -  \frac{8}{x_1}
          + \frac{4}{x_1^0}
         \bigg) 
          
  + \int dx_1 \int dx_2 ~\frac{2~ H_{gg,12}(x_1, x_2 , \mu_F^2) }{(x_1-x_1^0) (x_2 - x_2^0)}  
  
  + \int dx_1 \int dx_2 \frac{ H_{gg}(x_1, x_2 , \mu_F^2)} {(x_1+x_1^0) (x_2+x_2^0) (x_1 x_2^0 + x_2 x_1^0)^4}
  
\Bigg\{
       \frac{1}{x_1^3}  \Big(
           4~ (x_1^0)^7 (x_2)^4
          + 8 ~(x_1^0)^7 (x_2)^3 x_2^0
          + 8 ~(x_1^0)^7 (x_2)^2 (x_2^0)^2
          + 8 ~(x_1^0)^7 x_2 (x_2^0)^3
          \Big)

       +  \frac{1}{x_1^2}   \Big(
           16 ~(x_1^0)^6 (x_2)^3 x_2^0
          + 32 ~(x_1^0)^6 (x_2)^2 (x_2^0)^2
          + 24 ~(x_1^0)^6 x_2 (x_2^0)^3
          + 16~ (x_1^0)^6 (x_2^0)^4
          \Big)

       +  \frac{1}{x_1}   \Big(
           4 ~(x_1^0)^5 (x_2)^4
          + 8 ~(x_1^0)^5 (x_2)^3 x_2^0
          + 32~ (x_1^0)^5 (x_2)^2 (x_2^0)^2
          + 48 ~(x_1^0)^5 x_2 (x_2^0)^3
          + 40 ~(x_1^0)^5 (x_2^0)^4
          \Big)

       +  \frac{1}{x_1 x_2}   \big(
           12~ (x_1^0 x_2^0)^5
          \big)

       +   \frac{1}{x_1^0}  \Big(
           8 ~x_1^5 x_2 (x_2^0)^3
          + 4 ~x_1^5 (x_2^0)^4
          \Big)  
          
          + x_1^4 \Big( (16 x_2 x_2^0)^2   + 4 x_2 (x_2^0)^3 + 4 (x_2^0)^4 \Big)  
          
        +  x_1^3 \Big( 12 x_2^3 x_1^0 x_2^0 + 8 x_1^0 x_2 (x_2^0)^3 + 28 x_1^0 (x_2^0)^4 \Big) 
          
         + x_1^2 \Big(20 (x_2 x_1^0 x_2^0)^2 +88 (x_1^0)^2  (x_2^0)^3 x_2 + 48 (x_1^0)^2  (x_2^0)^4 \Big) 
         
         + x_1 \Big(68 (x_1^0 x_2^0)^3 x_2^0 \Big)
         
        +  x_1 x_2 \Big(64 (x_1^0 x_2^0)^3 \Big)
    
        + 40 (x_1^0 x_2^0)^4
          \Bigg\}   \Bigg] + \Big( 1 \leftrightarrow 2 \Big)\,.

\end{autobreak}
\end{align}
For the $qg$ and $gq$ channels, we obtain 
\begin{align}
\begin{autobreak}

   \frac{d\sigma_{qg}^{A,(1)}}{dY} = 
   C_F \Bigg[
    
     \int dx_1   H_{qg}(x_1, x_2^0 , \mu_F^2) \Bigg\{

         \bigg(
          \frac{2 x_1^0}{x_1^2} 
          \bigg)

       +   \bigg(
          - \frac{4}{x_1}
          + \frac{4}{x_1^0}
          + \frac{2 x_1^0}{x_1^2}
          \bigg)  K_a^1 \Bigg\}
  
  + \int dx_1 \int dx_2 \frac{ H_{qg,2}(x_1, x_2 , \mu_F^2)} {(x_2-x_2^0)}
  
\Bigg\{ \frac{2 x_1^0} {x_1^2} 
       - \frac{4}{x_1}   
       + \frac{4}{x_1^0}  
        \Bigg\}

        + \int dx_1 \int dx_2 \frac{ H_{qg}(x_1, x_2 , \mu_F^2)} {(x_1+x_1^0) (x_2+x_2^0) (x_1 x_2^0 + x_2 x_1^0)^4}

  \Bigg\{    \frac{1}{x_1^2}    \Big(
          - 2 (x_1^0)^5 x_2^3
          - 4 (x_1^0)^5  x_2^2 x_2^0
          - 4 (x_1^0)^5 x_2 (x_2^0)^2
          \Big)

       +  \frac{1}{x_1}   \Big(
           2 (x_1^0)^4 x_2^3
          - 2 (x_1^0)^4  x_2^2 x_2^0
          - 4 (x_1^0)^4  x_2 (x_2^0)^2
          \Big)

       +  \frac{1}{x_1^0}  \Big(
           8 x_1^4 x_2 (x_2^0)^2
          + 4 x_1^4 (x_2^0)^3
          \Big)

      +  \Big(
           4 x_1^3 x_2^2 x_2^0
          + 2 x_1^3 x_2 (x_2^0)^2
          + 2 x_1^2 x_1^0 x_2^3
          - 6 x_1^2 x_1^0 x_2 (x_2^0)^2
          - x_1^2 x_1^0 (x_2^0)^3
          + 3 x_1 (x_1^0)^2 x_2 (x_2^0)^2
          + x_1 (x_1^0)^2 (x_2^0)^3
          + 5 (x_1^0)^3x_2^2 x_2^0
          + 7 (x_1^0)^3 x_2 (x_2^0)^2
          \Big)

          \Bigg\}
          \Bigg]\,,
\end{autobreak}
\end{align}

\begin{align}
\frac{d\sigma_{qg}^{A,(1)}}{dY} =
\frac{d\sigma_{gq}^{A,(1)}}{dY} \Bigg|_{1 \leftrightarrow 2}    \,.
\end{align}
For the $q \bar q$-channel, we find
\begin{align}
\begin{autobreak}
 
 \frac{d\sigma_{q\bar q}^{A,(1)}}{dY} = 
 
 \frac{(N^2-1) C_F}{N}  \Bigg[ \int dx_1~ dx_2 ~H_{q\bar q}(x_1, x_2 , \mu_F^2) \bigg\{ 
 
 \frac{1}{x_1^2}  \Big(   4 (x_1^0)^3 (x_1^0 x_2^0 )^3 x_2  
 
 + 4 (x_1^0 x_2^0 )^4 (x_1^0)^2 \Big)
 
 + \frac{1}{x_1} \Big( 4 (x_1^0)^5 (x_2 x_2^0)^2 
 
 +  8 (x_1^0)^5 x_2 (x_2^0)^3 +  4 (x_1^0)^5 (x_2^0)^4 \Big)

 + x_1^4 \Big( 4 (x_2 x_2^0)^4 +4 x_2 (x_2^0)^3  \Big)
 
 + x_1^3 \Big(4 x_1^0 (x_2 x_2^0)^2 + 8 x_1^0 x_2 (x_2^0)^3 + x_1^0 (x_2^0)^4 \Big)

 + x_1^2 \Big( -8 (x_1^0 x_2^0 x_2)^2 + x_2 (x_1^0)^2  
 (x_2^0)^3  + (x_1^0 x_2^0)^2 (x_2^0)^2 \Big)

 + x_1 x_2 \Big( -16 (x_1^0 x_2^0 )^3 \Big) 
 
 + x_1 \Big( -12 (x_1^0 x_2^0)^3 x_2^0 \Big) - 8 (x_1^0 x_2^0)^4 
 \bigg\} \Bigg]
 + \Big( 1 \leftrightarrow 2 \Big) \,.
 \end{autobreak}
\end{align}

In the above equations, we have introduced the following abbreviations
\begin{eqnarray}
H_{ab,12}(x_1,x_2,\mu_F^2)&=&H_{ab}(x_1,x_2,\mu_F^2)
-H_{ab}(\xo,x_2,\mu_F^2)-H_{ab}(x_1,\xt,\mu_F^2)
\nonumber\\[2ex]&&
+H_{ab}(\xo,\xt,\mu_F^2)\,,
\nonumber\\[2ex]
H_{ab,1}(x_1,z,\mu_F^2)&=&H_{ab}(x_1,z,\mu_F^2)
-H_{ab}(\xo,z,\mu_F^2)\,,
\nonumber\\[2ex]
H_{ab,2}(z,x_2,\mu_F^2)&=&H_{ab}(z,x_2,\mu_F^2)
-H_{ab}(z,\xt,\mu_F^2)\,,
\label{Hab}
\end{eqnarray}
where
\begin{eqnarray}
H_{q \bar q}(x_1,x_2,\mu_F^2)&=&
f_q(x_1,\mu_F^2)~ 
f_{\bar q}(x_2,\mu_F^2)
+f_{\bar q}(x_1,\mu_F^2)~ 
f_q(x_2,\mu_F^2)\,,
\nonumber
\\[2ex]
H_{g q}(x_1,x_2,\mu_F^2)&=&
f_(x_1,\mu_F^2) ~
\Big(f_q(x_2,\mu_F^2)
+f_{\bar q}(x_2,\mu_F^2)\Big)\,,
\nonumber
\\[2ex]
H_{q g}(x_1,x_2,\mu_F^2)&=&
H_{g q}(x_2,x_1,\mu_F^2)\,,
\nonumber
\\[2ex]
H_{g g}(x_1,x_2,\mu_F^2)&=&
f_g(x_1,\mu_F^2)~ 
f_g(x_2,\mu_F^2)\,.
\end{eqnarray}

\begin{eqnarray}
{\cal K}_{a_1}&=&\ln\left(
{2 Q^2 (1-\xt)(x_1-\xo) \over
\mu_F^2 (x_1+\xo)  \xt}
\right)\,,
\quad \quad
{\cal K}_{b_1}=\ln\left(
{Q^2 (1-\xt)(x_1-\xo) \over
\mu_F^2  \xo \xt}
\right)\,,
\nonumber\\[2ex]
{\cal K}_{c_1}&=&\ln\left(
{ 2 \xo \over
x_1+\xo}
\right)\,,
\quad \quad
{\cal K}(\xo,\xt)=\ln\left(
{ (1-\xo) (1-\xt) \over
\xo \xt}
\right)\,.
\label{Kab}
\end{eqnarray}
The ${\cal K}_{a_2}$, ${\cal K}_{b_2}$ and ${\cal K}_{c_2}$ can be obtained
from ${\cal K}_{a_1}$, ${\cal K}_{b_1}$ and ${\cal K}_{c_1}$
by using $1 \leftrightarrow 2$ symmetry.
\section{Anomalous dimensions} \label{App:ano}

\begin{align}
\label{gmg}
  \gamma^A_{g,1}&={11 \over 3 } C_A - {2 \over 3 } n_f \, ,
           \nonumber \\[0.5ex]
  \gamma^A_{g,2}&={34 \over 3 } C_A^2- 2 n_f C_F -{10 \over 3} n_f C_A \, ,
           \nonumber \\[0.5ex]
  \gamma^A_{g,3}&={2857 \over 54} C_A^3 
           -{1415 \over 54} C_A^2 n_f
           +{79 \over 54} C_A n_f^2
           +{11 \over 9} C_F n_f^2
           -{205 \over 18} C_F C_A n_f
           + C_F^2 n_f \,.
\end{align}

\begin{align}
  A^A_{g,1} &= 4 C_A \,, 
 \nonumber \\
  A^A_{g,2} &= 8 C_A^2 \Bigg\{ \frac{67}{18} - \zeta_2 \Bigg\} + 8 C_A n_f \Bigg\{ -\frac{5}{9} \Bigg\} \,,
 \nonumber \\
  A^A_{g,3} &= 16 C_A^3 \Bigg\{ \frac{245}{24} - \frac{67}{9} \zeta_2  + \frac{11}{6} \zeta_3
                              + \frac{11}{5} \zeta_2^2 \Bigg\}
             + 16 C_A C_F n_f \Bigg\{ - \frac{55}{24} + 2 \zeta_3 \Bigg\}
 \nonumber\\
 &
             + 16 C_A^2 n_f \Bigg\{ - \frac{209}{108} + \frac{10}{9} \zeta_2 - \frac{7}{3} \zeta_3 \Bigg\}
             + 16 C_A n_f^2 \Bigg\{ - \frac{1}{27} \Bigg\} \,.
 \end{align}
 
 \begin{align}
 B^A_{g,1} &=
       n_f   \bigg\{  - \frac{2}{3} \bigg\} + C_A   \bigg\{ \frac{11}{3} \bigg\}\,,
\nonumber \\
   B^A_{g,2} &=
        C_F n_f   \bigg\{  - 2 \bigg\}

       + C_A n_f   \bigg\{  - \frac{8}{3} \bigg\}

       + C_A^2   \bigg\{ \frac{32}{3} + 12 \zeta_3 \bigg\}\,,
\nonumber \\
   B^A_{g,3} &=
        C_F n_f^2   \bigg\{ \frac{11}{9} \bigg\}

       + C_F^2 n_f 

       + C_A n_f^2   \bigg\{ \frac{29}{18} \bigg\}

       + C_A C_F n_f   \bigg\{  - \frac{241}{18} \bigg\}

       + C_A^2 n_f   \bigg\{  - \frac{233}{18} \nonumber \\& 
       - \frac{80}{3} \zeta_3 - \frac{8}{3} \zeta_2 - \frac{4}{3} \zeta_2^2 \bigg\}

       + C_A^3   \bigg\{ \frac{79}{2} - 80 \zeta_5 + \frac{536}{3} \zeta_3 + \frac{8}{3} \zeta_2 - 16 \zeta_2 \zeta_3
          + \frac{22}{3} \zeta_2^2 \bigg\}\,.
\end{align}

\begin{align}
 f_{g,1}^A &= 0 \,,
\nonumber \\
 f_{g,2}^A &= C_A^2  \Bigg\{ -\frac{22}{3} {\zeta_2} - 28 {\zeta_3} + \frac{808}{27} \Bigg\}
        + C_A n_f \Bigg\{ \frac{4}{3} {\zeta_2} - \frac{112}{27} \Bigg\} \,,
\nonumber \\
 f_{g,3}^A &= {C_A^3}  \Bigg\{ \frac{352}{5} {\zeta_2}^2 + \frac{176}{3} {\zeta_2} {\zeta_3}
        - \frac{12650}{81} {\zeta_2} - \frac{1316}{3} {\zeta_3} + 192 {\zeta_5}
        + \frac{136781}{729}\Bigg\}
\nonumber \\
&
        + {C_A^2}  {n_f} \Bigg\{ - \frac{96}{5} {\zeta_2}^2 
        + \frac{2828}{81} {\zeta_2}
        + \frac{728}{27} {\zeta_3} - \frac{11842}{729} \Bigg\} 
\nonumber \\
&
        + {C_A} {C_F} {n_f} \Bigg\{ \frac{32}{5} {\zeta_2}^2 + 4 {\zeta_2} 
        + \frac{304}{9} {\zeta_3} - \frac{1711}{27} \Bigg\}
        + {C_A} {n_f}^2 \Bigg\{ - \frac{40}{27} {\zeta_2} + \frac{112}{27} {\zeta_3}
        - \frac{2080}{729} \Bigg\} \, .
\end{align}

\begin{align}

    \mathbf{D}_{d,g,1}^A &= 0 \,,
   \nonumber \\
    \mathbf{D}_{d,g,2}^A &=

        C_A n_f   \bigg\{
           \frac{112}{27}
          - \frac{8}{3} \zeta_2
          \bigg\}

       + C_A^2   \bigg\{
          - \frac{808}{27}
          + 28 \zeta_3
          + \frac{44}{3} \zeta_2
          \bigg\}\,,
\nonumber \\
     \mathbf{D}_{d,g,3}^A &=

        C_A n_f^2   \bigg\{
          - \frac{1856}{729}
          - \frac{32}{27} \zeta_3
          + \frac{160}{27} \zeta_2
          \bigg\}

       + C_A C_F n_f   \bigg\{
           \frac{1711}{27}
          - \frac{304}{9} \zeta_3
          - 8 \zeta_2
          - \frac{32}{5} \zeta_2^2
          \bigg\}
\nonumber \\&
       + C_A^2 n_f   \bigg\{
           \frac{62626}{729}
          - \frac{536}{9} \zeta_3
          - \frac{7760}{81} \zeta_2
          + \frac{208}{15} \zeta_2^2
          \bigg\}

       + C_A^3   \bigg\{
          - \frac{297029}{729}
          - 192 \zeta_5
          + \frac{14264}{27} \zeta_3 \nonumber \\&
          + \frac{27752}{81} \zeta_2
          - \frac{176}{3} \zeta_2 \zeta_3
          - \frac{616}{15} \zeta_2^2
          \bigg\}\,.

\end{align}
\section{Results of SV rapidity distribution to third order} \label{app:SV}

\begin{align}
\label{DeltaSV1}    
\Delta_{d,g,1}^{A,SV} &=
\delta \overline{\delta}    \bigg\{C_A \Big( 4 + 6 \zeta_2 \Big)
\bigg\} 
+ \overline{\mathcal{\D}}_0 \mathcal{\D}_0  \bigg\{  C_A   \Big( 2 \Big) \bigg\}
+ \delta \overline{\mathcal{\D}}_1   \bigg\{  C_A \Big( 4 \Big) \bigg\} + \big( z_1 \leftrightarrow z_2 \big)\,,
 \end{align}

\begin{align}
\label{DeltaSV2}    
\Delta_{d,g,2}^{A,SV} &=    

\delta \overline{\delta} \Bigg[ \bigg\{  C_F n_f  \bigg(
          - \frac{80}{3}
          + 6 \ln \left(\frac{\mu_R^2}{m_t^2}\right) 
          + 8 \zeta_3  \bigg)
          \bigg\}

       +   \bigg\{ C_A n_f \bigg(
          - \frac{41}{3}
          - 4 \zeta_3
          - \frac{20}{3} \zeta_2
          \bigg) \bigg\} \nonumber \\&

       +   \bigg\{ C_A^2 \bigg(
          \frac{247}{3}
          - 22 \zeta_3
          + \frac{278}{3} \zeta_2
          + \frac{126}{5} \zeta_2^2 \bigg)
          \bigg\} \Bigg]

       + \overline{\mathcal{\D}}_0 \mathcal{\D}_0   \Bigg[ \bigg\{ C_A n_f  \bigg(
          - \frac{20}{9}
       \bigg)    \bigg\} \nonumber \\&

       +  \bigg\{ C_A^2    \bigg(
           \frac{278}{9}
          + 4 \zeta_2
         \bigg)  \bigg\} \Bigg]

       + \overline{\mathcal{\D}}_1 \mathcal{\D}_0  \Bigg[  \bigg\{ C_A n_f  \bigg(
           \frac{8}{3}
         \bigg)  \bigg\}

       +  \bigg\{ C_A^2    \bigg(
          - \frac{44}{3}
         \bigg)  \bigg\} \Bigg]

  \nonumber \\&

       + \mathcal{\D}_0 \overline{\mathcal{\D}}_2 \bigg\{ C_A^2    \bigg(
           24
         \bigg)  \bigg\}

       + \overline{\mathcal{\D}}_1 \mathcal{\D}_1 \bigg\{ C_A^2    \bigg(
           24
        \bigg)   \bigg\}

       + \delta \overline{\mathcal{\D}}_0   \Bigg[\bigg\{ C_A n_f  \bigg(
        + \frac{112}{27}
          - \frac{8}{3} \zeta_2
         \bigg)  \bigg\}

 \nonumber \\&
 
       +  \bigg\{ C_A^2    \bigg(
          - \frac{808}{27}
          + 60 \zeta_3
          + \frac{44}{3} \zeta_2
         \bigg)  \bigg\} \Bigg]

       + \delta \overline{\mathcal{\D}}_1   \Bigg[\bigg\{ C_A n_f  \bigg(
          - \frac{40}{9}
          \bigg) \bigg\}

       +  \bigg\{ C_A^2    \bigg(
          \frac{556}{9}   \nonumber \\&
          + 8 \zeta_2
        \bigg)   \bigg\}\Bigg]

       + \delta \overline{\mathcal{\D}}_2  \Bigg[ \bigg\{ C_A n_f  \bigg(
           \frac{4}{3}
        \bigg)   \bigg\}

       +  \bigg\{ C_A^2    \bigg(
          - \frac{22}{3}
        \bigg)   \bigg\}\Bigg]

       + \delta \overline{\mathcal D}_3 \bigg\{ C_A^2    \Big(
           8
       \Big)    \bigg\} + \big( z_1 \leftrightarrow z_2 \big)\,,

\end{align}


\begin{align}
\label{DeltaSV3}    
\Delta_{d,g,3}^{A,SV} &=

        \delta \overline{\delta}    \Bigg[\bigg\{ n_f \Big(
          - 2 C_J^{(2)} \Big)
          \bigg\}

       +    \bigg\{  C_F n_f^2
          \bigg( \frac{749}{9}
          - \frac{8}{9} \zeta_4
          - \frac{112}{3} \zeta_3
          - \frac{20}{9} \zeta_2 \bigg)
          \bigg\} \nonumber \\&

       +   \bigg\{  C_F^2 n_f 
           \bigg( \frac{457}{6}
          - 160 \zeta_5
          + 104 \zeta_3 \bigg)
          \bigg\}

       +   \bigg\{ C_A n_f^2  
          \bigg( \frac{3457}{81}
          - \frac{152}{9} \zeta_4
          + \frac{56}{3} \zeta_3
          - 8 \zeta_2 \nonumber \\&
          - \frac{8}{45} \zeta_2^2 \bigg)
          \bigg\}

       +    \bigg\{ C_A C_F n_f 
          \bigg( - \frac{1797}{2}
          + 48 \ln \left(\frac{\mu_R^2}{m_t^2}\right) 
          + 80 \zeta_5
          + \frac{44}{9} \zeta_4
          + \frac{904}{3} \zeta_3
          - \frac{3205}{9} \zeta_2
         \nonumber \\&
         + 72 \zeta_2 \ln \left(\frac{\mu_R^2}{m_t^2}\right) 
          + 144 \zeta_2 \zeta_3
         \bigg)  \bigg\}

       +   \bigg\{ C_A^2 n_f  
          \bigg( - \frac{56683}{81}
          + \frac{596}{9} \zeta_5
          + \frac{5258}{27} \zeta_4
          + \frac{188}{27} \zeta_3
          \nonumber \\&
          - \frac{5138}{27} \zeta_2
          - 136 \zeta_2 \zeta_3
          - \frac{824}{15} \zeta_2^2 \bigg)
          \bigg\}

       +    \bigg\{ C_A^3 
         \bigg(  \frac{57284}{27}
          - \frac{4006}{3} \zeta_6
          + \frac{682}{9} \zeta_5
          - \frac{15257}{27} \zeta_4  \nonumber \\&
          - \frac{33098}{27} \zeta_3
          + \frac{800}{3} \zeta_3^2
          + \frac{40235}{27} \zeta_2
          - 44 \zeta_2 \zeta_3
          + \frac{25982}{45} \zeta_2^2
          + \frac{12688}{35} \zeta_2^3 \bigg)
          \bigg\}\Bigg]

 \nonumber \\&

       + \delta \overline{\mathcal{\D}}_0  \bigg\{ C_A n_f^2   \bigg(
          - \frac{1856}{729}
          - \frac{32}{27} \zeta_3
          + \frac{160}{27} \zeta_2
        \bigg)  \bigg\}

       + \delta \overline{\mathcal{\D}}_0   \bigg\{ C_A C_F n_f 
         \bigg(  \frac{1711}{27}
          - \frac{304}{9} \zeta_3
          - 8 \zeta_2  \nonumber \\&
          - \frac{32}{5} \zeta_2^2
         \bigg) \bigg\}

       + \delta \overline{\mathcal{\D}}_0  \bigg\{ C_A^2 n_f   \bigg(
           \frac{86818}{729}
          - \frac{392}{3} \zeta_3
          - \frac{8144}{81} \zeta_2
          + \frac{16}{5} \zeta_2^2
       \bigg)   \bigg\}

       + \delta \overline{\mathcal{\D}}_0   \bigg\{ C_A^3  \bigg(
          - \frac{471557}{729} \nonumber \\&
          + 192 \zeta_5
          + \frac{40088}{27} \zeta_3
          + \frac{27560}{81} \zeta_2
          - \frac{608}{3} \zeta_2 \zeta_3
          + \frac{88}{5} \zeta_2^2
      \bigg)    \bigg\}

       + \delta \overline{\mathcal{\D}}_1  \bigg\{ C_A n_f^2   \bigg(
           \frac{400}{81}
          - \frac{32}{9} \zeta_2
        \bigg)  \bigg\}
\nonumber \\&
       + \delta \overline{\mathcal{\D}}_1   \bigg\{ C_A C_F n_f  \bigg(
          - 250
          + 48 \ln \left(\frac{\mu_R^2}{m_t^2}\right) 
          + 96 \zeta_3
      \bigg)    \bigg\}

       + \delta \overline{\mathcal{\D}}_1  \bigg\{ C_A^2 n_f   \bigg(
          - \frac{19940}{81}
          + 32 \zeta_3  \nonumber \\&
          + \frac{64}{3} \zeta_2
        \bigg)  \bigg\}

       + \delta \overline{\mathcal{\D}}_1   \bigg\{ C_A^3 
        \bigg(   \frac{103654}{81}
          - 704 \zeta_3
          + \frac{680}{9} \zeta_2
          - \frac{64}{5} \zeta_2^2
        \bigg)  \bigg\}

       + \delta \overline{\mathcal{\D}}_2  \bigg\{ C_A n_f^2  
        \bigg(  - \frac{80}{27}
          \bigg\}
 \nonumber \\&
       + \delta \overline{\mathcal{\D}}_2   \bigg\{ C_A C_F n_f 
           4
        \bigg)  \bigg\}

       + \delta \overline{\mathcal{\D}}_2  \bigg\{ C_A^2 n_f  
          \bigg( \frac{2116}{27}
          - \frac{112}{3} \zeta_2
         \bigg) \bigg\}

       + \delta \overline{\mathcal{\D}}_2   \bigg\{ C_A^3 
        \bigg(  - \frac{9992}{27}
          + 488 \zeta_3
           \nonumber \\&
          + \frac{616}{3} \zeta_2
         \bigg) \bigg\}

       + \delta \overline{\mathcal D}_3  \bigg\{ C_A n_f^2  \bigg(
           \frac{16}{27}
        \bigg)  \bigg\}

       + \delta \overline{\mathcal D}_3  \bigg\{ C_A^2 n_f   \bigg(
          - \frac{656}{27}
        \bigg)  \bigg\}

       + \delta \overline{\mathcal D}_3   \bigg\{ C_A^3 
          \bigg( \frac{5428}{27}
          - 64 \zeta_2
         \bigg) \bigg\}

 \nonumber \\&

       + \delta \overline{\mathcal{\D}}_4  \bigg\{ C_A^2 n_f  
         \bigg(  \frac{40}{9} \bigg)
          \bigg\}

       + \delta \overline{\mathcal{\D}}_4   \bigg\{ C_A^3 
         \bigg( - \frac{220}{9}
         \bigg) \bigg\}

       + \delta \overline{\mathcal{\D}}_5   \bigg\{ C_A^3 
         \Big(  8 \Big)
          \bigg\}

       + \overline{\mathcal{\D}}_0 \mathcal{\D}_0  \bigg\{ C_A n_f^2  
          \bigg( \frac{200}{81}
          - \frac{16}{9} \zeta_2
          \bigg) \bigg\}
 \nonumber \\&
      + \overline{\mathcal{\D}}_0 \mathcal{\D}_0   \bigg\{ C_A C_F n_f 
         \bigg( - 125
          + 24 \ln \left(\frac{\mu_R^2}{m_t^2}\right) 
          + 48 \zeta_3 \bigg)
          \bigg\}

       + \overline{\mathcal{\D}}_0 \mathcal{\D}_0  \bigg\{ C_A^2 n_f  
         \bigg( - \frac{9970}{81}
          + 16 \zeta_3
           \nonumber \\&
          + \frac{32}{3} \zeta_2 \bigg)
          \bigg\}

       + \overline{\mathcal{\D}}_0 \mathcal{\D}_0   \bigg\{ C_A^3 
        \bigg(   \frac{51827}{81}
          - 352 \zeta_3
          + \frac{340}{9} \zeta_2
          - \frac{32}{5} \zeta_2^2 \bigg)
          \bigg\}

       + \overline{\mathcal{\D}}_1 \mathcal{\D}_0  \bigg\{ C_A n_f^2  
        \bigg(  - \frac{160}{27} \bigg)
          \bigg\}

 \nonumber \\&

       + \overline{\mathcal{\D}}_1 \mathcal{\D}_0   \bigg\{ C_A C_F n_f 
          \Big( 8 \Big)
          \bigg\}

       + \overline{\mathcal{\D}}_1 \mathcal{\D}_0  \bigg\{ C_A^2 n_f  
         \bigg(  \frac{4232}{27}
          - \frac{224}{3} \zeta_2 \bigg)
          \bigg\}

       + \overline{\mathcal{\D}}_1 \mathcal{\D}_0   \bigg\{ C_A^3 
         \bigg( - \frac{19984}{27}
          + 976 \zeta_3
           \nonumber \\&
           
          + \frac{1232}{3} \zeta_2 \bigg)
          \bigg\}

       + \mathcal{\D}_0 \overline{\mathcal{\D}}_2  \bigg\{ C_A n_f^2  
          \bigg(  \frac{16}{9} \bigg)
          \bigg\}

       + \mathcal{\D}_0 \overline{\mathcal{\D}}_2  \bigg\{ C_A^2 n_f  
         \bigg( - \frac{656}{9} \bigg)
          \bigg\}

       + \mathcal{\D}_0 \overline{\mathcal{\D}}_2   \bigg\{ C_A^3 
        \bigg(   \frac{5428}{9}
          - 192 \zeta_2 \bigg)
          \bigg\}

 \nonumber \\&

       + \mathcal{\D}_0 \overline{\mathcal D}_3  \bigg\{ C_A^2 n_f  
        \bigg(   \frac{160}{9} \bigg)
          \bigg\}

       + \mathcal{\D}_0 \overline{\mathcal D}_3   \bigg\{ C_A^3 
        \bigg(  - \frac{880}{9} \bigg)
          \bigg\}

       + \mathcal{\D}_0 \overline{\mathcal{\D}}_4   \bigg\{ C_A^3 
          \Big( 40 \Big)
          \bigg\}

       + \overline{\mathcal{\D}}_1 \mathcal{\D}_1  \bigg\{ C_A n_f^2  
          \bigg( \frac{16}{9} \bigg)
          \bigg\}
 \nonumber \\&
       + \overline{\mathcal{\D}}_1 \mathcal{\D}_1  \bigg\{ C_A^2 n_f  
        \bigg(  - \frac{656}{9} \bigg)
          \bigg\}

       + \overline{\mathcal{\D}}_1 \mathcal{\D}_1   \bigg\{ C_A^3 
          \bigg( \frac{5428}{9}
          - 192 \zeta_2 \bigg)
          \bigg\}

       + \mathcal{\D}_1 \overline{\mathcal{\D}}_2  \bigg\{ C_A^2 n_f  
         \bigg( \frac{160}{3} \bigg)
          \bigg\}
 \nonumber \\&
       + \mathcal{\D}_1 \overline{\mathcal{\D}}_2   \bigg\{ C_A^3 
          \bigg(- \frac{880}{3} \bigg)
          \bigg\}

       + \mathcal{\D}_1 \overline{\mathcal D}_3   \bigg\{ C_A^3 
         \Big(  160 \Big)
          \bigg\}

       + \overline{\mathcal{\D}}_2 \mathcal{\D}_2   \bigg\{ C_A^3 
        \Big(   120 \Big)
          \bigg\} + \big( z_1 \leftrightarrow z_2 \big) \,.
\end{align}
Here, ${\cal D}_j=\Bigg[{\ln^j(1-z_1) \over (1-z_1)}\Bigg]_+$, $\overline{{\cal D}}_j=\Bigg[{\ln^j(1-z_2) \over (1-z_2)}\Bigg]_+$, $\delta = \delta(1-z_1)$ and $\overline{\delta} = \delta(1-z_2)$.
\section{The Resummation constant $\tilde{g}_{d,g,0}^A$} \label{app:g0t}
\begin{align}
    \tilde g^A_{d,g,0}(a_s(\mu_R^2)) = \sum_{i=0}^\infty a_s^i(\mu_R^2) \tilde g^A_{d,g,0,i}\quad \,.
\end{align}
\begin{align}

\begin{autobreak}

  \tilde g^A_{d,g,0,0} =
       1
         \,,
\end{autobreak}\\
  \begin{autobreak}
 \tilde g^A_{d,g,0,1}   =

        n_f   \bigg\{
           \frac{4}{3} L_{fr}
          \bigg\}

       + C_A   \bigg\{
           8
          + 16 \zeta_2
          - \frac{22}{3} L_{fr}
          \bigg\}

       + \gamma_E C_A   \bigg\{
          - 8 L_{qr}
          + 8 L_{fr}
          \bigg\}

       + \gamma_E^2 C_A   \bigg\{
           8
          \bigg\}\,,
\end{autobreak}\\
  \begin{autobreak}
   \tilde g^A_{d,g,0,2} =

        n_f C_F   \bigg\{
          - \frac{160}{3}
          + 12 L_{rmt}
          + 16 \zeta_3
          + 4 L_{fr}
          \bigg\}

       + n_f^2   \bigg\{
           \frac{4}{3} L_{fr}^2
          \bigg\}

       + C_A n_f   \bigg\{
          - \frac{82}{3}
          - \frac{88}{9} \zeta_3
          - \frac{160}{9} \zeta_2
          + \frac{20}{3} L_{qr}
          + \frac{32}{3} L_{qr} \zeta_2
          + 16 L_{fr}
          + \frac{64}{3} L_{fr} \zeta_2
          - \frac{44}{3} L_{fr}^2
          \bigg\}

       + C_A^2   \bigg\{
           \frac{494}{3}
          - \frac{308}{9} \zeta_3
          + \frac{2224}{9} \zeta_2
          + 92 \zeta_2^2
          - \frac{92}{3} L_{qr}
          + 24 L_{qr} \zeta_3
          - \frac{176}{3} L_{qr} \zeta_2
          - 80 L_{fr}
          - 24 L_{fr} \zeta_3
          - \frac{352}{3} L_{fr} \zeta_2
          + \frac{121}{3} L_{fr}^2
          \bigg\}

       + \gamma_E C_A n_f   \bigg\{
          - \frac{224}{27}
          + \frac{80}{9} L_{qr}
          - \frac{8}{3} L_{qr}^2
          - \frac{80}{9} L_{fr}
          - \frac{32}{3} L_{fr} L_{qr}
          + \frac{40}{3} L_{fr}^2
          \bigg\}

       + \gamma_E C_A^2   \bigg\{
           \frac{1616}{27}
          - 56 \zeta_3
          - \frac{1112}{9} L_{qr}
          - 112 L_{qr} \zeta_2
          + \frac{44}{3} L_{qr}^2
          + \frac{1112}{9} L_{fr}
          + 112 L_{fr} \zeta_2
          + \frac{176}{3} L_{fr} L_{qr}
          - \frac{220}{3} L_{fr}^2
          \bigg\}

       + \gamma_E^2 C_A n_f   \bigg\{
          - \frac{80}{9}
          + \frac{16}{3} L_{qr}
          + \frac{32}{3} L_{fr}
          \bigg\}

       + \gamma_E^2 C_A^2   \bigg\{
           \frac{1112}{9}
          + 112 \zeta_2
          - \frac{88}{3} L_{qr}
          + 32 L_{qr}^2
          - \frac{176}{3} L_{fr}
          - 64 L_{fr} L_{qr}
          + 32 L_{fr}^2
          \bigg\}

       + \gamma_E^3 C_A n_f   \bigg\{
          - \frac{32}{9}
          \bigg\}

       + \gamma_E^3 C_A^2   \bigg\{
           \frac{176}{9}
          - 64 L_{qr}
          + 64 L_{fr}
          \bigg\}

       + \gamma_E^4 C_A^2   \bigg\{
           32
          \bigg\}\,,
\end{autobreak}\\
  \begin{autobreak}
   \tilde g^A_{d,g,0,3} =

        n_f   \bigg\{
          - 4 C_J^{(2)}
          \bigg\}

       + n_f C_F^2   \bigg\{
           \frac{457}{3}
          - 320 \zeta_5
          + 208 \zeta_3
          - 2 L_{fr}
          \bigg\}

       + n_f^2 C_F   \bigg\{
           \frac{1498}{9}
          - \frac{16}{9} \zeta_4
          - \frac{224}{3} \zeta_3
          - \frac{40}{9} \zeta_2
          - \frac{640}{9} L_{qr}
          + 16 L_{qr} L_{rmt}
          + \frac{64}{3} L_{qr} \zeta_3
          - \frac{662}{9} L_{fr}
          + 16 L_{fr} L_{rmt}
          + \frac{64}{3} L_{fr} \zeta_3
          + \frac{28}{3} L_{fr}^2
          \bigg\}

       + n_f^3   \bigg\{
           \frac{32}{27} L_{fr}^3
          \bigg\}

       + C_A n_f C_F   \bigg\{
          - 1797
          + 96 L_{rmt}
          + \frac{88}{9} \zeta_4
          + 160 \zeta_5
          + \frac{1792}{3} \zeta_3
          - \frac{8660}{9} \zeta_2
          + 192 \zeta_2 L_{rmt}
          + 384 \zeta_2 \zeta_3
          + \frac{3628}{9} L_{qr}
          - 88 L_{qr} L_{rmt}
          - \frac{352}{3} L_{qr} \zeta_3
          + 32 L_{qr} \zeta_2
          + \frac{4049}{9} L_{fr}
          - 88 L_{fr} L_{rmt}
          - \frac{352}{3} L_{fr} \zeta_3
          + 64 L_{fr} \zeta_2
          - \frac{154}{3} L_{fr}^2
          \bigg\}

       + C_A n_f^2   \bigg\{
           \frac{6914}{81}
          - \frac{304}{9} \zeta_4
          + \frac{3344}{81} \zeta_3
          - \frac{896}{81} \zeta_2
          - \frac{64}{45} \zeta_2^2
          - \frac{976}{27} L_{qr}
          - \frac{352}{27} L_{qr} \zeta_3
          - \frac{640}{27} L_{qr} \zeta_2
          + \frac{40}{9} L_{qr}^2
          + \frac{64}{9} L_{qr}^2 \zeta_2
          - \frac{119}{3} L_{fr}
          - \frac{352}{27} L_{fr} \zeta_3
          - \frac{640}{27} L_{fr} \zeta_2
          + \frac{80}{9} L_{fr} L_{qr}
          + \frac{128}{9} L_{fr} L_{qr} \zeta_2
          + \frac{212}{9} L_{fr}^2
          + \frac{64}{3} L_{fr}^2 \zeta_2
          - \frac{176}{9} L_{fr}^3
          \bigg\}

       + C_A^2 n_f   \bigg\{
          - \frac{113366}{81}
          + \frac{10516}{27} \zeta_4
          + \frac{808}{9} \zeta_5
          - \frac{7336}{81} \zeta_3
          - \frac{50768}{81} \zeta_2
          - \frac{784}{3} \zeta_2 \zeta_3
          - \frac{8576}{45} \zeta_2^2
          + \frac{13064}{27} L_{qr}
          - \frac{736}{27} L_{qr} \zeta_3
          + \frac{14288}{27} L_{qr} \zeta_2
          + 120 L_{qr} \zeta_2^2
          - \frac{404}{9} L_{qr}^2
          + 16 L_{qr}^2 \zeta_3
          - \frac{704}{9} L_{qr}^2 \zeta_2
          + \frac{4397}{9} L_{fr}
          + \frac{2144}{27} L_{fr} \zeta_3
          + \frac{14864}{27} L_{fr} \zeta_2
          + \frac{376}{3} L_{fr} \zeta_2^2
          - \frac{808}{9} L_{fr} L_{qr}
          + 32 L_{fr} L_{qr} \zeta_3
          - \frac{1408}{9} L_{fr} L_{qr} \zeta_2
          - \frac{2146}{9} L_{fr}^2
          - 48 L_{fr}^2 \zeta_3
          - \frac{704}{3} L_{fr}^2 \zeta_2
          + \frac{968}{9} L_{fr}^3
          \bigg\}

       + C_A^3   \bigg\{
           \frac{114568}{27}
          - \frac{8012}{3} \zeta_6
          - \frac{30514}{27} \zeta_4
          + \frac{3476}{9} \zeta_5
          - \frac{158620}{81} \zeta_3
          + 96 \zeta_3^2
          + \frac{345064}{81} \zeta_2
          - \frac{2024}{3} \zeta_2 \zeta_3
          + \frac{6224}{3} \zeta_2^2
          + \frac{6080}{7} \zeta_2^3
          - \frac{36064}{27} L_{qr}
          - 160 L_{qr} \zeta_5
          + \frac{21608}{27} L_{qr} \zeta_3
          - \frac{54256}{27} L_{qr} \zeta_2
          + 352 L_{qr} \zeta_2 \zeta_3
          - 660 L_{qr} \zeta_2^2
          + \frac{1012}{9} L_{qr}^2
          - 88 L_{qr}^2 \zeta_3
          + \frac{1936}{9} L_{qr}^2 \zeta_2
          - \frac{13115}{9} L_{fr}
          + 160 L_{fr} \zeta_5
          - \frac{8056}{27} L_{fr} \zeta_3
          - \frac{58288}{27} L_{fr} \zeta_2
          - 352 L_{fr} \zeta_2 \zeta_3
          - \frac{2068}{3} L_{fr} \zeta_2^2
          + \frac{2024}{9} L_{fr} L_{qr}
          - 176 L_{fr} L_{qr} \zeta_3
          + \frac{3872}{9} L_{fr} L_{qr} \zeta_2
          + \frac{5390}{9} L_{fr}^2
          + 264 L_{fr}^2 \zeta_3
          + \frac{1936}{3} L_{fr}^2 \zeta_2
          - \frac{5324}{27} L_{fr}^3
          \bigg\}

       + \gamma_E C_A n_f C_F   \bigg\{
          - \frac{3422}{27}
          + \frac{608}{9} \zeta_3
          + \frac{64}{5} \zeta_2^2
          + 500 L_{qr}
          - 96 L_{qr} L_{rmt}
          - 192 L_{qr} \zeta_3
          - 8 L_{qr}^2
          - 500 L_{fr}
          + 96 L_{fr} L_{rmt}
          + 192 L_{fr} \zeta_3
          - 32 L_{fr} L_{qr}
          + 40 L_{fr}^2
          \bigg\}

       + \gamma_E C_A n_f^2   \bigg\{
           \frac{3712}{729}
          + \frac{64}{9} \zeta_3
          - \frac{800}{81} L_{qr}
          + \frac{160}{27} L_{qr}^2
          - \frac{32}{27} L_{qr}^3
          - \frac{992}{81} L_{fr}
          + \frac{320}{27} L_{fr} L_{qr}
          - \frac{32}{9} L_{fr} L_{qr}^2
          - \frac{160}{9} L_{fr}^2
          - \frac{32}{3} L_{fr}^2 L_{qr}
          + \frac{416}{27} L_{fr}^3
          \bigg\}

       + \gamma_E C_A^2 n_f   \bigg\{
          - \frac{173636}{729}
          + \frac{1808}{27} \zeta_3
          - \frac{9104}{81} \zeta_2
          - \frac{32}{5} \zeta_2^2
          + \frac{39880}{81} L_{qr}
          + \frac{704}{9} L_{qr} \zeta_3
          + \frac{2240}{9} L_{qr} \zeta_2
          - \frac{4328}{27} L_{qr}^2
          - \frac{352}{3} L_{qr}^2 \zeta_2
          + \frac{352}{27} L_{qr}^3
          - \frac{17096}{81} L_{fr}
          - \frac{2048}{9} L_{fr} \zeta_3
          - \frac{2240}{9} L_{fr} \zeta_2
          - \frac{5920}{27} L_{fr} L_{qr}
          - 64 L_{fr} L_{qr} \zeta_2
          + \frac{352}{9} L_{fr} L_{qr}^2
          + \frac{3416}{9} L_{fr}^2
          + \frac{544}{3} L_{fr}^2 \zeta_2
          + \frac{352}{3} L_{fr}^2 L_{qr}
          - \frac{4576}{27} L_{fr}^3
          \bigg\}

       + \gamma_E C_A^3   \bigg\{
           \frac{943114}{729}
          + 384 \zeta_5
          - \frac{36752}{27} \zeta_3
          + \frac{64784}{81} \zeta_2
          - \frac{2336}{3} \zeta_2 \zeta_3
          - \frac{176}{5} \zeta_2^2
          - \frac{207308}{81} L_{qr}
          + \frac{5632}{9} L_{qr} \zeta_3
          - \frac{23072}{9} L_{qr} \zeta_2
          - \frac{2752}{5} L_{qr} \zeta_2^2
          + \frac{16912}{27} L_{qr}^2
          - 192 L_{qr}^2 \zeta_3
          + \frac{1936}{3} L_{qr}^2 \zeta_2
          - \frac{968}{27} L_{qr}^3
          + \frac{136204}{81} L_{fr}
          + \frac{1760}{9} L_{fr} \zeta_3
          + \frac{23072}{9} L_{fr} \zeta_2
          + \frac{2752}{5} L_{fr} \zeta_2^2
          + \frac{22448}{27} L_{fr} L_{qr}
          + 384 L_{fr} L_{qr} \zeta_3
          + 352 L_{fr} L_{qr} \zeta_2
          - \frac{968}{9} L_{fr} L_{qr}^2
          - \frac{13120}{9} L_{fr}^2
          - 192 L_{fr}^2 \zeta_3
          - \frac{2992}{3} L_{fr}^2 \zeta_2
          - \frac{968}{3} L_{fr}^2 L_{qr}
          + \frac{12584}{27} L_{fr}^3
          \bigg\}

       + \gamma_E^2 C_A n_f C_F   \bigg\{
          - 500
          + 96 L_{rmt}
          + 192 \zeta_3
          + 16 L_{qr}
          + 32 L_{fr}
          \bigg\}

       + \gamma_E^2 C_A n_f^2   \bigg\{
           \frac{800}{81}
          - \frac{320}{27} L_{qr}
          + \frac{32}{9} L_{qr}^2
          - \frac{320}{27} L_{fr}
          + \frac{64}{9} L_{fr} L_{qr}
          + \frac{32}{3} L_{fr}^2
          \bigg\}

       + \gamma_E^2 C_A^2 n_f   \bigg\{
          - \frac{39880}{81}
          - \frac{704}{9} \zeta_3
          - \frac{2240}{9} \zeta_2
          + \frac{9008}{27} L_{qr}
          + \frac{448}{3} L_{qr} \zeta_2
          - \frac{992}{9} L_{qr}^2
          + \frac{64}{3} L_{qr}^3
          + \frac{1856}{9} L_{fr}
          + \frac{448}{3} L_{fr} \zeta_2
          + 64 L_{fr} L_{qr}
          + \frac{64}{3} L_{fr} L_{qr}^2
          - \frac{1696}{9} L_{fr}^2
          - \frac{320}{3} L_{fr}^2 L_{qr}
          + 64 L_{fr}^3
          \bigg\}

       + \gamma_E^2 C_A^3   \bigg\{
           \frac{207308}{81}
          - \frac{5632}{9} \zeta_3
          + \frac{23072}{9} \zeta_2
          + \frac{2752}{5} \zeta_2^2
          - \frac{13376}{9} L_{qr}
          + 640 L_{qr} \zeta_3
          - \frac{2464}{3} L_{qr} \zeta_2
          + 840 L_{qr}^2
          + 384 L_{qr}^2 \zeta_2
          - \frac{352}{3} L_{qr}^3
          - \frac{16144}{27} L_{fr}
          - 640 L_{fr} \zeta_3
          - \frac{2464}{3} L_{fr} \zeta_2
          - \frac{11248}{9} L_{fr} L_{qr}
          - 768 L_{fr} L_{qr} \zeta_2
          - \frac{352}{3} L_{fr} L_{qr}^2
          + \frac{9496}{9} L_{fr}^2
          + 384 L_{fr}^2 \zeta_2
          + \frac{1760}{3} L_{fr}^2 L_{qr}
          - 352 L_{fr}^3
          \bigg\}

       + \gamma_E^3 C_A n_f C_F   \bigg\{
          - \frac{32}{3}
          \bigg\}

       + \gamma_E^3 C_A n_f^2   \bigg\{
          \frac{640}{81}
          - \frac{128}{27} L_{qr}
          - \frac{128}{27} L_{fr}
          \bigg\}

       + \gamma_E^3 C_A^2 n_f   \bigg\{
          - \frac{16928}{81}
          - \frac{128}{3} \zeta_2
          + \frac{5248}{27} L_{qr}
          - 64 L_{qr}^2
          - \frac{2432}{27} L_{fr}
          - \frac{128}{3} L_{fr} L_{qr}
          + \frac{320}{3} L_{fr}^2
          \bigg\}

       + \gamma_E^3 C_A^3   \bigg\{
           \frac{79936}{81}
          - 448 \zeta_3
          + \frac{704}{3} \zeta_2
          - \frac{43424}{27} L_{qr}
          - 768 L_{qr} \zeta_2
          + 352 L_{qr}^2
          - \frac{256}{3} L_{qr}^3
          + \frac{35680}{27} L_{fr}
          + 768 L_{fr} \zeta_2
          + \frac{704}{3} L_{fr} L_{qr}
          + 256 L_{fr} L_{qr}^2
          - \frac{1760}{3} L_{fr}^2
          - 256 L_{fr}^2 L_{qr}
          + \frac{256}{3} L_{fr}^3
          \bigg\}

       + \gamma_E^4 C_A n_f^2   \bigg\{
          \frac{64}{27}
          \bigg\}

       + \gamma_E^4 C_A^2 n_f   \bigg\{
          - \frac{2624}{27}
          + \frac{640}{9} L_{qr}
          + \frac{128}{9} L_{fr}
          \bigg\}

       + \gamma_E^4 C_A^3   \bigg\{
           \frac{21712}{27}
          + 384 \zeta_2
          - \frac{3520}{9} L_{qr}
          + 256 L_{qr}^2
          - \frac{704}{9} L_{fr}
          - 512 L_{fr} L_{qr}
          + 256 L_{fr}^2
          \bigg\}

       + \gamma_E^5 C_A^2 n_f   \bigg\{
          - \frac{256}{9}
          \bigg\}

       + \gamma_E^5 C_A^3   \bigg\{
           \frac{1408}{9}
          - 256 L_{qr}
          + 256 L_{fr}
          \bigg\}

       + \gamma_E^6 C_A^3   \bigg\{
          + \frac{256}{3}
          \bigg\}\,.
          \end{autobreak}
 \end{align}
                                                    
\section{NSV Resummation exponents $   \overline{g}_{d,g,i}^A(\omega)$} \label{app:gbdN}
\begin{align}
\begin{autobreak}

  \overline{g}_{d,g,1}^A(\omega) = 0 
  
  \quad \,,
   \overline{g}_{d,g,2}^A(\omega) =

        \frac{1}{\beta_0}C_A   \bigg\{
          2 L_\omega
          \bigg\}\,,
   
\end{autobreak}\\
  \begin{autobreak}
   \overline{g}_{d,g,3}^A(\omega) =

       \frac{\beta_1}{\beta_0^2} C_A   \bigg\{
           2 \omega
          + 2 L_\omega
          \bigg\}

       +\frac{1}{\beta_0}C_A n_f   \bigg\{
           \frac{20}{9} \omega
          \bigg\}

       +\frac{1}{\beta_0}C_A^2   \bigg\{
          - \frac{134}{9} \omega
          + 4 \omega \zeta_2
          \bigg\}

       + C_A   \bigg\{
          - 2
          + 2 L_{qr}
          - 2 L_{fr}
          + 2 L_{fr} \omega
          - 4 \gamma_E
          \bigg\}\,,
\end{autobreak}\\
  \begin{autobreak}

   \overline{g}_{d,g,4}^A(\omega) =

       \frac{\beta_1^2}{\beta_0^3} C_A   \bigg\{
          \omega^2
          - L_\omega^2
          \bigg\}

       + \frac{\beta_2}{\beta_0^2} C_A   \bigg\{
          - \omega^2
          \bigg\}

       + \frac{\beta_1}{\beta_0^2} C_A n_f   \bigg\{
          - \frac{20}{9} \omega
          + \frac{10}{9} \omega^2
          - \frac{20}{9} L_\omega
          \bigg\}

       + \frac{\beta_1}{\beta_0^2} C_A^2   \bigg\{
           \frac{134}{9} \omega
          - 4 \omega \zeta_2
          - \frac{67}{9} \omega^2
          + 2 \omega^2 \zeta_2
          + \frac{134}{9} L_\omega
          - 4 L_\omega \zeta_2
          \bigg\}

       +\frac{1}{\beta_0}C_A n_f^2   \bigg\{
           \frac{8}{27} \omega
          - \frac{4}{27} \omega^2
          \bigg\}

       +\frac{1}{\beta_0}C_A C_F n_f   \bigg\{
           \frac{55}{3} \omega
          - 16 \omega \zeta_3
          - \frac{55}{6} \omega^2
          + 8 \omega^2 \zeta_3
          \bigg\}

       +\frac{1}{\beta_0}C_A^2 n_f   \bigg\{
           \frac{418}{27} \omega
          + \frac{56}{3} \omega \zeta_3
          - \frac{80}{9} \omega \zeta_2
          - \frac{209}{27} \omega^2
          - \frac{28}{3} \omega^2 \zeta_3
          + \frac{40}{9} \omega^2 \zeta_2
          \bigg\}

       +\frac{1}{\beta_0}C_A^3   \bigg\{
          - \frac{245}{3} \omega
          - \frac{44}{3} \omega \zeta_3
          + \frac{536}{9} \omega \zeta_2
          - \frac{88}{5} \omega \zeta_2^2
          + \frac{245}{6} \omega^2
          + \frac{22}{3} \omega^2 \zeta_3
          - \frac{268}{9} \omega^2 \zeta_2
          + \frac{44}{5} \omega^2 \zeta_2^2
          \bigg\}

       + \frac{\beta_1}{\beta_0} C_A   \bigg\{
           2 L_\omega
          - 2 L_\omega L_{qr}
          + 4 L_\omega \gamma_E
          \bigg\}

       + C_A n_f   \bigg\{
           \frac{116}{27}
          - \frac{2}{3} \zeta_2
          - \frac{20}{9} L_{qr}
          + \frac{20}{9} L_{fr}
          - \frac{40}{9} L_{fr} \omega
          + \frac{20}{9} L_{fr} \omega^2
          + \frac{40}{9} \gamma_E
          \bigg\}

       + C_A^2   \bigg\{
          - \frac{806}{27}
          + 14 \zeta_3
          + \frac{23}{3} \zeta_2
          + \frac{134}{9} L_{qr}
          - 4 L_{qr} \zeta_2
          - \frac{134}{9} L_{fr}
          + 4 L_{fr} \zeta_2
          + \frac{268}{9} L_{fr} \omega
          - 8 L_{fr} \omega \zeta_2
          - \frac{134}{9} L_{fr} \omega^2
          + 4 L_{fr} \omega^2 \zeta_2
          - \frac{268}{9} \gamma_E
          + 8 \gamma_E \zeta_2
          \bigg\}

       + \beta_0 C_A   \bigg\{
          - \zeta_2
          + 2 L_{qr}
          - L_{qr}^2
          + L_{fr}^2
          - 2 L_{fr}^2 \omega
          + L_{fr}^2 \omega^2
          - 4 \gamma_E
          + 4 \gamma_E L_{qr}
          - 4 \gamma_E^2
          \bigg\}\,,
\end{autobreak}
\end{align}

\section{NSV Resummation exponents $  h^A_{d,g,ij}(\omega)$ and $  \tilde{h}^A_{d,g,ij}(\omega,\omega_l)$  } \label{app:hdN}
\begin{align}
\begin{autobreak}  
   h^A_{d,g,00}(\omega) =

        \frac{1}{\beta_0}  C_A   \bigg\{
          - 4 L_\omega
          \bigg\} \quad 
   
   h^A_{d,g,01}(\omega) = 0\,,
   
\end{autobreak}\\ 
  \begin{autobreak}
   h^A_{d,g,10}(\omega) =

       \frac{1}{2 \beta_0^2 (\omega -1)} \Bigg[    \beta_1 C_A   \bigg\{
           8 \omega
          + 8 L_\omega
          \bigg\}

       + \beta_0 C_A n_f   \bigg\{
           \frac{80}{9} \omega
          \bigg\}

       + \beta_0 C_A^2   \bigg\{
          - \frac{536}{9} \omega
          + 16 \omega \zeta_2
          - 32 \gamma_E \omega
          \bigg\}

       + \beta_0^2 C_A   \bigg\{
          - 4
          - 8 L_{fr}
          + 8 L_{fr} \omega
          + 8 L_{qr}
          - 16 \gamma_E 
          \bigg\}\Bigg]\,,

 \end{autobreak}\\ 
  \begin{autobreak}
  \tilde{h}^A_{d,g,11}(\omega, \omega_l) =
  \frac{ C_A^2}{\beta_0 }   \bigg\{
           -\frac{4 \omega_l}{(\omega -1)^2}
          - \frac{16 \omega}{(\omega -1)} 
          \bigg\}\,,
   


  \end{autobreak}\\ 
  \begin{autobreak}
   h^A_{d,g,20}(\omega) =

      \frac{1}{2 \beta_0^3 (\omega -1)^2} \Bigg[     \beta_1^2 C_A   \bigg\{
          - 4 \omega^2
          + 4 L_\omega^2
          \bigg\}

       + \beta_0 \beta_2 C_A   \bigg\{
           4 \omega^2
          \bigg\}

       + \beta_0 \beta_1 C_A n_f   \bigg\{
           \frac{80}{9} \omega
          - \frac{40}{9} \omega^2
          + \frac{80}{9} L_\omega
          \bigg\}

       + \beta_0 \beta_1 C_A^2   \bigg\{
          - \frac{536}{9} \omega
          + 16 \omega \zeta_2
          + \frac{268}{9} \omega^2
          - 8 \omega^2 \zeta_2
          - 32 \gamma_E \omega
          + 16 \gamma_E \omega^2
          - \frac{536}{9} L_\omega
          + 16 L_\omega \zeta_2
          - 32 L_\omega \gamma_E
          \bigg\}

       + \beta_0^2 C_A n_f^2   \bigg\{
          - \frac{32}{27} \omega
          + \frac{16}{27} \omega^2
          \bigg\}

       + \beta_0^2 C_A C_F n_f   \bigg\{
          - \frac{172}{3} \omega
          + 64 \omega \zeta_3
          + \frac{86}{3} \omega^2
          - 32 \omega^2 \zeta_3
          \bigg\}

       + \beta_0^2 C_A^2 n_f   \bigg\{
          - \frac{1096}{27} \omega
          - \frac{224}{3} \omega \zeta_3
          + \frac{320}{9} \omega \zeta_2
          + \frac{548}{27} \omega^2
          + \frac{112}{3} \omega^2 \zeta_3
          - \frac{160}{9} \omega^2 \zeta_2
          - \frac{640}{9} \gamma_E \omega
          + \frac{320}{9} \gamma_E \omega^2
          \bigg\}

       + \beta_0^2 C_A^3   \bigg\{
           \frac{724}{3} \omega
          - \frac{112}{3} \omega \zeta_3
          - \frac{2144}{9} \omega \zeta_2
          + \frac{352}{5} \omega \zeta_2^2
          - \frac{362}{3} \omega^2
          + \frac{56}{3} \omega^2 \zeta_3
          + \frac{1072}{9} \omega^2 \zeta_2
          - \frac{176}{5} \omega^2 \zeta_2^2
          + \frac{4288}{9} \gamma_E \omega
          - 128 \gamma_E \omega \zeta_2
          - \frac{2144}{9} \gamma_E \omega^2
          + 64 \gamma_E \omega^2 \zeta_2
          \bigg\}

       + \beta_0^2 \beta_1 C_A   \bigg\{
           8 \omega
          - 4 \omega^2
          - 4 L_\omega
          + 8 L_\omega L_{qr}
          - 16 L_\omega \gamma_E
          \bigg\}

       + \beta_0^3 C_A n_f   \bigg\{
          - \frac{272}{27}
          + \frac{32}{3} \zeta_2
          - \frac{80}{9} L_{fr}
          + \frac{160}{9} L_{fr} \omega
          - \frac{80}{9} L_{fr} \omega^2
          + \frac{80}{9} L_{qr}
          - \frac{148}{9} \gamma_E
          \bigg\}

       + \beta_0^3 C_A^2   \bigg\{
           \frac{1808}{27}
          - 56 \zeta_3
          - \frac{224}{3} \zeta_2
          + \frac{536}{9} L_{fr}
          - 16 L_{fr} \zeta_2
          - \frac{1072}{9} L_{fr} \omega
          + 32 L_{fr} \omega \zeta_2
          + \frac{536}{9} L_{fr} \omega^2
          - 16 L_{fr} \omega^2 \zeta_2
          - \frac{536}{9} L_{qr}
          + 16 L_{qr} \zeta_2
          + \frac{1060}{9} \gamma_E
          - 32 \gamma_E \zeta_2
          + 32 \gamma_E L_{fr}
          - 64 \gamma_E L_{fr} \omega
          + 32 \gamma_E L_{fr} \omega^2
          - 32 \gamma_E L_{qr}
          + 56 \gamma_E^2
          \bigg\}

       + \beta_0^4 C_A   \bigg\{
           16 \zeta_2
          - 4 L_{fr}^2
          + 8 L_{fr}^2 \omega
          - 4 L_{fr}^2 \omega^2
          - 4 L_{qr}
          + 4 L_{qr}^2
          + 8 \gamma_E
          - 16 \gamma_E L_{qr}
          + 16 \gamma_E^2
          \bigg\}\Bigg]\,,
   
\end{autobreak}\\ 
  \begin{autobreak}
   h^A_{d,g,21}(\omega) =

  \frac{1}{2 \beta_0^2 (\omega -1)^2} \Bigg[
        \beta_1 C_A^2   \bigg\{
          - 32 \omega
          + 16 \omega^2
          - 32 L_\omega
          \bigg\}

       + \beta_0 C_A^2 n_f   \bigg\{
          - \frac{640}{9} \omega
          + \frac{320}{9} \omega^2
          \bigg\}

       + \beta_0 C_A^3   \bigg\{
           \frac{4288}{9} \omega
          - 128 \omega \zeta_2
          - \frac{2144}{9} \omega^2
          + 64 \omega^2 \zeta_2
          \bigg\}

       + \beta_0^2 C_A n_f   \bigg\{
           \frac{4}{3}
          \bigg\}

       + \beta_0^2 C_A^2   \bigg\{
          - \frac{4}{3}
          + 32 L_{fr}
          - 64 L_{fr} \omega
          + 32 L_{fr} \omega^2
          - 32 L_{qr}
          + 48 \gamma_E
          \bigg\} \Bigg]\,,

\end{autobreak}\\ 
  \begin{autobreak}


           \tilde{h}^g_{d,22}(\omega, \omega_l) =
        \frac{\omega_l}{\beta_0 (\omega -1)^3} \Bigg[ C_A^2 n_f   \bigg\{
          \frac{32}{27} 
          \bigg\} + 
         C_A^3   \bigg\{
          - \frac{176}{27}
          \bigg\} \Bigg] \,,

\end{autobreak}
\end{align}
where $\gamma_E$ is the Euler-Mascheroni constant. Here, ${L}_{\omega}=\ln(1-\omega)$ with $\omega = \beta_0 a_s(\mu_R^2) \ln N_1 N_2$, $\omega_l = \beta_0 a_s(\mu_R^2) \ln N_l$ with $l=1,2$, $L_{qr} = \ln \big(\frac{q^2}{\mu_R^2}\big)$ and $L_{fr} = \ln \big(\frac{\mu_F^2}{\mu_R^2}\big)$.
\bibliographystyle{JHEP}
\bibliography{main.bib}

\end{document}